\documentclass[twocolumn]{aastex7}
\usepackage{booktabs}
\usepackage{amsmath}

\newcommand{\rc}[1]{{\color{black}#1}}
\newcommand{\rcnew}[1]{{\color{black}#1}}

\begin{document}

\title{TESS Planet Occurrence Rates Reveal the Disappearance of the Radius Valley Around Mid-to-Late M Dwarfs}

\author[orcid=0009-0000-4040-6628,sname='Gillis']{Erik Diego Gillis}
\affiliation{Department of Physics \& Astronomy, McMaster University, 1280 Main St West, Hamilton, ON, L8S 4L8, Canada}
\email[show]{gillie1@mcmaster.ca}  

\author[orcid=0000-0001-5383-9393]{Ryan Cloutier}
\affiliation{Department of Physics \& Astronomy, McMaster University, 1280 Main St West, Hamilton, ON, L8S 4L8, Canada}
\email{ryan.cloutier@mcmaster.ca}  

\submitjournal{AJ
}\accepted{February 26 2026}
\revised{March 27 2026}

\author[orcid=0000-0002-1533-9029]{Emily K. Pass}
\altaffiliation{Juan Carlos Torres Postdoctoral Fellow}
\affiliation{Kavli Institute for Astrophysics and Space Research, Massachusetts Institute of Technology, Cambridge, MA 02139, USA}
\email{epass@mit.edu}  

\newcommand{\ntargets}{8134 }
\newcommand{\SEocc}{$0.954\pm0.147$ }
\newcommand{\SNocc}{$0.148\pm0.045$ }
\newcommand{\cumocc}{$1.10\pm0.16$ }
\defcitealias{cloutier_evolution_2020}{CM20}
\defcitealias{Dressing_2015}{DC15}
\defcitealias{ment_occurrence_2023}{MC23}
\defcitealias{fulton_california-kepler_2017}{F17}
\defcitealias{FP2018}{FP18}

\begin{abstract}
We present the deepest systematic search for planets around mid-to-late M dwarfs to date. We have surveyed \ntargets mid-to-late M dwarfs observed by TESS with a custom built pipeline and recover 77 vetted transiting planet candidates. We characterize the sensitivity of our survey via injection--recovery and measure the occurrence rate of planets as a function of orbital period, instellation, and planet radius. We measure a cumulative occurrence rate of \cumocc planets per star with radii $>1\, R_\oplus$ orbiting within 30 days. \rcnew{This value is consistent with the cumulative occurrence rate around early M dwarfs, making M dwarfs collectively the most prolific hosts of small close-in planets}. Unlike the bimodal Radius Valley exhibited by close-in planet population around FGK and early M dwarfs, we recover a unimodal planet radius distribution peaking at $1.25\pm0.05 \, R_\oplus$. We additionally find \SEocc super-Earths and \SNocc sub-Neptunes per star, with super-Earths outnumbering sub-Neptunes 5.5:1, firmly demonstrating that the Radius Valley disappears around the lowest mass stars. The dearth of sub-Neptunes around mid-to-late M dwarfs is consistent with predictions from water-rich pebble accretion models that predict a fading Radius Valley with decreasing stellar mass. Our results support the emerging idea that the sub-Neptune population around M dwarfs is composed of water-rich worlds. \rcnew{We find no hot Jupiters in our survey and set an upper limit of 0.012 hot Jupiters per mid-to-late M dwarf within 10 days.}
\end{abstract}


\section{Introduction}
 The exoplanet community has recently surpassed the discovery of over 6000 exoplanets. This collection is mostly comprised from thousands of exoplanets uncovered by large-scale transit surveys, which have revolutionized our understanding of the planet population. The bulk of confirmed transiting planets have been discovered by the Kepler space telescope, which has provided a clear picture of the rates of close-in planets around FGK and early M type stars \citep{howard2012, Dressing_2015, fulton_california-kepler_2017, cloutier_evolution_2020}. Despite Kepler's unprecedented and significant contributions, the telescope's focus on a single field and relatively blue bandpass \citep[420-900 nm;][]{kepler_mission} limited Kepler's ability to provide detailed characterization of the planet population around mid-to-late M dwarfs due to their red spectra. These stars have $\lesssim 0.4\,R_\odot$, and represent the most common type of star in the galaxy \citep{winters_most_common}. Previous works have provided constraints on the cumulative sub-Neptune and super-Earth occurrence rates around mid-to-late M dwarfs using data from Kepler and K2 \citep{hardegree2019,HU2025}, but a more detailed characterization of the planet population around these remains limited with Kepler.

In 2018, the Transiting Exoplanet Survey Satellite \citep[TESS;][]{tess_ricker} launched. TESS is enabling the investigation of the planet population around mid-to-late M dwarfs in two key ways. Its large field-of-view and full sky access allows it to observe nearly every nearby mid-to-late M dwarf whereas Kepler and K2 were limited to roughly one thousand mid-to-late M dwarfs with stellar mass $<0.4\,M_\odot$. Additionally, TESS's bandpass is significantly redder than Kepler's (600-1000 nm), which improves the photometric precision by better covering the spectral energy distributions of these cool, red stars. This has enabled more detailed investigations into the occurrence rates of planets around the smallest stars, most recently by \citet[][hereafter \citetalias{ment_occurrence_2023}]{ment_occurrence_2023} who characterized the exoplanet population around mid-to-late M dwarfs out to 7 days.

\rc{Kepler and TESS have provided a detailed window into the occurrence of planets with radii larger than Earth.} This planet population around Sun-like stars \citep{fulton_california-kepler_2017}, late K and early M dwarfs alike \citep[][hereafter \citetalias{cloutier_evolution_2020}]{cloutier_evolution_2020} is dominated by super-Earths and sub-Neptunes, with a dearth separating these two populations known as the Radius Valley. The ubiquity of this feature across spectral types has motivated the development of planet formation theories to explain the Radius Valley's properties and the evolution of its features---most notably the slope separating the planet populations in period--radius space. Around FGK stars, thermally driven atmospheric escape through photoevaporation \citep{owen_evaporation_2017} and core-powered mass loss  \citep{gupta_sculpting_2019} have been shown to explain the Radius Valley's properties. Conversely, there is a growing body of observational evidence that the Radius Valley around early M dwarfs is instead sculpted by the effects of water-rich formation and migration \citep[e.g.][]{cloutier_evolution_2020, diamond-lowe_2022, luque_2022, cherubim_2023, piaulet_2023, gaidos_2024, icy_ho_2024}. The planet population and any subsequent Radius Valley remains to be robustly characterized around the lowest mass main-sequence stars beyond planets orbiting within 7 days \citepalias{ment_occurrence_2023}.

Here we present the results of a large-scale transit survey of \ntargets mid-to-late M dwarfs with TESS to characterize the occurrence rates of terrestrial to Jovian-sized objects around mid-to-late M dwarfs. We survey our targets using a custom-built pipeline and employ detailed injection--recovery tests to characterize our survey's sensitivity. By combining our sample of found planets with the completeness of our survey, we derive a robust understanding of the planet population around the most common stars. 

Our paper is structured as follows: In Section \ref{sec:stel-samp}, we present our sample of stars, the cuts we apply to ensure that our sample is free from contamination, and our homogeneous stellar characterization. In Section \ref{sec:pipeline} we present our custom transit search pipeline to find and characterize transiting planets, along with the injection--recovery tests necessary to characterize our pipeline's sensitivity. In Section \ref{sec:planet_search}, we present our sample of recovered planets and in Section \ref{sec:occrate} we combine this sample with the completeness of our survey to recover the occurrence rate of planets around mid-to-late M dwarfs out to 30 days. Finally, we present this work in context and highlight our conclusions with Sections \ref{sec:discussion} and \ref{sec:concl}.

\section{Target Sample and Stellar Characterization}\label{sec:stel-samp}
Our initial stellar sample for our survey follows from the targets proposed in TESS Guest Investigator (GI) Programs 
G03274\footnote{TESS Cycle 3, \textit{Understanding The Physical Origin Of The Rocky/Non-Rocky Transition Around Mid-To-Late M Dwarfs With TESS}, PI: Cloutier.},
G04214\footnote{TESS Cycle 4, \textit{The Physical Origin of the Rocky/Enveloped Transition Around Mid-To-Late M Dwarfs}, PI: Cloutier.},
and G05152\footnote{TESS Cycle 5, \textit{Establishing The Physical Origin of the Rocky/Enveloped Transition Around Mid-To-Late M Dwarfs with TESS}, PI: Cloutier.}.
These programs aimed to characterize the demographics of the exoplanet population around mid-to-late M dwarfs to determine the dominant physical mechanism responsible for sculpting the Radius Valley in this stellar mass regime. The original survey sample requested 2-minute cadence light curves of  11,103 mid-to-late M dwarfs. Their target list comprised the set of known mid-to-late M dwarfs within 100 pc with mass $M <  0.45 M_\odot$ derived with the relation of \citet{mann2019} and expected $\mathrm{S/N}>7.3$ for a fiducial transiting Radius Valley planet using the expected noise described in \citet{sullivan_tess_noise}. The mass cut in the initial target selection is conservative compared to the final mass cut in our survey, and the signal-to-noise cut effectively acts as a stellar radius dependent magnitude cut.

\begin{figure}[h]
    \centering
    \includegraphics[width=\linewidth]{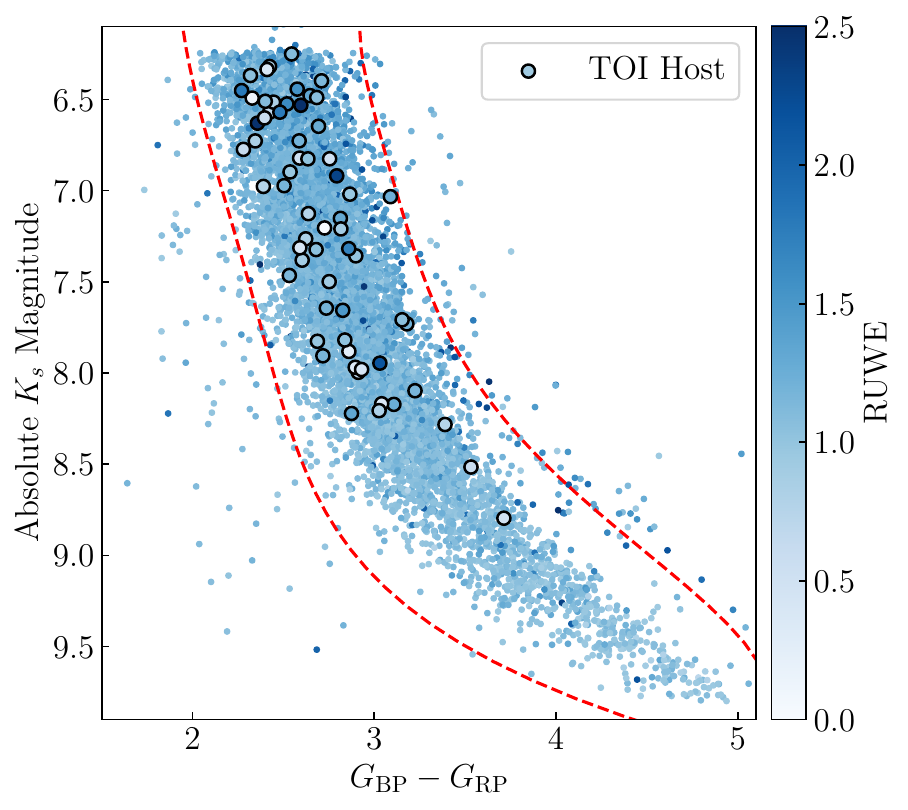}
    \caption{$M_{K_s}$/$G_{\mathrm{BP}}-G_{\mathrm{RP}}$ color-magnitude diagram of our original target sample, with color cuts plotted with the red dashed lines. Targets hosting TOIs are encircled.}
    \label{fig:t-hr_diag}
\end{figure}

We begin by refining the original target list using updated astrometric and photometric observations from Gaia DR3. We retrieve the $G_{\mathrm{BP}}-G_{\mathrm{RP}}$ color, parallax, Gaia $G$ magnitude and the Renormalised Unit Weight Error (RUWE) of each source observed by the aforementioned TESS GI programs. RUWE is a measure of the quality of a star's astrometric fit in Gaia assuming that it is a single star. A high RUWE is characteristic of unresolved binary systems, and we reject any targets with RUWE $>2.5$ because we cannot reliably derive stellar (nor planetary) parameters for unresolved binary systems. We retrieve each target's 2MASS apparent $K_s$ magnitude, and derive the Absolute $K_s$ Magnitude ($M_{K_s}$) using its Gaia parallax and the distance modulus. We then attempt to remove white dwarfs and over-luminous sources from the sample using the main sequence in the $G_{\mathrm{BP}}-G_{\mathrm{RP}}/M_{K_s}$ color-magnitude diagram (see Figure \ref{fig:t-hr_diag}). We reject all targets with a $G_{\mathrm{BP}}-G_{\mathrm{RP}}$ color more than 2.5 standard deviations away from the median per $M_{K_s}$. This cut preserves all confirmed planets and known TESS Objects of Interest (TOI) in this sample, which are transiting planet candidate signals identified by the TESS Science Processing and Operations Center (SPOC). 

Next, we derive stellar parameters for all remaining stars in our refined sample. We retrieve each star's TESS magnitude ($T_{\mathrm{mag}}$) and apparent $K_s$ from the TIC. We calculate the stellar radii of each star using the empirically calibrated radius-$K_s$-band luminosity relationship for M dwarfs from \cite{mann2015} and the mass of each target using the empirically calibrated mass-$M_{K_s}$-band luminosity relationship for M dwarfs from \cite{mann2019}. We retain all targets $M_{K_s} > 6.25$, corresponding to all M stars less massive than 0.4 $M_\odot$, concretely focusing our survey on mid-to-late M dwarfs. This cut implicitly removes any giant M stars. These cuts in mass, color and RUWE leave \ntargets mid-to-late M dwarfs for our survey. 

We calculate the bolometric luminosities of our targets using two bolometric corrections from the $K_s$-band and $V$-band respectively. For both methods, we begin by estimating the $V$-band magnitude using the $G-V$--$G_{\mathrm{BP}}-G_{\mathrm{RP}}$ color relations from \citet{evans_photometry}. We then compute a $V$-band bolometric correction $BC_V$ using the $V-K$ color/$V$-band bolometric correction relationship provided by \cite{pecault2013}. This $V$-band bolometric correction, and our calculated $V$-band magnitude provides our first bolometric luminosity estimate using

\begin{equation}
    L_\text{bol}/L_\odot = 10^{-0.4(M_V-BC_V-4.74)},
    \label{eq:bc}
\end{equation}

\noindent where 4.74 is the Sun's absolute bolometric magnitude. We derive a second bolometric correction with the $K_s$-band magnitude, computing the bolometric correction in the $K$ band from the $V-K_s$--$K_s$-band bolometric correction provided in \cite{mann2015}. This $K_s$ band bolometric correction and the observed $K_s$ band magnitude provides our second bolometric luminosity estimate as in equation \ref{eq:bc}. \rc{The bolometric luminosities derived through the $V$-band magnitude is typically 6\% higher than the bolometric luminosity calculated by through the $K_s$-band magnitude. We adopt the average of the two $L_\text{bol}$ values as the star's bolometric luminosity which have a typical error of 3\%.} We combine the computed radii with the derived $L_\text{bol}$ values to calculate the effective temperature through the Stefan-Boltzmann law.

\begin{figure}
    \centering
    \includegraphics[width=\linewidth]{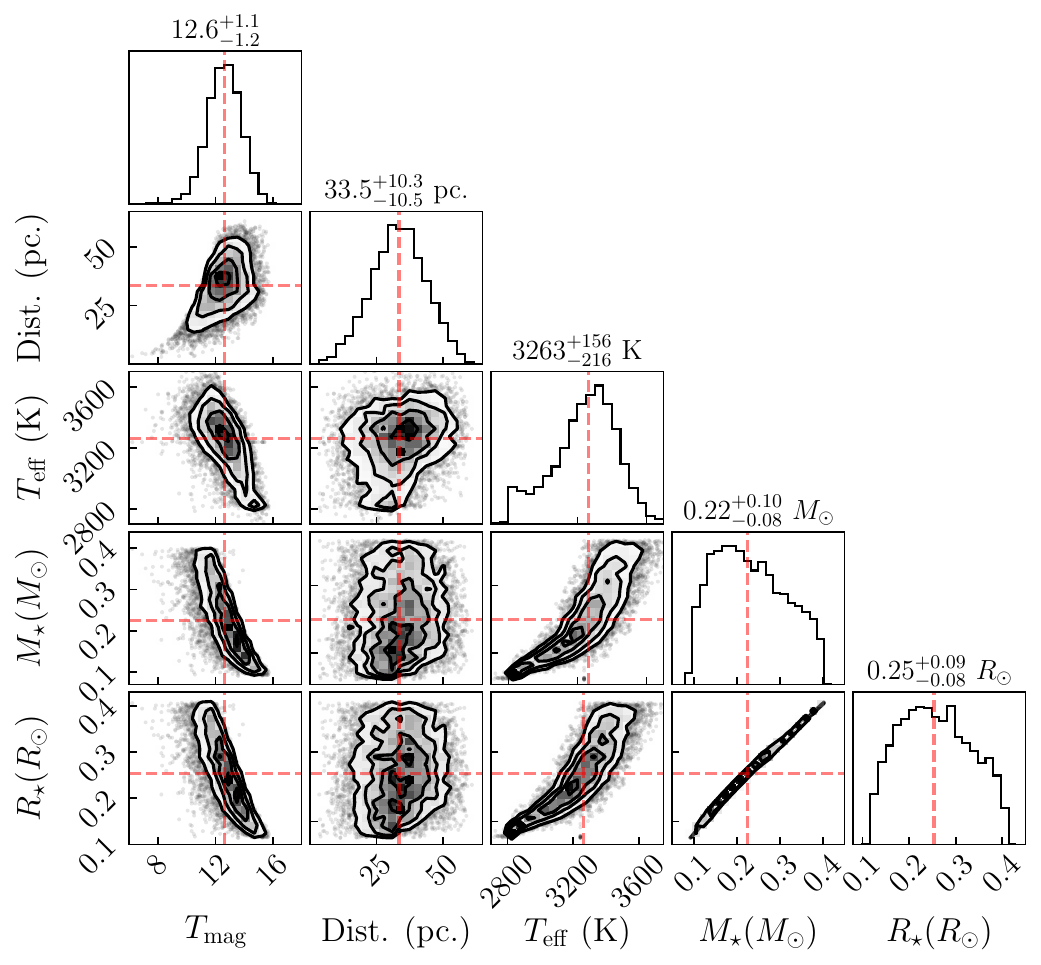}
    \caption{Distributions of $T_{\mathrm{mag}}$, distance, effective temperatures, radii, masses and of our \ntargets targets. The median of each column and row is marked with a dashed red line, which is also reported in the column titles along with each parameters' 16th and 84th percentiles.}
    \label{fig:target_hist}
\end{figure}

The distributions of $T_{\mathrm{mag}}$, distance, effective temperature and mass and radius are plotted in Figure \ref{fig:target_hist} with their respective medians marked.

\section{Transit Search Pipeline} \label{sec:pipeline}
\subsection{Light Curve Data}
We conduct our transit survey using all available TESS data for each of our targets from sectors 1 though 86 spanning six years of observation from July 25 2018 to December 18 2024. TESS images the sky with four cameras spanning a total $24\times96$ square degree field-of-view, and each sector of observation spans an approximately 28 day long pointing of the telescope. TESS records the brightness of each target using Simple Aperture Photometry (SAP) using an aperture of pixels subtended by the star on TESS's detector. SAP derives the flux from a target by summing the dark-subtracted, debiased pixels within the TESS predetermined photometric aperture. This flux is subject to instrumental variations which are corrected for in the Presearch Data Conditioning Corrected SAP \citep[PDCSAP;][]{pdcsap_smith_2012, pdcsap_stumpe_2012, pdscap_stumpe_2014} . The PDCSAP uses principal component analysis over all light curves from the detector to remove large-scale correlated variations exhibited in all the sources on the detector. The PDCSAP flux attempts to correct for large scale systematics while preserving sources of astrophysical variability in each target such as flares, transits, and stellar rotation signatures. For each sector in which a given target was observed, our pipeline retrieves the PDCSAP light curve and supplementary data products, which comprise the following timeseries: the BJD timestamp of each observation, the PDCSAP flux, the 1$\sigma$ error on the PDCSAP flux, the exposure time, the data quality flags, and the sector of the observation. Any data with a nonzero quality flag (indicating some issue) are excised from the time series. 

Our pipeline saves these data by concatenating sectors of consecutive observations as a series of light curves, up to a maximum of four consecutive sectors ($\sim 112$ days). While a long baseline is ideal to detect shallow transits because many transits can be stacked to build S/N, the computational expense of a transit search scales exponentially with baseline duration, making transit searches of targets with many consecutive sectors of observation prohibitively expensive.
Our imposed maximum light curve duration balances the sensitivity of our search over our period range with the computational expense of the pipeline employed to detect transit-like signals. 

In our survey, we define our period search range of interest to span 0.2--30 days. The lower period limit of 0.2 days is motivated by the expected Roche limit for rocky planets around our target stars. The upper limit of our search is motivated by the number of targets with at least two or more consecutive TESS sectors of observation, yielding at least one light curve of 56 days or longer. If we optimistically assume that the occurrence rate of planets at 30 days is 0.1, we ought to find 3 planets in our survey at 30 days if we have perfect sensitivity to transiting planets around the stars with 2 more more consecutive sectors. The breakdown of targets with $N$ consecutive sectors of observations is summarized in Table \ref{tab:consectut}.

\begin{table}[h]
    \centering
    \begin{tabular}{lc}
    \toprule
No. of Consecutive Sectors & Target Count (Fraction) \\ \midrule
        No Consecutive Sectors &  4650 (57\%) \\
        Two Consecutive Sectors & 2457 (30\%) \\
        Three Consecutive Sectors & 648 (8\%) \\
        $\geq$ Four Consecutive Sectors & 379 (5\%) \\
    \bottomrule
    \end{tabular}
    \caption{Number of targets with observations in consecutive TESS sectors.}
    \label{tab:consectut}
\end{table}

\subsection{Cleaning and Detrending Light Curves} \label{subsec:detrend}
The methods we employ to find and characterize transiting planet candidates (PCs) with our pipeline are only reliable when run on light curves that have been cleaned of sporadic outliers, excess stellar variability, and residual systematics. Prior to our transit search, we clean each light curve following the methods described below.

Any data with a non-zero quality flag are excised from the time series. \rc{76\% of light curves have no flagged data, and in 95\% of light curves, these flagged data comprise less than 0.1\%.} Mid-to-late M dwarfs are known to exhibit sporadic flaring events that we seek to identify and remove from our light curves. Our flare removal procedure follows \cite{chang_flares}, where flares are identified by any sequence of three or more consecutive flux measurements that are three or more global median absolute deviations above the median of the light curve. When we detect these consecutive outliers are detected, the 30 data points on either side of the consecutive outliers are taken as part of the series which may contain the flare. With this sequence of data we apply the change-point algorithm described in \citet{chang_flares}. With a sequence of flux measurements ${x_1,x_2,...,x_n}$ with some mean $\bar{x}$ we define the $k$\textsuperscript{th} cumulative sum as 
\begin{equation}
    S_k = \sum_{i=0}^{k} \left(x_i - \bar{x}\right).
\end{equation}
A change-point corresponds to the maximum value of $|S_k|$ and its significance is determined by boostrapping and comparing its value to the cumulative sum from shuffled ${x_i}$ series. If its magnitude exceeds 90\% of the boostrapped change-points, it is considered significant, and the series is divided into two sets on either side of the change-point. This process continues recursively on those subsets until no new change-points are identified. Identifying these change-points allows the pipeline to identify the extent of a flare. The identified change-points, along with the data in between them, are excised from the timeseries. 

Many of our targets show strong quasi-periodic variation due to stellar rotational signals. In cases where this rotation signal is significant, we use a Gaussian process to remove the variation. We begin characterizing the rotation signal by computing the generalized Lomb-Scargle periodogram of each light curve \citep{gls_py} to identify the best-fit sinusoid with a period of $P_\mathrm{GLS}$, \rc{where $0.1\,\mathrm{days}< P_\mathrm{GLS} < T/2$ and $T$ as the baseline of the light curve.} To test whether the periodic signal is significant we perform a model comparison between this sinusoidal model and a constant flux model equal to the median of the light curve. We compare these two models using the Bayesian Information Criterion
\begin{equation}
    \text{BIC} = - 2 \ln(\mathcal{L}) +  k\ln{(n)},
\end{equation}
where $\mathcal{L}$ is the likelihood of a model, $k$ is the number of free parameters in the model (3 and 1 for the sinusoidal and constant models respectively), and $n$ is the number of data points. The difference between the BICs for the sinusoidal and constant flux models quantifies how well one fits over the other and is expressed as the $\Delta$BIC. \rc{We consider a periodic signal to be significant if the $\Delta$BIC test favors the sinusoidal model by a score of 50 or more. Additionally, for light curves that exhibit rotational signals with $P_\mathrm{GLS}>1$ day, we also require that the amplitude of the rotation signal is greater than the median absolute deviation (MAD) of the light curve.} The latter condition for longer period light curves is necessary because while Gaussian processes are efficient in removing rotation signals, they are also known to degrade transit signals \citep{nuance}. We are able to characterize this effect using our injection--recovery tests, and outline the dilution in section \ref{subsec:injrec}. Out of our \ntargets targets, 1,586 show significant signs of rotation according to this criterion. \rc{Since we require light curves with periods longer than 1 day to show greater amplitude modulation than the MAD of the light curve, our fraction of rotating targets (16\%) is smaller than the fraction of rotating targets found by \citetalias{ment_occurrence_2023} in their sample of mid-to-late M dwarfs.}

Light curves without a significant rotational signal do not require a quasi-periodic model to detrend any rotational variability. We detrend these light curves using a running median filter with a 12-hour window. Each light curve with significant rotation is detrended using a Gaussian process. Our Gaussian process (GP) employs the rotation term kernel provided by \texttt{celerite2} \citep{celerite2}. 
This kernel is a combination of two simple harmonic oscillator kernels, with two modes in Fourier space: one at the recovered rotation period given by the peak of the GLS periodigram, and one at one-half the recovered rotation period. The power spectral density of the simple harmonic oscillator is 

\begin{equation}
    S(\omega) = \sqrt{\frac{\pi}{2}}\frac{S_0 {\omega_0}^4}{\left(\omega^2-{\omega_0}^2\right)^2+ \omega^2{\omega_0}^2/Q^2},
\end{equation}

\noindent where $\omega_0=1/P_\text{GLS}$ is the undamped angular frequency, $Q$ is the quality factor, and $S_0$ characterizes the power at ${\omega_0}^2$. The parameters used in this model are further detailed in Table \ref{tab:gp_params}.  The rotation term kernel combines two harmonic oscillators as $R(\omega) = S_1(\omega) + S_2(\omega)$. The respective parameters of $S_1$ and $S_2$ are 

\begin{align}\label{eq:os1}
   Q_1 &= 1/2 + Q_0 + \Delta Q \\
   \omega_{0,1}  &= \frac{4\pi Q_1}{P\sqrt{4{Q_1}^2-1}}\\
   S_{0,1} &= \frac{\sigma^2}{(1+f)\omega_{0,1}Q_1}
\end{align}

\noindent and 

\begin{align}
   Q_2 &= 1/2 + Q_0 \\
   \omega_{0,2}  &= \frac{8\pi Q_2}{P\sqrt{4{Q_2}^2-1}}\\
   S_{0,2} &= \frac{f\sigma^2}{(1+f)\omega_{0,2}Q_2}. \label{eq:os2}
\end{align}
The free parameters in Eqs.~\ref{eq:os1}--\ref{eq:os2} are the hyper-parameters of the rotation term kernel, and are outlined in Table \ref{tab:gp_params}. 

\begin{table*}
    \centering
    \begin{tabular}{clcc}
        \multicolumn{2}{l}{Gaussian Process Hyper-parameters:} & & \\
        \toprule
        Parameter & Explanation & Prior & Unit  \\ 
        \midrule
         $\mu$ & Light curve mean & $\mathcal{N}(0, 10)$ & Normalized Flux \\
         $\log(\sigma_\text{jitter})$ & Excess white noise & $\mathcal{U}(-3, 2)$ & Normalized Flux \\
         $\log(\sigma_\text{gp})$ & GP Amplitude & $\mathcal{U}(-3, \log(\sigma_\text{rms}))$ & Normalized Flux \\
         $\sigma_{\text{lc}}$ & Measurement uncertainty & $\log{\mathcal{N}}(\log(\bar{\sigma}), 0.1)$ & Normalized Flux \\
         $P_\text{rot}$ & Rotation period & $\mathcal{N}(P_\text{LS}, 0.1)$ & Days \\
         $Q_0$ & Quality factor & $\mathcal{N}(7.5, 2)$ & Unitless \\
         $\Delta Q$ & Difference between modes' quality factors & $\log{\mathcal{N}}(0,1)$ & Unitless \\
         $f$ & Secondary mode fractional amplitude & $\mathcal{U}(0.8, 1)$& Unitless \\
         \bottomrule
         & & \\
         \multicolumn{2}{l}{Planet Candidate Parameters and Priors:} & & \\
        \toprule
        Parameter & Explanation & Prior & Unit \\ 
        \midrule    
        $T_0$ & Mid-transit epoch of the final observed transit & $\mathcal{N}(T_{0,\text{TLS}}, \sigma T_{0,\text{TLS}})$& Days \\
        $P$ & Period of the transiting planet & $\mathcal{N}(P_{\text{TLS}}, \sigma P_{\text{TLS}})$& Days \\
        $R_p/R_\star$ & Planet-to-star radius ratio & $\log\mathcal{N}(0.005, 0.5)$ & Unitless \\
        $b$ & Impact Parameters &  $\mathcal{U}(0, 0.95)$& Unitless \\
        $F_0$ & Out of transit flux offset & $\mathcal{N}(0, \sigma_\text{transit})$ & Normalized Flux \\
        \bottomrule
        \end{tabular}
    \caption{
    Top: Summary of the parameters used in the Gaussian Process used to remove the strong stellar rotation signals from light curves. Here $\mathcal{N}$ denotes a normal distribution, $\mathcal{U}$ denotes a uniform distribution,  $\log{\mathcal{N}}$ denotes a log-normal distribution, $\sigma_\text{rms}$ is the standard-deviation of the light curve, $P_\text{LS}$ is the period found by the Lomb-Scargle periodigram and $\bar{\sigma}$ is the median flux error. Bottom: Summary of parameters and priors used in the Monte-Carlo Markov chain used to fit planet parameters. Here, $T_{0,TLS}$ is the mid-transit epoch fit by the highest SDE TLS, $\sigma T_{0,TLS}$ is the uncertainty on $T_{0,TLS}$, $P_{TLS}$ is the period found  by the highest SDE TLS, $\sigma P_{TLS}$ is the uncertainty on $P_{TLS}$, and $\sigma_\text{transit}$ is standard deviation of the light curve within two durations of the mid-transit.}
    \label{tab:gp_params}
\end{table*} 
\begin{figure}
    \centering
    \includegraphics[width=\linewidth]{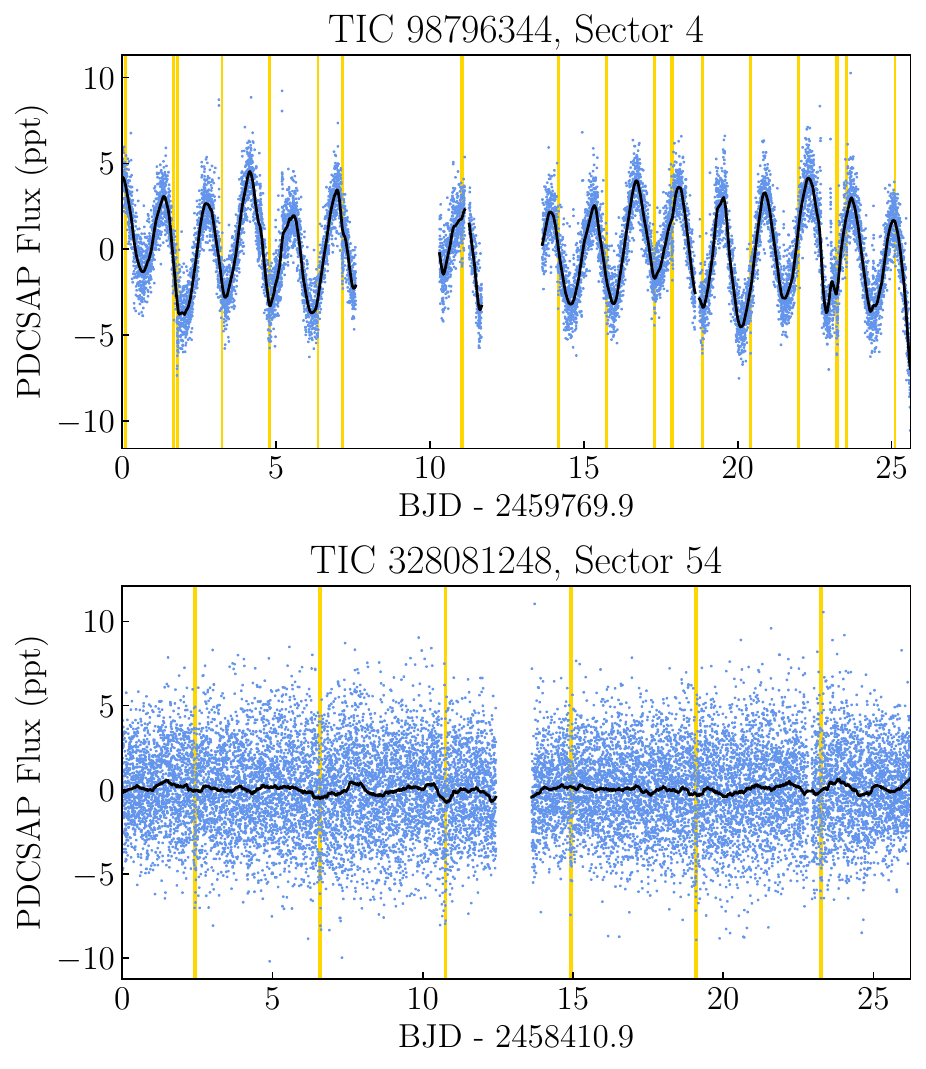}
    \caption{Selected light curves from TICs 98796344 (top) and 328081248 (bottom) respectively after flare removal. The trends recovered by a Gaussian process (top) and median filter (bottom) are overplotted in black, and the transit events in each light curve are highlighted in gold.}
    \label{fig:lt_with_trend}
\end{figure}

\subsection{Transiting Planet Search}
We search for periodic transit-like signals in our detrended light curves using the Transit Least Squares algorithm  \citep[TLS;][]{tls_hippke}. TLS shares many similarities with the more traditional Box Least Squares \citep[BLS;][]{bls_kovaks} algorithm, but instead of a boxcar function to fit transit-like signals, TLS fits a physically motivated transit model that incorporates the known stellar mass, radius, and quadratic limb-darkening coefficients to improve the signal-to-noise ratio of recovered signals and runtime efficiency. The TLS searches a range of transit periods, mid-transit epochs, and a set of transit durations, the latter of which encompasses all known exoplanets as of \citet{tls_hippke}. TLS produces a signal detection efficiency (SDE) spectrum across a set of periods characterizing the goodness-of-fit of the best fit model marginalized over all other parameters.  This process must be run iteratively to recover multi-planet systems in our sample. We identify transit-like signals from the TLS as periodicities with an \rc{$\mathrm{SDE} > 6$.} We then run a least-squares optimization to fit a physical transit model using the \texttt{batman} \citep{batman} Python package to the TLS signal. The model is parameterized by the mid-transit epoch $T_0$, the orbital period $P$, the planet-to-star radius ratio $R_p/R_\star$, the impact parameter $b$, and the quadratic limb-darkening parameters retrieved from the TIC. While the limb-darkening parameters could in principle be calculated manually given stellar parameters, any differences are expected to have a negligible effect on our model fits. We do not directly fit the transit parameter $a/R_\star$ and instead calculate it from $P$, $M_\star$, and $R_\star$.
\rcnew{We assume circular orbits by fixing the eccentricity to zero.} \footnote{\rcnew{We ran preliminary tests with non-zero eccentricities and confirmed that the resulting eccentricity posteriors are consistent with zero for all but two planet candidates in our sample. Our fit planet parameters for all of our planet candidates are only negligibly affected. The two planet candidates for which we measure a non-zero eccentricities are LTT 3780 c ($0.36\pm0.16$) and LP 261-75 C ($0.38\pm0.07$). The former is known to have an orbit consistent with circular from joint transit and radial velocity fitting \citep{LTT_3780}, and the latter is a known brown dwarf \citep{LP-271-72}. This alternate model is only able to weakly constrain eccentricity for our other planet candidates with a typical 1$\sigma$ upper limit of 0.36.}} We subtract the best fit model centered at zero flux from the light curve to remove the detected signal. We continue this process iteratively until the TLS fails to find a signal with $\mathrm{SDE} > 6$ or a maximum number of four iterations are run. We choose four iterations as our maximum because no targets in our sample have more than three confirmed planets or TOIs.

\subsection{Planet Candidate Characterization}

Each light curve yields a (possibly empty) set of transit candidate events (TCEs) and fit batman models. Each TCE must pass the following five vetting conditions:
\begin{enumerate}
    \item The transit depth $\Delta$ from the best-fit transit model must result in a signal-to-noise ratio (S/N) greater than 3. We define S/N as:
    \begin{equation}
        \text{S/N} = \frac{\Delta}{\sigma_\text{lc}}\sqrt{{N_\text{transits}}},
    \end{equation} where $\Delta$ is the TCE depth, $N_\text{transits}$ is the number of transits observed, and $\sigma_\text{lc}$ is the photometric rms of the light curve binned to the TCE's duration.\footnote{We note that $\sigma_\mathrm{lc}$ serves as proxy for the Combined Differential Photometric Precision (CDPP), which was only recently released by \citet{twicken2025} and was not available at the time of our code development.} This intermediary cut does not represent the final S/N cutoff in our survey.
    \item The period of the TCE cannot match the period of another higher S/N TCE from the same light curve within a tolerance of 0.1\%. 
    \item The odd and even transit depths computed by the TLS must be consistent within 3$\sigma$.
    \item The TLS result cannot have the same period, or be an integer multiple of the star’s rotation period within a 0.3\% tolerance. We note that this condition is only applicable to stars with strong evidence for rotation (see section \ref{subsec:detrend}). TLS results with $\mathrm{SDE}>15$ are exempt from this condition due to the significance of the result.
    \item  The TCE's period must match the period of results from half the light curves or more, rounded down if the target has four of more light curves. TLS results with  $\mathrm{SDE}>15$ and targets with ten or more sectors of observation are exempt from this condition.
\end{enumerate}

We build a set of PCs by combining the TCEs with common properties from each light curve. Each of these PCs will be characterized by a joint fit using all available data at once, and further vetted with a second set of vetting conditions. From the sets of TCEs that our iterative search generates, we collect sets of TCEs with common depths within 20\%, durations within 40\%, and periods whose ratio is within 0.01\% of an integer. Transit signals can be picked up as period aliases by the TLS motivating the need for a tolerance in period ratio rather than period, and the large ranges on durations and depth are necessary to account for these possibly misfolded signals.

To best characterize the PC's period, we run a narrow TLS search within the period uncertainty of the result with the highest SDE using the full concatenated light curve. The pipeline then uses the folded light curve to fit the remaining transit parameters $T_0$, $R_p/R_\star$, $b$ as well as an out of transit flux parameter $F_0$ to better characterize the transit baseline using \texttt{scipy.optimize.curve\_fit} \citep{2020SciPy-NMeth} with \texttt{batman} transit models.

At this stage, we require that the best-fit over all sectors of TESS data must have an $\text{S/N} > 6$. The pipeline further characterizes these PCs by running a Markov Chain Monte Carlo (MCMC) using the \texttt{emcee} package \citep{emcee} to sample the posterior distributions of the five transit model parameters. These quantities and their respective priors are further outlined in Table \ref{tab:gp_params}. We also adopt quadratic limb-darkening parameters from the TIC as above.

To be considered a validated planet and therefore included in our forthcoming calculations of planet occurrence, each PC must pass a final set of statistical vetting conditions, which are calibrated to remove false alarms (i.e. statistical false positive signals that appear in the light curve and do not have an astrophysical origin) through searches of light curves without planet signals.

\begin{enumerate}
    \item The five-parameter fit model must have a reduced $\chi^2$ value of 1.5 or less.
    \item  A $\Delta$BIC test must favor the fitted transit model over a flat line at the median of the folded light curve within 1 transit duration of the transit midpoint by -10 or more.
    \item A second $\Delta$BIC test must favor the fitted transit model over a flat line at $F_0$ within 1 transit duration of the transit midpoint by -10 or more.
\end{enumerate}

 To further protect our search from astrophysical false positives (i.e.\ non-transit astrophysical signals that mimic TCEs), we employ \texttt{TRICERATOPS} \citep{triceratops_code, triceratops_paper} which characterizes the likelihood that a given signal is an astrophysical false positive. \texttt{TRICERATOPS} conducts this analysis by considering all stars that might pollute the flux in the TESS photometric aperture, and compares the transit model to other phenomena which may produce periodic transit-like signals (e.g.\ background eclipsing binaries). We find that among the TOIs the pipeline is able to recover, none have a false positive probability (FPP) greater than 8\%. In our survey, we choose to reject any signals with a \texttt{TRICERATOPS} FPP greater than 10\%. 
 
\begin{figure}
     \centering
     \includegraphics[width=\linewidth]{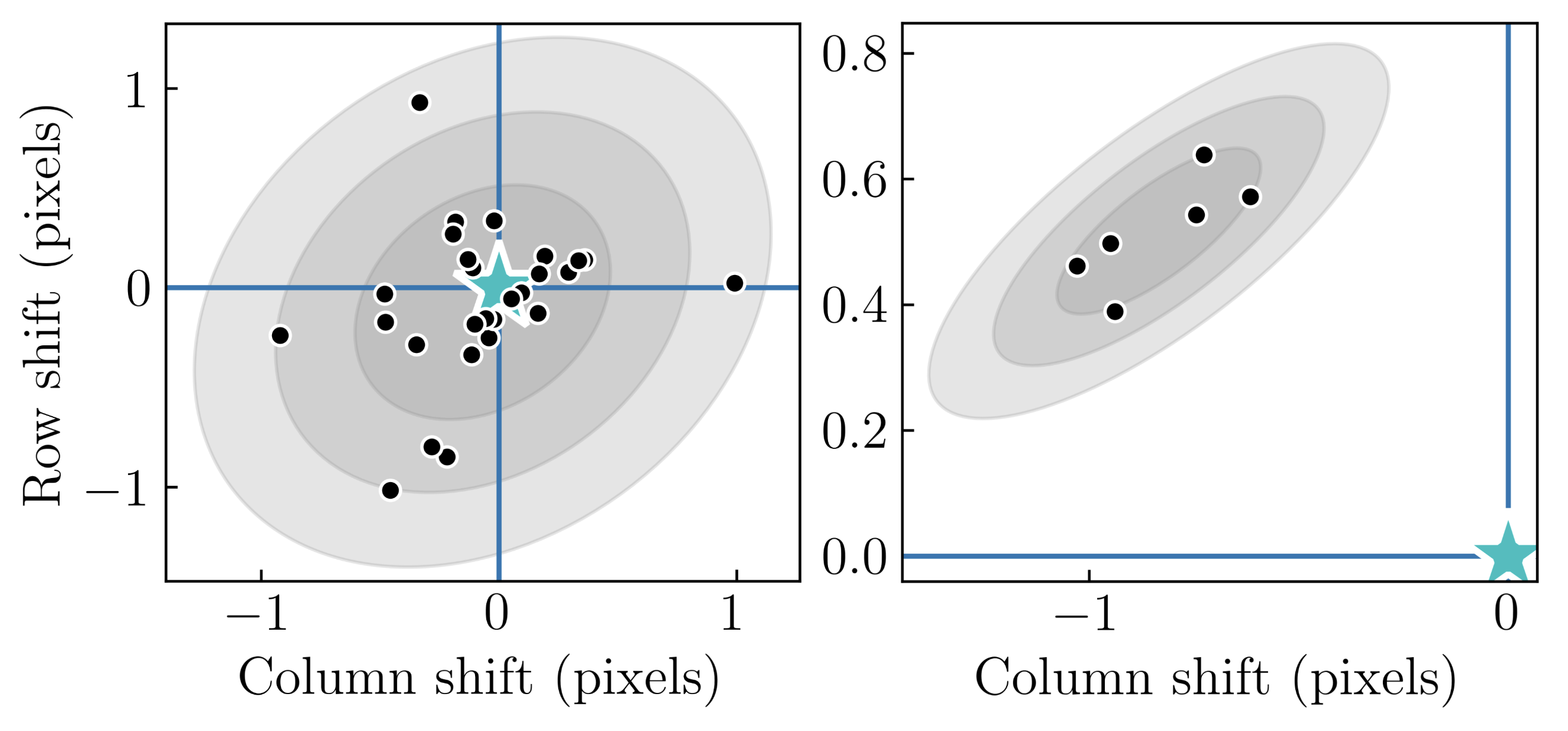}
     \caption{Example in-transit centroid offsets for TOIs 562.01 (left) and 3494.01 (right). The star marker in each panel marks the out-of-transit centroid and the black points mark the relative in-transit centroid positions of each TCE. The contours mark the 1, 2, and $3\sigma$ confidence intervals of the estimated in-transit centroid distribution. TOI-562.01 has a 97\% on-target probability and TOI-3494.01 has a 0\% on-target probability.}
     \label{fig:centroids}
\end{figure}

 As an additional check for astrophysical FPs in our set of PCs, we search for evidence of in-transit centroid offsets, which are indicative of off-target transit-like signals or eclipsing binaries. We employ \texttt{exovetter} \citep{exovetter} to this end, which compares the centroid of each in-transit event to the out-of-transit baseline to characterize the likelihood that the signal contributing to each transit is on target. Figure \ref{fig:centroids} illustrates an on-target and an off-target centroid from two PCs for our search. We reject signals with an on-target probability over all observed transits of $<1$\%. We perform this analysis and the aforementioned \texttt{TRICERATOPS} vetting using the TESS full-frame images retrieved for the in- and out-of-transit epochs using \texttt{lightkurve} \citep{lightkurve}. 

It bears mentioning that there has been extensive effort to validate PCs through the TESS Follow-up Observing Program (TFOP) by conducting strategic follow-up observations of many TOIs. While these data would certainly improve our ability to validate PCs found by our pipeline, these observations represent a heterogeneous dataset that cannot be consistently applied to all PCs. We therefore ignore the availability of any TFOP observations in our PC characterization and vetting.

\subsection{Detection Sensitivity Characterization} \label{subsec:injrec}
We have run a suite of injection--recovery tests to characterize the detection sensitivity of our pipeline as a function period and $R_p/R_\star$. Fully characterizing our pipeline's sensitivity to transiting planets on a per-light curve basis is beyond the scope of the computational resources available for this project. Consequently, we consolidate this effort by calculating the average detection sensitivity within six sample bins. First targets were separated based on their magnitude with bins of $T_{\mathrm{mag}}$ $<\!12$, $12\!-\!13$ and $>\!13$ respectively, with $T_{\mathrm{mag}}$ each bin containing roughly one third of our full sample. With this subdivision of our sample in $T_{\mathrm{mag}}$, we aim to capture nonlinearities in photometric noise as a function of target brightness. We further subdivide these three magnitude bins into rotating and non-rotating subsamples based on the rotation criteria used in our pipeline. Rotating and non-rotating targets must be treated separately because the two different detrending methods used may introduce different noise properties in the resulting detrended light curves. The target counts in these bins are summarized in Table \ref{tab:injrec_counts}.

\begin{table}
    \centering
    \begin{tabular}{ccc}\toprule
         $T_{\mathrm{mag}}$ & Rotating? &  Count (Fraction) \\ \midrule
         \textless 12 & Yes & 366 (4.5\%) \\
         \textless 12 & No &  2043 (25\%) \\
         12-13 & Yes & 402 (4.9\%) \\
         12-13 & No & 2261 (28\%)\\
         \textgreater 13 & Yes & 562 (6.9\%) \\
         \textgreater 13 & No & 2500 (31\%)\\
         \bottomrule
    \end{tabular}
    \caption{Target counts in each of our injection--recovery bins used to characterize our pipeline's sensitivity.}
    \label{tab:injrec_counts}
\end{table}

 We inject signals with periods spanning 0.2--30 days, and injected $R_p/R_\star$ from 0.01--0.2. The sampled period range corresponds to a $R_p$ range of 0.3--5.5 $R_\oplus$ around an average star in our sample. These signals are forward modeled using \texttt{batman} and injected into the PDCSAP light curve of a sampled target from the magnitude/rotation class bin under consideration. We inject this transit signal into the light curve of a randomly sampled target from the bin, before the light curve is processed in any way by or pipeline. Prior to injection any known TOI signals are masked using white noise matching the mean, local slope and scatter of the light curve at each transit epoch. We run our full transit search and claim a successful recovery of an injected signal if our transit search pipeline recovers a vetted PC whose recovered period $P'$ matches any of: the injected period $P$, $Pn$, or $P/n$ where $n\in\{2,3,4,5,6\}$ within a tolerance of 0.2\%. 

Our strategy to efficiently sample the period--$Rp/R\star$ parameter space with injections follows \citetalias{ment_occurrence_2023} over the period--$R_p/R_\star$ space. The space is initially sectioned into four leaves using a quadtree, with a minimum 50 injections in each leaf of the tree. Within each leaf, $P$ and $R_p/R_\star$ values are sampled from log-uniform distributions, with $T_0$ and $b$ drawn uniformly from 0--$P$ and 0--1, respectively. Once each leaf of the tree is populated with at least 50 injections, we determine whether more injections are required to accurately resolve the space. If two neighboring leaves of the tree in $P-R_p/R_\star$ space differ in cumulative recovery rate by more than 15\%, we recurse into the leaf by rebinning the leaf into four smaller leaves. We perform a maximum of five recursions with this method.

\begin{figure*}[ht]
    \centering
    \includegraphics[width=\linewidth]{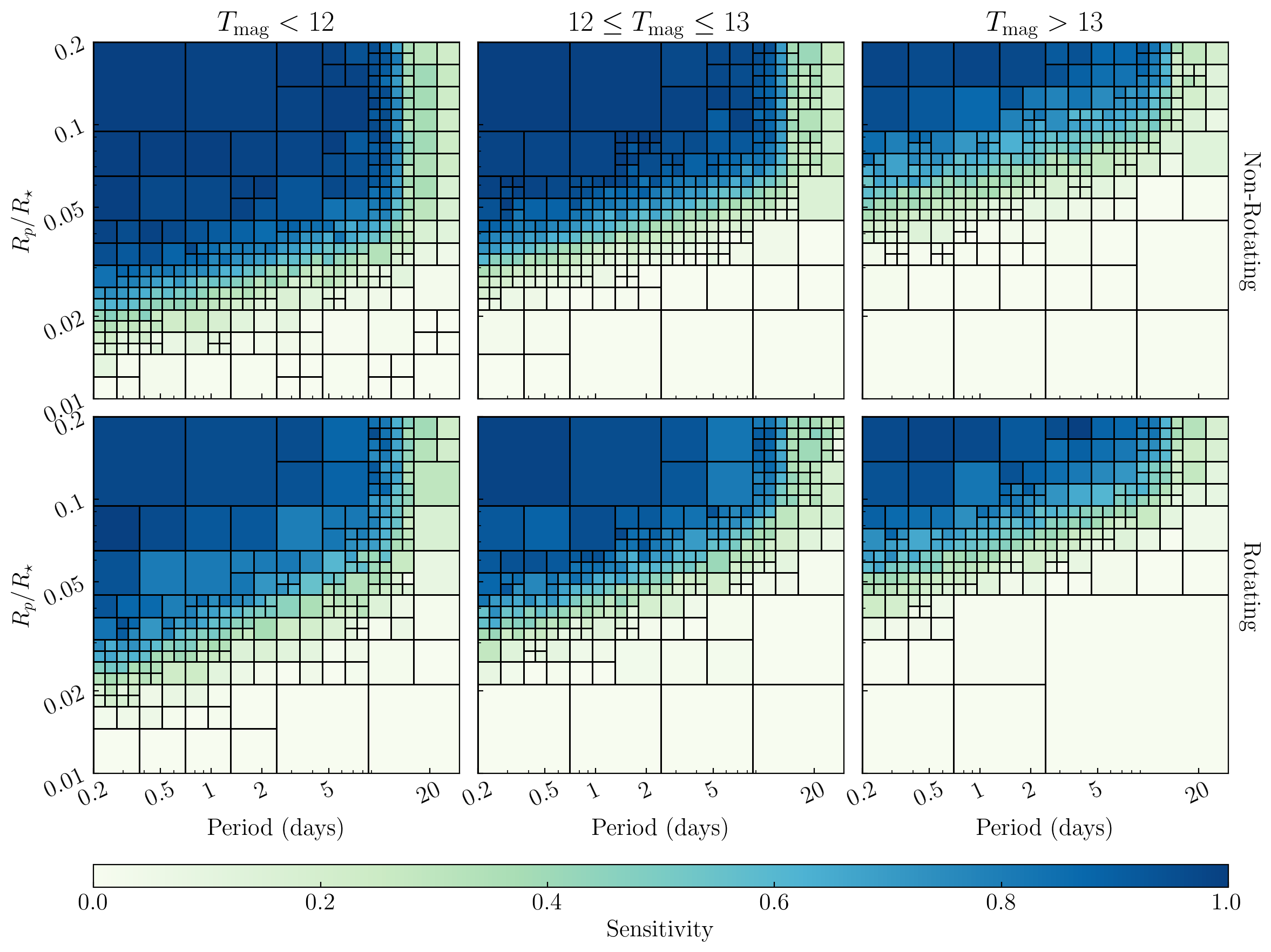}
    \caption{Injection/recovery maps over each of our six injection recovery target bins that span different TESS magnitudes (columns) and non-rotating (top row) and rotating targets (bottom row) showing our pipeline's sensitivity to transiting planet signals across period--$R_p/R_\star$ space. Each cell in this recursive structure contains at least 50 injected planets with the black grid denoting the boundaries of each cell.}
    \label{fig:injecrec_map}
\end{figure*}

Our injection--recovery maps for each of the six bins considered are depicted in Figure \ref{fig:injecrec_map}. In comparing our rotating versus non-rotating TESS magnitude bins, we find worse sensitivity for shallow signals as TESS magnitude increases. We also see a sharp falloff beyond 14-day periods over all bins. This is expected, as 4,650 of our targets lack any consecutive sectors of observations, which makes detecting planets with orbital periods \textgreater14 days impossible. We also see that our pipeline's sensitivity is degraded around rotating stars compared to non-rotating stars of the same magnitude. This isn't unexpected, as Gaussian processes are known to degrade transit signals \citep{nuance}. This drop in sensitivity may also indicate that active targets have a higher noise floor than non-active targets because active stars exhibit higher levels of uncorrelated jitter that cannot be removed through detrending.

Our injection--recovery tests also offer us insight into how reliable our detrending methods are in preserving the radius of injected planets by comparing the ratio of the injected radius to the recovered radius. Over our set of non-rotating targets where a median filter was used in detrending, we find that $R_\mathrm{p,rec}/R_\mathrm{p,inj} = 0.96 \pm 0.05$. Over the set of rotating targets where a GP was used for detrending, we find that $R_\mathrm{p,rec}/R_\mathrm{p,inj} = 0.86^{+0.10}_{-0.20}$, which is not surprising because it is known that GPs can degrade transit depths \citep{nuance}. 

\begin{figure}[h]
    \centering
    \includegraphics[width=\linewidth]{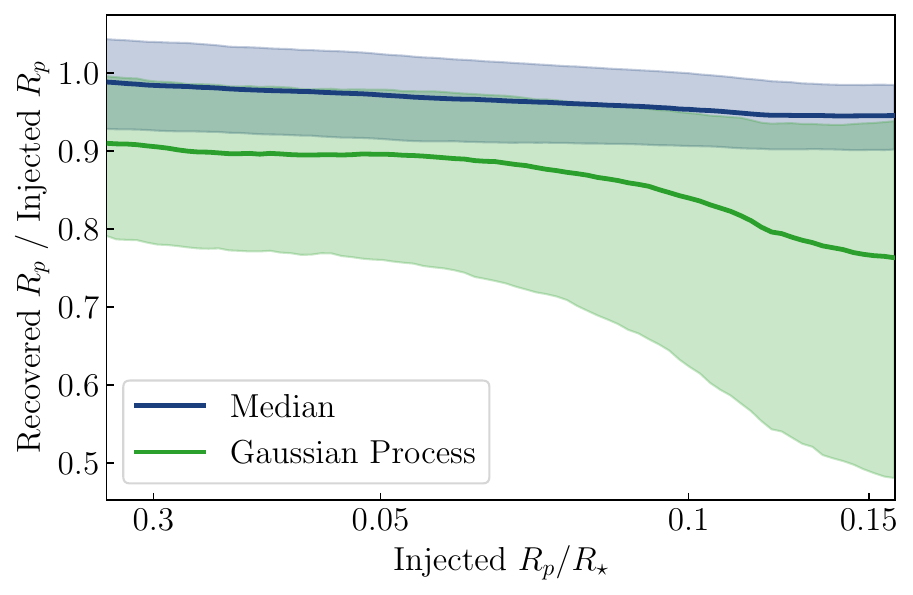}
    \caption{The ratio of recovered planet radii to injected planet radii as a function of the injected $R_p/R_\star$  injected. The curves depict the biases on our inferred planetary radii under our median filter (blue) and GP (green) detrending models. The shaded regions depict the $\pm1\sigma$ intervals around each curve.}
    \label{fig:gp_recovery}
\end{figure}

In Figure~\ref{fig:gp_recovery} we illustrate $R_\mathrm{p,rec}/R_\mathrm{p,inj}$ as a function of injected $R_p/R_\star$ for our median-detrended and GP-detrended samples along with their respective $1\sigma$ ranges. We clearly see that the degradation of the transit depth a our GP models is less impactful for small $R_p/R_\star$. Considering transit injections into the light curves of rotating stars with $R_p/R_\star < 0.05$ (corresponding to a radius of $1.2\,R_\oplus$ around our median stellar radius of $0.25\,R_\odot$), we find $R_\mathrm{p,rec}/R_\mathrm{p,inj} = 0.90^{+0.08}_{-0.12}$. We will revisit this effect in Section~\ref{sec:planet_search} and note that the degradation of the transit depths by our GP model does not have a strong effect on our planet sample due to the small fraction of found transiting planets orbiting stars that exhibiting rotation. 

We convert the full sensitivity of our pipeline from period--$R_p/R_\star$ space to period--$R_p$ space and instellation--$R_p$ space as we are ultimately interested in characterizing planetary occurrence as a function of these quantities. We convert injected orbital period to instellation, spanning $1-1000S_\oplus$, using the expression 
\begin{equation}
S = S_\oplus \left(\frac{T_\text{eff}}{5777\, \text{K}}\right)^4\left(\frac{P}{365.25 \, \text{d}}\right)^{-\frac{4}{3}}\left(\frac{R_*}{R_\odot}\right)^2\left(\frac{M_*}{M_\odot}\right)^{-\frac{2}{3}}.
\end{equation}
\noindent Similarly, our converted planetary radii span $0.5-6.5\, R_\oplus$. In our conversions, we use the calculated stellar parameters associated with each target. 

\begin{figure}
    \centering
    \includegraphics[width=0.9\linewidth]{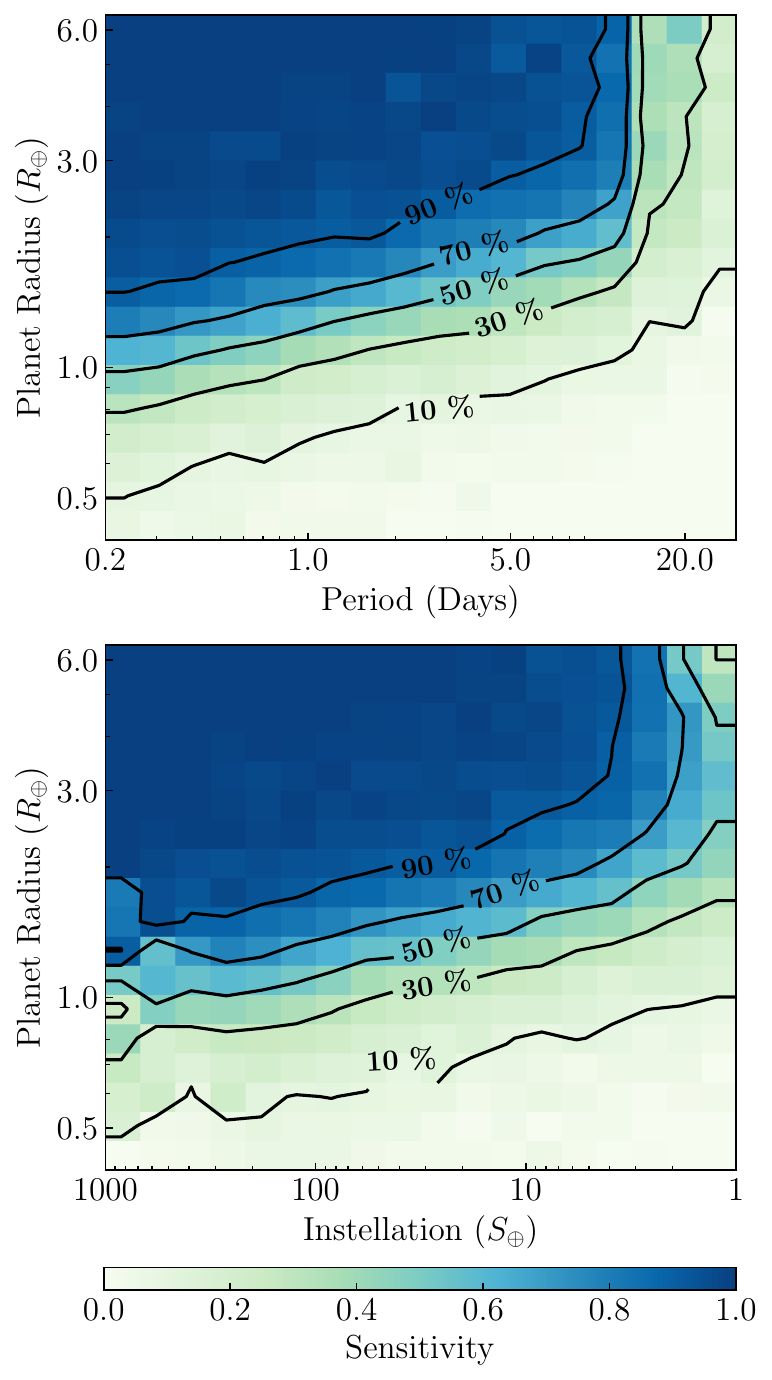}
    \caption{The average sensitivity of our survey as a function of period, planet radius, radius and instellation from $0.2-30$ days, $0.5-6.5\, R_\oplus$, and $1-1000\, S_\oplus$, respectively.}
    \label{fig:full_completeness}
\end{figure}

Figure \ref{fig:full_completeness} depicts the resulting average detection sensitivity maps as functions of $P$--$R_p$ and $S$--$R_p$, respectively. These maps are computed from the weighted average of the six $T_{\mathrm{mag}}$/rotation bin maps, where each contributing map is weighted by the fraction of our sample targets that fall within that bin. We note that because all injections have been transformed to radius and instellation spaces, dependent on the parameters of each target, the full extent of our sensitivity characterization can be extended beyond the ranges depicted in Figure \ref{fig:full_completeness}. Our pipeline is sensitive to 50\% of planets with $R_p=1\,R_\oplus$ out to 0.5 days and 50\% of planets with $R_p=2\, R_\oplus$ out to 8 days. Our survey has limited sensitivity to sub-Earth-sized planets because of the wide range of targets' $T_\mathrm{mag}$ surveyed and the characteristic unsuitability of our faint targets to the detection of sub-Earths. We also have very limited sensitivity to planets within the habitable zone (HZ). For example, beyond a conservative inner HZ edge of $0.9 S_\oplus$ based on the water-loss limit \citep{hab_zone_koppararu}, we have 30\% sensitivity to planets with $R_p=3\, R_\oplus$, which degrades to 12\% sensitivity for Earth-size planets. Our sensitivity drops steeply toward lower instellation values due to the fact that we only search out to a maximum period of 14 days for 57\% of our targets with zero consecutive sectors of observation. These relatively poor sensitivity limits imply that we will only be positioned to make weak statements regarding the occurrence rate of sub-Earth and the population of HZ planets around mid-to-late M dwarfs. The low transit probability of relatively far out HZ planets compounds with our low sensitivity, making HZ planet detections exceedingly unlikely. 

We note that our injection--recovery maps, when converted from $R_p/R_\star$ to $R_p$, do not extend to planets larger than $\sim 6.5\, R_\oplus$. This being said, our survey sensitivity to these large planets is high ($>93$\%) and appears roughly constant per orbital period. In Section \ref{sec:occrate} we will use this property of our sensitivity calculations to extrapolate this high sensitivity to constrain the occurrence rate of close-in Jovian planets around mid-to-late M dwarfs. 

\subsection{Transit Survey Completeness}
Occurrence rate inferences depend on the survey completeness, which is the product of the detection sensitivity computed from our injection--recovery tests and the geometric transit probability. For a given injected planet, we adopt a transit probability of
\begin{equation}
    P_{\rm tr} = \frac{R_\star}{a}
\end{equation}
\noindent where $a$ is the semi-major axis calculated from the planet's orbital period and host star mass using Kepler's third law. The transit probability is independent of the host star's $T_{\rm mag}$ and rotation period such that we compute the average $P_{\rm tr}$ over our full sample of stars in the $P-R_p$ and $S-R_p$ parameter spaces. The product of our sample's average transit probability and survey sensitivity (c.f.\ Figure~\ref{fig:full_completeness}) yields our survey completeness maps, which we denote as $c(P,R_p)$ and $c(S,R_p)$.

\section{Planet Search Results} \label{sec:planet_search}

Using our pipeline described in Section~\ref{sec:pipeline}, we systematically search each of our \ntargets targets for transiting planets from 0.2 to 30 days. Our pipeline identifies, vets, and subsequently characterizes TCEs that pass each of our vetting and statistical validation conditions. 

\texttt{TRICERATOPS} and \texttt{exovetter} are incorporated into our vetting procedure to filter out probable false positives.

To converge on our final population of planets that can be used for occurrence rate calculations, our recovered set of PCs must be corrected for a non-zero false alarm rate to avoid overestimating occurrence rates. We note that our injection--recovery module is designed to calculate the false negative rate (i.e. detection sensitivity) but is not suited to characterizing the false alarm rate because only physical signals (i.e. true positives) are injected. We therefore inspect the PCs vetted by our fully automated transit search to attempt to identify and remove probable false alarms. This vetting step involves the detailed inspection of each PC's TLS results, the pre- and post-detrending light curves, the phase-folded light curve and transit fit results from our MCMC analysis. Common false alarms include imperfect detrending of rotation signals and unflagged light curve edge effects. Following this step we proceed by assuming that our sample is free from false alarms. 

\begin{figure*}
    \centering
    \includegraphics[width=\linewidth]{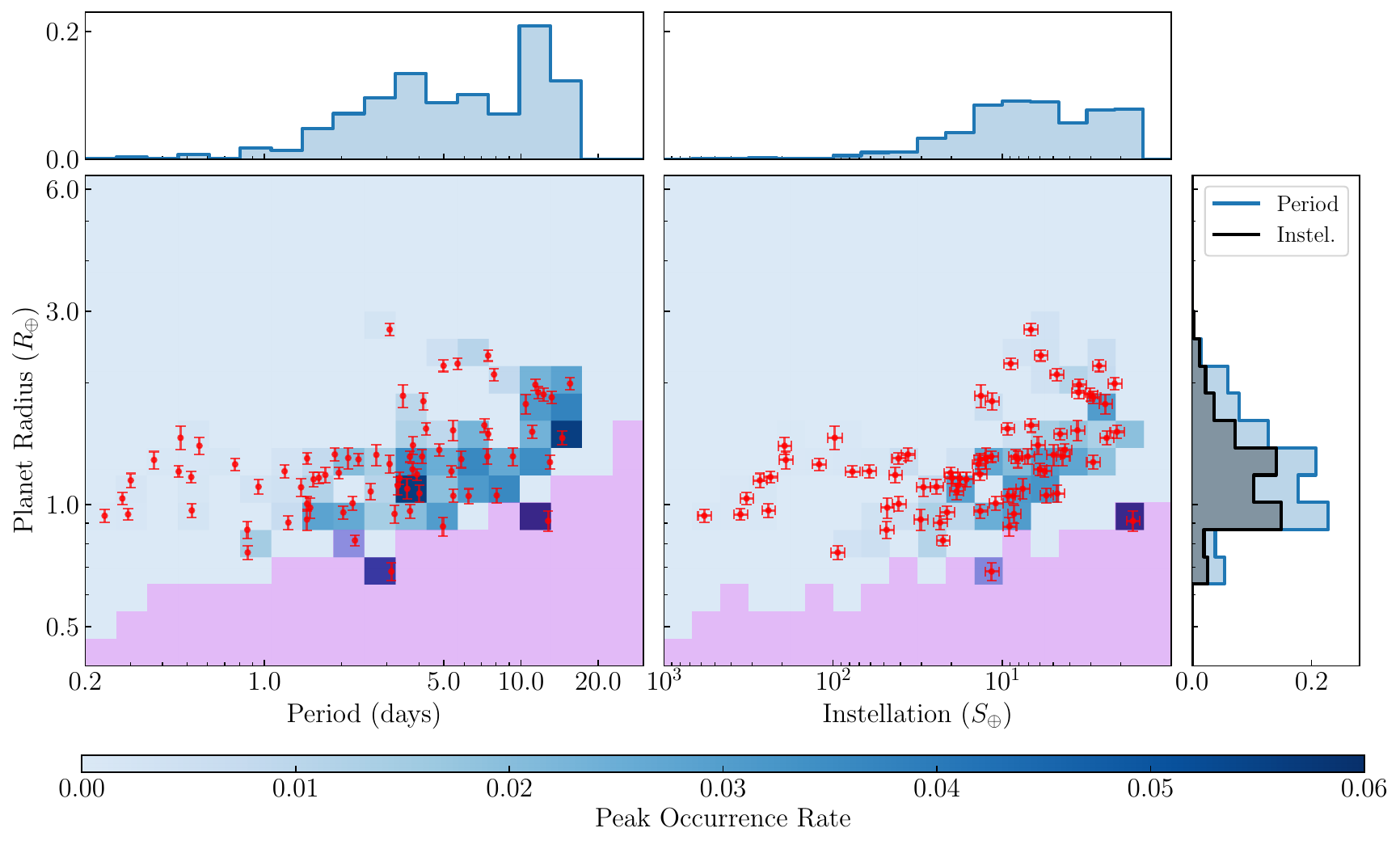}
\caption{Distribution of our found planets in period--radius space (left) and instellation--radius space (right). Each planet and its associated errors are are marked with the red points, while the background grid shows the peak occurrence rate $f_\mathrm{peak}$ in each bin of the grid. Regions where our average sensitivity is $<10\%$ are highlighted in pink. The top and leftmost panels show the cumulative $f_\mathrm{peak}$ distributions from each map. We note that these distributions are not representative of the full occurrence rates, and differ because planets may find themselves in different regions of sensitivity based on their instellation and period.}
    \label{fig:2d_occurrence}
\end{figure*}

Over our full set of survey targets, \rc{532 TCEs are identified that pass our pipeline's statistical vetting conditions. \texttt{TRICERATOPS} rejects 427 (80.2\%) of these with an FPP $>10\%$. Of the 105 remaining, \texttt{exovetter} rejects 5, and 23 are rejected by our manual vetting.} After statistical and manual vetting, our pipeline recovers a total of 77 transiting PCs around 65 different stars. The periods, radii and instellations of these PCs are overplotted in Figure~\ref{fig:2d_occurrence} (see Section~\ref{sec:2d_occ}). These PCs represent the set of planets that we will use to calculate planetary occurrence rates in Section~\ref{sec:occrate}. The remaining PC recovered by our pipeline is a newly discovered PC around TIC 149927512, which we detail below. Among these 77 PCs, 76 have transit parameters consistent with a corresponding TOIs.

\rc{Notably absent from our PC sample are planets with orbital periods between 20--30 days. According to the NASA Exoplanet Archive \citep{exoplanet_archive} queried on December 18 2025, there are six such confirmed transiting planets around mid-to-late M dwarfs. All but one of these planets' host stars\footnote{\rc{TRAPPIST-1 h \citep{trappist1h}, Kepler-1512 b \citep{k1512b}, Kepler-54 d \citep{k54d}, K2-72 e \citep{k272e}, TOI-1227 b \citep{toi1227b}.}} are absent from our stellar sample. The one exception is LHS 1140 \citep{LHS1140,LHS1140RV}, which hosts the sub-Neptune LHS 1140 b with a period of 24.7 days. LHS 1140 was observed by TESS in sectors 3 and 30, and as such it was not observed in consecutive sectors. By the design of our planet search, we can only detect planets out to 14 days in single sectors making LHS 1140 b undetectable. We note that we do detect LHS 1140 c \citep{LHS1140c}.}

Out of the 77 vetted PCs detected, only 10 are found around targets with strong signals of rotation where a GP was used to detrend the light curve. Additionally only two of these ten have $R_p/R_\star > 0.05$. Since these planets do not represent a significant fraction of our planet sample, and because we have established in Figure~\ref{fig:gp_recovery} that the detrimental impact of the GP on measured transit depth is relatively small ($\sim10\%$ for $R_p/R_\star < 0.05$), we do not attempt to correct the recovered radii of planets detected in light curves detrended by a GP model. 

Our full set of recovered planetary parameters is available in machine readable form and we present a subset of this table in Appendix~\ref{app:pc_tab}. 

We detect a previously unidentified signal around TIC 149927512, which does not appear to have been previously identified as a known planet, a TOI, or a Community TOI (CTOI). This PC passes all of our vetting conditions. We recovered this transit-like signal around TIC 149927512 ($M_\star = 0.25\,M_\odot, R_\star = 0.27\,R_\odot, T_{\rm eff} = 3269$ K) with an orbital period of $P=1.4646834^{+0.0000028}_{-0.0000024}$ days and a radius of $R_p=0.917^{+0.070}_{-0.067}\, R_\oplus$. TIC 149927512 was observed by TESS in sectors 14, 21, 41, 48, 74 and 75, and we detected the planet candidate in the TLS results of sector 21 and the combined sectors 74 and 75. The corresponding in-transit centroid offsets, TLS results, PDCSAP light curves, phase-folded light curve with fitted transit model, and transiting planet parameters are outlined in Appendix~\ref{app:tic149927512}. We note that the inclusion or omission of this PC in our subsequent occurrence rate calculation does not significantly impact our derived occurrence rates.

\subsection{Comparison to known TOIs}
\begin{figure}
    \centering
    \includegraphics[width=\linewidth]{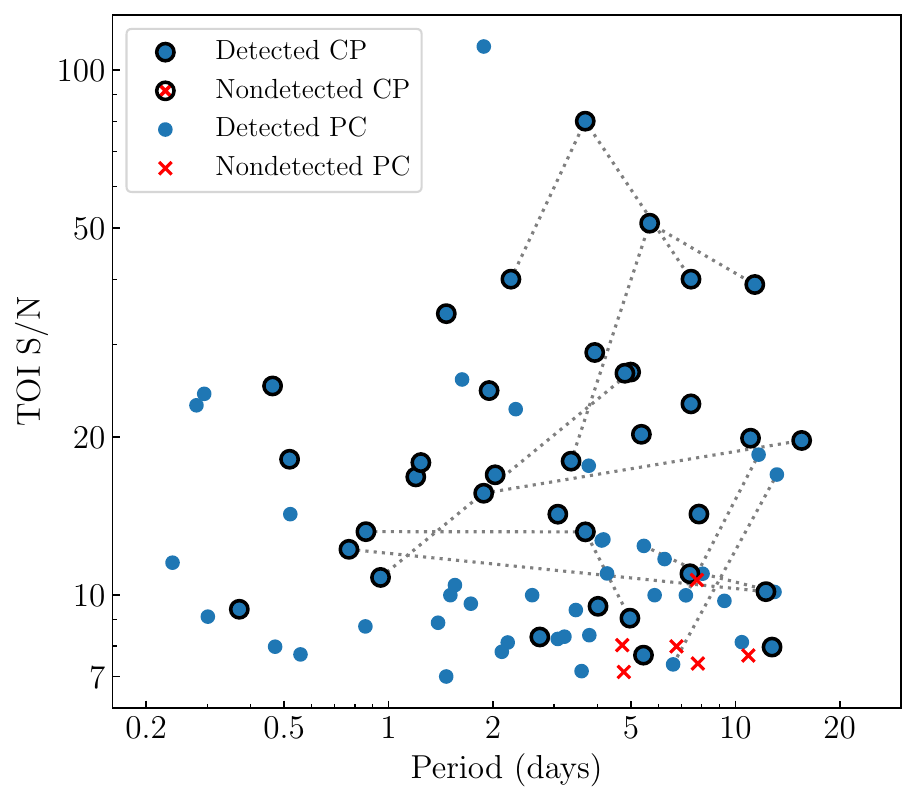}
    \caption{Detected and nondetected TOIs (as of October 2024) among stars in our survey sample as a function of orbital period and transit S/N. Detected and nondetected TOIs are depicted by circle markers and ``X'' markers, respectively. Confirmed planet (CP) TOIs are encircled while planet candidate (PC) TOIs are not. Known false alarms and false positives are excluded.   Multiplanet systems are connected with dashed lines.}
    \label{fig:toi-finds}
\end{figure}

To complement the detailed characterization of our pipeline's detection sensitivity via injection--recovery tests (see Section~\ref{subsec:injrec}), here we assess the recovery rate of known TOIs in our survey sample. Figure~\ref{fig:toi-finds} depicts the sample of TOIs in our survey sample as of October 2024 that have not been vetted as false alarms or false positives, and includes confirmed/known planets as well as TOIs whose dispositions remain as PCs. \rc{The set of recovered TOIs have a median S/N of 12.98 and a minimum S/N of 6.76 (TOI 6258.01).}  We note that our pipeline finds and accurately characterizes all 36 confirmed planets in our sample and 74/80 (93\%) of all TOIs (including confirmed planets) in our sample as of October 2024. 

Here we detail why our pipeline fails to recover six TOIs, each of which are currently considered to be non-confirmed (i.e. planet candidates) and therefore may ultimately be revealed to be false positives. Our pipeline rejects TOI-3494.01 (S/N 10.7 at 7.74 days) during vetting due to the in-transit centroids showing a systematic offset from the target and an on-target probability of less than 1\%. We therefore interpret TOI-3494.01 as a probable false positive. TOIs 2495.01, 5716.01, 6002.01, 6714.01, and 6716.01 were likely missed by our pipeline because our TLS searches did not produce any TCEs with periods that closely match the TOIs. All of these signals exhibit a low transit $\mathrm{S/N} < 8$. These TOIs may have been missed by our pipeline because unlike the TESS Transiting Planet Search \ \citep[TPS;][]{Jenkins2002,Jenkins2010}, we did not conduct our transit search over all available TESS data simultaneously due to runtime limitations, which limits our ability to recover low S/N signals.

\section{Occurrence Rate Calculation} \label{sec:occrate}

\subsection{Planetary Occurrence versus Planet Radius, Orbital Period, and Instellation} \label{sec:2d_occ}
We calculate the planetary occurrence rate distribution over planet radius--period space and planet radius--instellation space, denoted $f(P,R_P)$ and $f(S,R_P)$, respectively. We bin these parameter spaces into discrete grids to calculate the cumulative occurrence rate within each grid cell. Following \citet[][hereafter \citetalias{Dressing_2015}]{Dressing_2015}, we measure the occurrence rate in each cell as a binomial distribution. This follows from the view that the occurrence rate measurement in each bin is the result of a Bernoulli experiment whose probability of success depends on the product of the occurrence rate $f$ and completeness $c$. For brevity, and without loss of generality, we outline the computation of $f(P,R_p)$ because the calculation of $f(S,R_p)$ is analogous. 

The occurrence rate $f(P,R_p)$ depends on the number of recovered planets $N_p(P,R_p)$ (see Section~\ref{sec:planet_search}) and our average survey completeness $c(P,R_p)$. We calculated the latter from the binned sensitivity maps shown in Figure~\ref{fig:injecrec_map} by weighting each map by the total number of targets therein. We note that we also tested calculating the occurrence rate map $f(P,R_p)$ using the individual sensitivity bins in Figure~\ref{fig:injecrec_map} by calculating six independent $f(P,R_p)$ maps (one for each injection--recovery bin), based on the subset of $N_p(P,R_p)$ whose host star properties fall with each bin of interest (i.e. $T_{\rm mag}$ and whether rotation is detected), and averaging the six maps. We confirm that our final occurrence rate inferences do not differ depending on the choice of using the average versus individual sensitivity maps. 

The number of trials of each Bernoulli experiment is the number of stars searched $N_\star$, and the number of successful experiments (i.e. planet detections) is the number of planets $N_p(P, R_p)$ found by our planet search. Dropping the function arguments $P$ and $R_p$ for brevity, the posterior probability distribution for the occurrence rate $f$ is given by
\begin{equation}
P(f|N_p,N_\star,c) = \genfrac{(}{)}{0pt}{}{N_\star}{N_p} ({fc})^{N_p} (1-fc)^{N_\star - N_p}.
     \label{eq:binomial_p}
\end{equation}

\noindent It follows that the most likely value of $f(P,R_p)$ is 

\begin{equation}
    f_\text{peak}(P,R_p) = \frac{N_p(P,R_p)}{N_\star\cdot c(P,R_p)}.
\end{equation}

When calculating the cumulative occurrence over a large range of radii, periods, or instellations, using this method will lead to an inaccurate posterior if the survey's completeness dynamically changes over the bins of interest. In these cases, we compute the occurrence over a log-log grid that spans the space of interest. We compute a cumulative occurrence $P(\text{Occ.}=f)$ distribution by summing samples from the grid where $N_p \geq 1$. In cases where the range of interest contains no found planets, we instead sample from all posteriors in the grid and report the average weighted by the span of each grid cell in log-log space.

\subsection{The 2D Occurrence Rate Distribution}
In Table \ref{tab:2docc_table}, we outline the occurrence rates of planets in period--radius and instellation--radius spaces with our peak occurrence rate maps $f_{\mathrm{peak}}(P,R_p)$ and $f_{\mathrm{peak}}(S,R_p)$ shown in Figure~\ref{fig:2d_occurrence}. The planetary parameters recovered in our survey are over-plotted. The panels in the top row and rightmost column of Figure~\ref{fig:2d_occurrence} depict the cumulative marginalized distributions of $f_{\mathrm{peak}}(P)$, $f_{\mathrm{peak}}(S)$, and $f_{\mathrm{peak}}(R_p)$, respectively. It should be noted that these distributions are not fully reflective of the cumulative planet occurrence rate as they are based solely on the most-likely occurrence rate values and do not account for the associated uncertainty in each cell. 

We also highlight the regions in each panel where our average detection sensitivity is $<10$\%. Any occurrence rate calculations over the highlighted regions of low sensitivity ($<10\%$) are not considered reliable for computing the occurrence rate of planets. We generally planets in this region are not counted towards most of our our subsequent occurrence rate calculations, save for the ranges in Tables \ref{tab:2docc_table}, \rcnew{\ref{tab:period_comparison} and \ref{tab:instel_comparison} (see section~\ref{sec:mc23_compare})}, and our habitable zone planet occurrence rates. Aside from $P>14$ days, where our sensitivity falls rapidly with increasing orbital period primarily because only 43\% of our targets are searched for planets beyond 14 days (c.f. Figure~\ref{fig:full_completeness}), the decline in our sensitivity occurs between $0.4-1\, R_\oplus$ depending on the period or instellation.

One planet of note that lies in this region is Gliese 12 b \citep[$P=12.76$ days, $R_p=0.95\, R_\oplus$][]{GJ12-2024-2, GJ12-2024-1, GJ12-2025-1, GJ12-2025-2}. This discovery has been rigorously confirmed \rc{and provides a rare example of a temperate Earth-sized planet.} Our \rc{survey-averaged} sensitivity to this planet falls below 10\% and so we do not trust its contribution to our occurrence rate inferences, \rc{although our sensitivity to Gliese 12 b in its injection--recovery bin is $>30$\% due to the brightness of its host star ($T_{\rm mag}=10.18$) such that its detection in our survey is not unexpected.}

The planetary occurrence rate calculations in Figure~\ref{fig:2d_occurrence} clearly shows that the planet population around mid-to-late M dwarfs is dominated by sub- and super-Earths ($R_p\lesssim 1.6\, R_\oplus$). With only five sub-Neptunes with $R_p\geq 2\,R_\oplus$ in our planet sample, the occurrence rate of sub-Neptunes is less than for terrestrial bodies at all periods out to 15 days. None of our detected sub-Neptunes with $R_p>2\, R_\oplus$ have orbital periods within 2 days, which may suggest that such planets, that would likely harbor a large mass fraction of volatiles (i.e. H$_2$/He or water) on their surface, are unstable to thermally-driven atmospheric loss. 
We also find no planets with $R_p \geq3\, R_\oplus$ implying that the occurrence rates of Neptunes and gas giants, both hot and temperate, are consistent with 
\rc{zero\footnote{Neptunes and Jovian planets have occasionally been found around mid-to-late M dwarfs \citep[e.g.][]{neptune_2, neptune_3, neptune_1, neptune_4}, none of which our in our sample of stars. We find that their occurrence rate is consistent with zero in our study, implying the rarity of these objects.}} around mid-to-late M dwarfs, with our upper limits across periods from 0.2-30 days outlined in Table \ref{tab:2docc_table}. The unimodality of the $f(P,R_p)$ and $f(S,R_p)$ distributions stand in contrast to period/instellation--radius distributions around earlier type stars, specifically the occurrence rate of planets around late K to mid-M dwarfs \citepalias{cloutier_evolution_2020} and the occurrence rate of planets around FGK stars \citep[e.g][]{fulton_california-kepler_2017}, which both show evidence for the bimodal Radius Valley in these 2-dimensional parameter spaces.

\begin{table*}[ht]
    \centering
    \begin{tabular}{cccccc}
    \toprule
         $R_p\,(R_\oplus)$ & $0.2-1$ days & $1-3$ days & $3-7$ days & $7-15$ days & $15-30$ days \\ \midrule
         $0.5-1.0$ & $0.035_{-0.014}^{+0.014}  (19.4\%)$ & $0.101_{-0.036}^{+0.035}  (9.9\%)$ & $0.343_{-0.143}^{+0.143}  (6.1\%)$ & $0.390_{-0.239}^{+0.259}  (3.2\%)$ & $<0.88  (1.2\%)$ \\
         $1.0-1.5$ & $0.021_{-0.005}^{+0.005}  (64.8\%)$ & $0.141_{-0.028}^{+0.028}  (41.9\%)$ & $0.401_{-0.094}^{+0.094}  (28.2\%)$ & $0.428_{-0.140}^{+0.140}  (17.1\%)$ & $<0.45  (6.5\%)$ \\
         $1.5-2.0$ & $<9.7\times10^{-4}   (91.3\%)$ & $<2.9\times10^{-3}  (77.7\%)$ & $0.047_{-0.017}^{+0.017}  (60.3\%)$ & $0.186_{-0.053}^{+0.053}  (40.7\%)$ & $0.058_{-0.038}^{+0.038}  (17.0\%)$ \\
         $2.0-3.0$ & $<9.2\times10^{-4} (97.4\%)$ & $<2.2\times10^{-3}  (94.1\%)$& $0.022_{-0.009}^{+0.009}  (87.2\%)$ & $0.025_{-0.012}^{+0.012}  (67.7\%)$ & $<0.09  (23.3\%)$ \\
         $3.0-6.0$ & $<9.8\times10^{-4}  (99.5\%)$ & $<2.9\times10^{-3}    (98.3\%)$ &  $<4.8\times10^{-3}  (96.8\%)$ & $<1.0\times10^{-2}  (80.2\%)$ & $<6.4\times10^{-2}  (28.8\%)$\\
         \bottomrule \\
         \toprule
         $R_p\,(R_\oplus)$ & $10^3-10^2  S_\oplus$ & $10^2- 50  S_\oplus$ & $50-10  S_\oplus$ & $10-5  S_\oplus$ & $5-1 S_\oplus$ \\ \midrule
         $0.5-1.0$ & $0.003_{-0.002}^{+0.002}  (23.4\%)$ & $0.009_{-0.006}^{+0.006}  (16.8\%)$ & $0.176_{-0.060}^{+0.060}  (11.1\%)$ & $0.059_{-0.029}^{+0.029}  (8.0\%)$ & $0.294_{-0.187}^{+0.196}  (5.2\%)$ \\
         $1.0-1.5$ & $0.008_{-0.002}^{+0.002}  (68.6\%)$ & $0.009_{-0.003}^{+0.003}  (55.6\%)$ & $0.115_{-0.023}^{+0.023}  (42.9\%)$ & $0.173_{-0.041}^{+0.041}  (30.8\%)$ & $0.178_{-0.053}^{+0.053}  (20.9\%)$ \\
         $1.5-2.0$ & $<9.6\times10^{-4}   (91.0\%)$ & $<1.2\times10^{-3}  (86.2\%)$ & $0.010_{-0.005}^{+0.005}  (77.3\%)$ & $0.018_{-0.008}^{+0.008}  (62.1\%)$ & $0.141_{-0.037}^{+0.037}  (43.6\%)$ \\
         $2.0-3.0$ & $<7.3\times10^{-4}  (89.2\%)$ & $<1.5\times10^{-3}   (95.8\%)$ & $<1.9\times10^{-3} (93.6\%)$ & $0.019_{-0.008}^{+0.008}  (86.4\%)$ & $0.020_{-0.010}^{+0.010}  (66.3\%)$\\
         $3.0-6.0$ &  $<7.8\times10^{-4}  (90.4\%)$ & $<2.7\times10^{-3}   (97.8\%)$ &  $<2.2\times10^{-3}   (98.3\%)$ & $<3.5\times10^{-3}  (96.8\%)$ & $<8.8\times10^{-3}  (73.4\%)$ \\
         \bottomrule
    \end{tabular}
    \caption{Top: Number of planets per star vs orbital period and planetary radius. Bottom: Number of planets per star vs instellation and planetary radius. In this table the numbers in parentheses are the target weighted percentage of all planets which were recovered in our injection--recovery tests in the given intervals.}
    \label{tab:2docc_table}
\end{table*} 

\subsection{The Planetary Radius Distribution}
We derive the planetary occurrence rate as function of radius $f(R_p)$ by marginalizing $f(P,R_p)$ (c.f.\ Figure~\ref{fig:2d_occurrence}) over orbital period.  Due to our low sensitivity to small planets, we only consider our cumulative radius distribution reliable for planets with $R_p>1\,R_\oplus$ where our average sensitivity out to 30 days is greater than 20\%. By sampling the distributions in each cell of Figure \ref{fig:2d_occurrence} from their associated posteriors parametrized by equation \ref{eq:binomial_p} and integrating over orbital period, we obtain many realizations of the radius distribution of planets. We do not consider bins where $N_p=0$ in this calculation, as samples from many distributions with non-zero tails adds an arbitrarily large amount of occurrence to the cumulative occurrence rate; however, the distributions in regions with zero occurrence can still be used to set an upper limit on the occurrence of planets therein.

We also independently estimate the underlying distribution of planets using a Gaussian kernel density estimator (KDE) with our completeness-weighted planet sample. We use the KDE to quantify the peak of the super-Earth distribution to compare the corresponding radius to the population of planets around different spectral types. We only consider planets with $R_p>1.0\, R_\oplus$ in this calculation because of the aforementioned sensitivity conditions. To determine the error on this calculation, we recompute the KDE using a bootstrapped sample with periods and radii drawn from each planet's posterior distribution. Each planet's sampled weight is also taken from a smooth completeness map where each injection is represented as a Gaussian width equal to one third of the width of the quadtree cell (see Figure \ref{fig:injecrec_map}) in which the injection was sampled. 

\begin{figure}
    \centering
    \includegraphics[width=\linewidth]{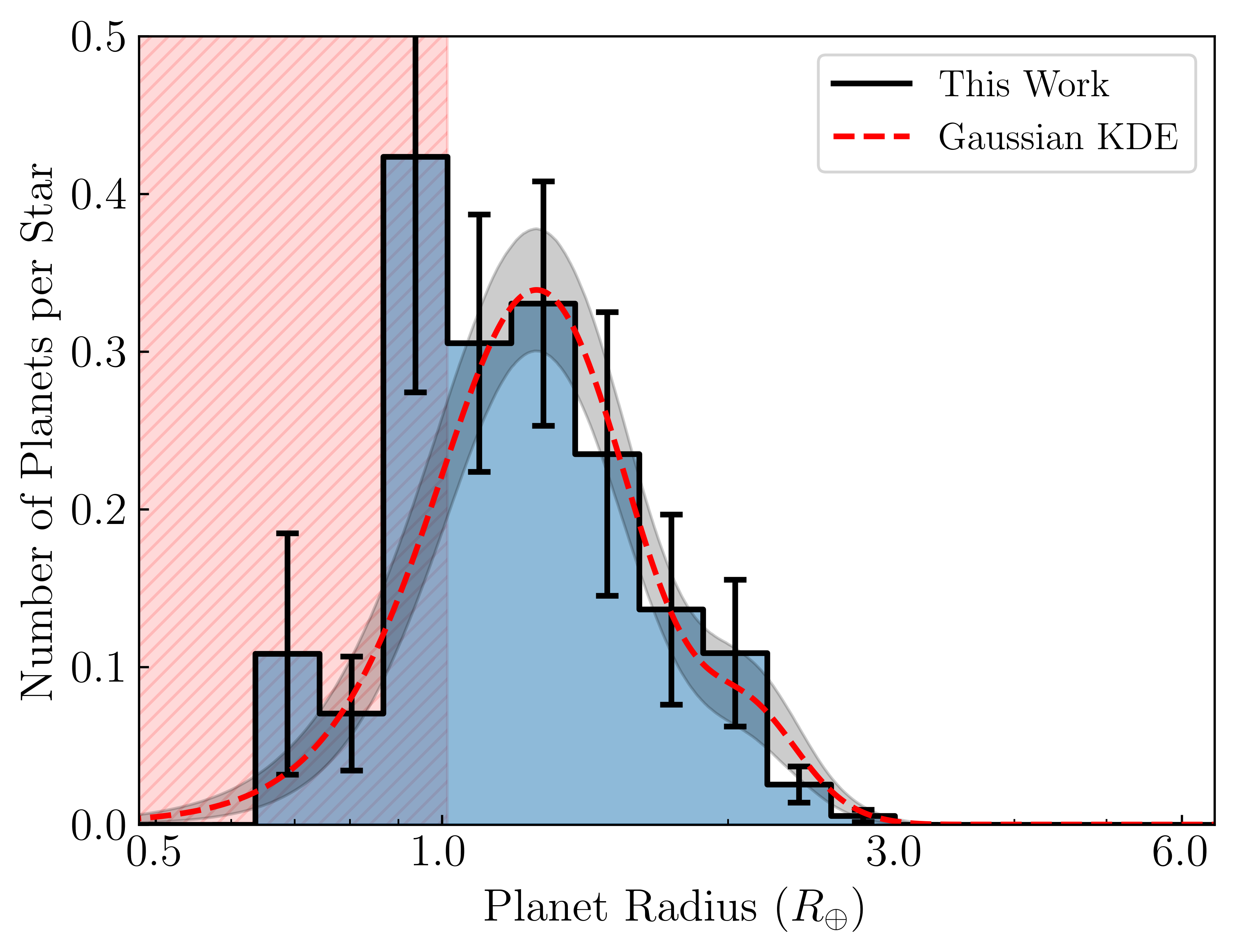}
    \caption{Occurrence rate of planets in radius space, realized by sampling the posterior distribution in each bin of Figure \ref{fig:2d_occurrence} with a Gaussian KDE of the occurrence rate overplotted. The red shaded region marks the area of low sensitivity where we consider our results unreliable.}
    \label{fig:radius_dist_err}
\end{figure}

These distributions of planet occurrence rates with their associated errors are shown in Figure \ref{fig:radius_dist_err}. From the distribution plotted in Figure \ref{fig:radius_dist_err} it is clear that the population of planets around mid-to-late M dwarfs orbiting within 30 days is unimodal without any sign of a Radius Valley. Through our bootstrapped KDE samples, we find that the super-Earth peak around mid-to-late M dwarfs lies at $1.25\pm0.05\, R_\oplus$.

\begin{figure}
    \centering
    \includegraphics[width=\linewidth]{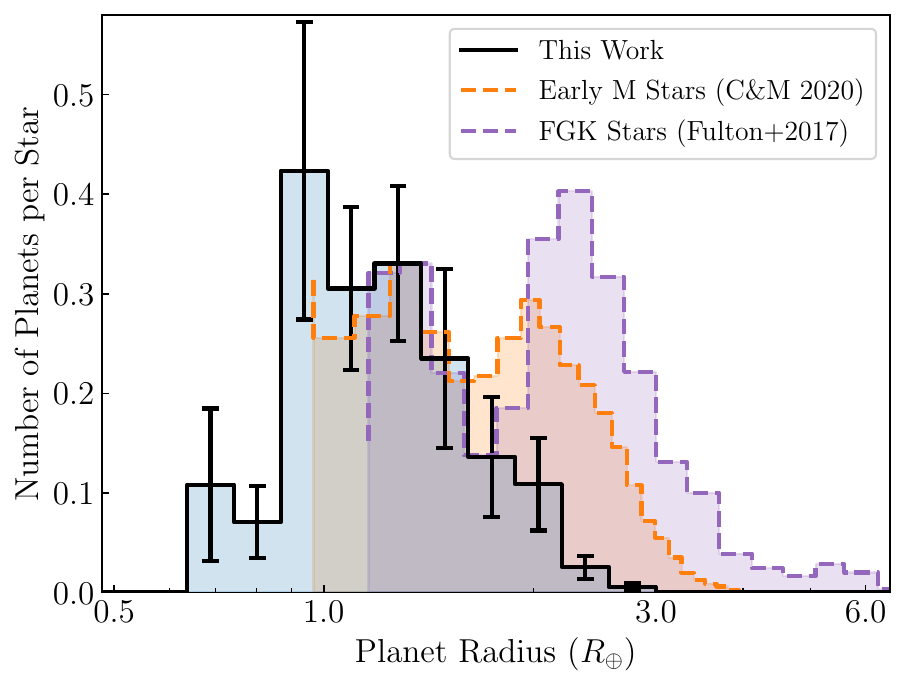}
    \caption{Comparison of the our occurrence rate to the scale occurrence rates found by \cite{fulton_california-kepler_2017} and \citetalias{cloutier_evolution_2020}, with each super-Earth peak scaled to match our super-Earth peak.}
    \label{fig:radius_dist_compare}
\end{figure}

In Figure \ref{fig:radius_dist_compare}, we also compare our radius distribution to similar distributions around early-M to late-K dwarfs \citepalias{cloutier_evolution_2020}, and FGK stars \citep{fulton_california-kepler_2017}. For convenience of comparison, their super-Earth peaks are scaled to our super-Earth peak. We note that both of these surveys searched for planets out to 100 days, and we are limited to 30 days. We also note that because our sample of M stars are dimmer, the range of instellations being probed is similar. These distributions show that early M dwarfs and FGK stars have substantial populations of sub-Neptunes, separated from super-Earths by a Radius Valley. Conversely, we only see a population of super-Earths, which is similarly shaped, with a sharp monotonic decline toward larger radii.

To determine the precise contributions of super-Earths and sub-Neptunes to the occurrence rate of planets around mid-to-late M dwarfs, we carry out the same binned occurrence rate calculation from 0.2--30 days over two radius ranges representing these two planet types.  We find a cumulative occurrence rate of super-Earths with $(R_p = 1  - 1.6\, R_\oplus)$ of \SEocc and a cumulative occurrence rate of sub-Neptunes with $(R_p =1.8-3\, R_\oplus)$ of \SNocc. While sub-Neptunes are represented in our sample of planets, these relative occurrence rates show that super-Earths dominate the planet population, outnumbering sub-Neptunes by a factor of $5.5\pm1.1$ to 1 around our sample of mid-to-late M dwarfs.

Without any Neptune-sized planets in our sample, we can only set an upper limit on the occurrence rate of Neptune-sized planets around mid-to-late M dwarfs. We sample the binomial distributions spanning periods from 0.2 to 30 days and radii spanning $R_p = 3-6.5\, R_\oplus$. We find a 95th percentile upper limit to the occurrence rate of Neptunes around mid-to-late M dwarfs of $\leq0.0341$ planets per star.

\subsection{The Occurrence Rate of Jovian-sized Objects}
In our transit search we detect one PC outside of the radius range spanned by our injection--recovery tests. TOI-1779.01, or LP 261-75 C, for which we measure a radius of $9.7 \pm 0.3\, R_\oplus$, a period of 1.88 days. This object is a known brown dwarf with a mass of $68.1 \pm 10\, M_\mathrm{Jup}$ \citep{Irwin_2018}. 

Our injection--recovery tests have not directly probed our pipeline's sensitivity to transiting planets with $R_p > 6.5\, R_\oplus$, but we can safely assume that our sensitivity to larger objects for the same orbital period is at least as good as what we calculate at the upper end of our radius range. We therefore posit that we have complete sensitivity to transiting planets with $R_p>6.5\,R_\oplus$ out to 2 days, which encapsulates the one Jovian-sized object found in our survey. Based on this single detection we measure a cumulative occurrence rate of $0.0028_{-0.0015}^{+0.0027}$ Jovian-sized objects per mid-to-late M dwarf orbiting within 10 days. Beyond this period limit our sensitivity to Jovians degrades by a factor of $\sim 2$. We emphasize that this cumulative occurrence rate is of Jovian-sized objects because it is based solely on our transit survey data, which is insufficient to distinguish giant planets from sub-stellar objects.

That being said, LP 261-75 C is not truly a planet. Determining its provenance is beyond the scope of our survey and its status as a stellar companion does not invalidate our measurement of the occurrence rate of Jovian-sized objects. With this additional information however, we can place an upper limit on the occurrence rate of hot Jupiters around mid-to-late M dwarfs. Assuming that our sensitivity to hot Jupiters is at least as good as our sensitivity to $6.5\, R_\oplus$ planets, we find a 95th percentile upper limit of 0.012 hot Jupiters per star orbiting within 10 days. \rc{This upper limit is consistent with the giant planet occurrence of $0.0029\pm0.0015$ around mid-to-late M dwarfs found by \citet{tess_giant_occurrence}.}

\subsection{Occurrence of Planets within the Habitable Zone} \label{sect:hz}
While our sensitivity to planets in the HZ is drops off steeply beyond $\sim1.5S_\oplus$, we can still place constraints on the occurrence rate of planets within the habitable zone of mid-to-late M dwarfs.

 \rc{We can assume empirical optimistic HZ limits of $0.2-2S_\oplus$. The inner edge is bounded by the ``recent Venus" scenario motivated by the inference that Venus had liquid water on its surface during its history \citep{early_venus}, and the outer edge emerges from the maximum greenhouse limit \citep{hab_zone_koppararu}.} In these bounds, we find a cumulative occurrence rate of $0.178_{-0.115}^{+0.116}$ planets per star with radii between $0.4-6.5\, R_\oplus$, and an occurrence rate of terrestrial planets between $0.8-1.5\, R_\oplus$ of $0.142_{-0.092}^{+0.093}$ planets per star. 

\citet{hab_zone_koppararu} provides a more conservative HZ range of $0.2-0.9 S_\oplus$ that corresponds to the water-loss inner edge and maximum greenhouse outer edge of the HZ. We do not find any planets in our survey in this instellation range, and our sensitivity is poor, but we can still place 95\textsuperscript{th} percentile upper limits on the occurrence rate of planets in this range. We find that the occurrence rate of HZ planets with radii between $0.4-6.5\, R_\oplus$ is $<0.19$ planets per star, and the occurrence rate of terrestrial ($0.8-1.5\, R_\oplus$) habitable planets is $<0.351$ planets per star.  We note that this latter upper limit is larger than the former despite spanning a smaller radius range because we have poorer average sensitivity to terrestrial planets compared to all planets from $0.4-6.5\, R_\oplus$. In essence, no more than 1-in-3 mid-to-late M dwarfs hosts a terrestrial planet in their habitable zone.

\section{Discussion} \label{sec:discussion}

\subsection{Comparison to Ment \& Charbonneau 2023} \label{sec:mc23_compare}
The most appropriate direct comparison we can make to this occurrence rate calculation is \citetalias{ment_occurrence_2023}  with their survey of 363 mid-to-late M dwarfs within 15 parsecs, searching for transiting planets out to 7 days. Their survey does provide constraints for the occurrence rate of super-Earths, but without any sub-Neptune detections, they could only set an upper limit on the occurrence of any second peak of the Radius Valley. By cutting the search space of our survey, we can directly compare the results of our survey to this previous occurrence rate calculation. In Tables \ref{tab:period_comparison} and \ref{tab:instel_comparison}, we outline our relative occurrence rates. 
\begin{table}
    \centering
    \begin{tabular}{cccc} \toprule
     \multicolumn{2}{c}{Source} & \citetalias{ment_occurrence_2023}  & This Work \\
     \multicolumn{2}{c}{Mass range} &  $0.1-0.3M_\odot$ &  $0.1-0.4M_\odot$ \\  
     \multicolumn{2}{c}{Median Mass} &  $0.17M_\odot$ &  $0.22M_\odot$ \\ \cmidrule{3-4} 
     Period (days) & $R_p$ ($R_\oplus$) & \multicolumn{2}{c}{Occurrence Rate by Source}\\ 
     \midrule
     0.2-1 & 0.5-4.0 & $0.047_{-0.015}^{+0.019}$ & $0.052\pm 0.014$ \\
     1-2.7 & & $0.159_{-0.050}^{+0.064}$ & $0.175_{-0.046}^{+0.045}$ \\
     2.7-7 & & $0.402_{-0.127}^{+0.162}$ & $0.530\pm 0.099$ \\
     7-30 & & -* & $0.510_{-0.152}^{+0.151}$ \\ \midrule
     0.2-30 & 1.0-1.5 &  $0.446^{+0.162}_{-0.118}$* & $0.785_{-0.144}^{+0.142}$ \\
     & 1.5-2.0 & $\leq0.073$* & $0.297\pm0.075$ \\
     & $>2.0$ & $\leq0.072$* & $0.040\pm0.013$ \\ \cmidrule{2-4}
     & 0.5-4.0 & $0.61^{+0.24}_{-0.19}*$ & $1.358_{-0.195}^{+0.195}$ \\ \bottomrule
    \end{tabular}
    \caption{Comparison of our occurrence rate calculation in planet radius and period to \citet{ment_occurrence_2023} (MC23). This table is an adapted version of Table 7 from \citet{ment_occurrence_2023}. *The occurrence rate calculation from \citet{ment_occurrence_2023} extends from 0.2-7 days.}
    \label{tab:period_comparison}
\end{table} 
\begin{table}
    \centering
    \begin{tabular}{cccc} \toprule
     \multicolumn{2}{c}{Source}  & \citetalias{ment_occurrence_2023}  & This Work \\
     \multicolumn{2}{c}{Mass range} &  $0.1-0.3M_\odot$ &  $0.1-0.4M_\odot$ \\  
     \multicolumn{2}{c}{Median Mass} &  $0.17M_\odot$ &  $0.22M_\odot$ \\ \cmidrule{3-4} 
     Instel. ($S_\oplus$) & $R_p$ ($R_\oplus$) & \multicolumn{2}{c}{Occurrence Rate by Source}\\ 
     \midrule
     4-10 & 0.5-4.0 & $0.303_{-0.095}^{+0.120}$ & $0.354\pm0.062$ \\
     10-50 & &  $0.162^{+0.064}_{-0.051}$ & $0.216\pm0.043$ \\
     50-200 & &  $0.030_{-0.010}^{+0.012}$ & $0.018\pm0.006$ \\ \midrule
     4-200 & 0.5-1.0 &   $0.107_{-0.055}^{+0.098}$ & $0.228\pm0.061^*$ \\
     & 1.0-1.5 &    $0.370^{+0.162}_{-0.118}$ & $0.319_{-0.055}^{+0.056}$\\
     & 1.5-2.0 & $\leq0.060$ & $0.298 \pm 0.070$ \\
     & $>2.0$ & $\leq0.056$ & $0.044 \pm 0.015$ \\ \cmidrule{2-4}
     & 0.5-4.0 & \rc{$0.49^{+0.19}_{-0.15}$} & $0.571_{-0.074}^{+0.075}$\\ \bottomrule
    \end{tabular}
    \caption{Comparison of our occurrence rate calculation in planet radius and instellation to \citet{ment_occurrence_2023} (MC23). This table is an adapted version of \rc{Table 8} from \citet{ment_occurrence_2023}. *Because of our low sensitivity, we consider occurrence rates for planets $<1R_\oplus$ unreliable.}
    \label{tab:instel_comparison}
\end{table} 
The results of our survey are in very close agreement with the findings of \citetalias{ment_occurrence_2023}, and the occurrence rate of sub-Neptunes lies below their lower limit. This agreement is expected, because our survey covers a very similar mass range, with our median masses differing by 0.05$M_\odot$. Our very low sub-Neptune occurrence rate means we can confidently say that the Radius Valley disappears in this regime. Because the stars in our sample are much fainter on average than the stars surveyed by \citetalias{ment_occurrence_2023} we consider their work to be more reliable in constraining the population of planets with $<1\, R_\oplus$ out to 7 days.

\subsection{Cumulative Planetary Occurrence Around Mid-to-Late M Dwarfs} 
We measure a cumulative occurrence rate of planets with radii $>1\, R_\oplus$ orbiting within 30 days of \cumocc planets per star in our sample of mid-to-late M dwarfs. To put this result into context and to illustrate the trend of cumulative occurrence over a range of stellar types, we have gathered four other transit surveys, characterizing occurrence rates in main sequence stars with effective temperatures from $2700\text{K}-7300\text{K}$. \citet{km_fgk} takes 96,280 Kepler stars and uses approximate Bayesian computation (ABC) to report the occurrence rate of planets around F $(6000\text{K}\leq T_\text{eff} \leq7300\text{K})$, G $(5300\text{K}\leq T_\text{eff} \leq6000\text{K})$ and K $(3900\text{K}\leq T_\text{eff} \leq5300\text{K})$ type stars. \citet{hsu2018} uses ABC to characterize the occurrence rate of planets around FGK type stars using the Kepler sample of planets, and \citet{hsu2020} applies same methodology to planets around M stars found by Kepler.
\citetalias{Dressing_2015} searches the four-year Kepler data set with a custom pipeline to place strong constraints on the cumulative occurrence rate of planets around a sample of M dwarfs with $2661\text{K}\leq T_\text{eff}\leq3999\text{K}$ and a mean effective temperature of $3746^{+217}_{-300}\,\mathrm{K}$, biasing their sample towards earlier type M stars than the stars characteristic of our sample. We also include the occurrence rate value from \citetalias{ment_occurrence_2023} as a lower limit because their survey only extends out to 7 days. We sample each of these surveys' reported occurrence rates from $1-4\, R_\oplus$ with $P<30$ days to derive a cumulative occurrence that can be readily compared to our occurrence rate. 

\begin{figure}
    \centering
    \includegraphics[width=\linewidth]{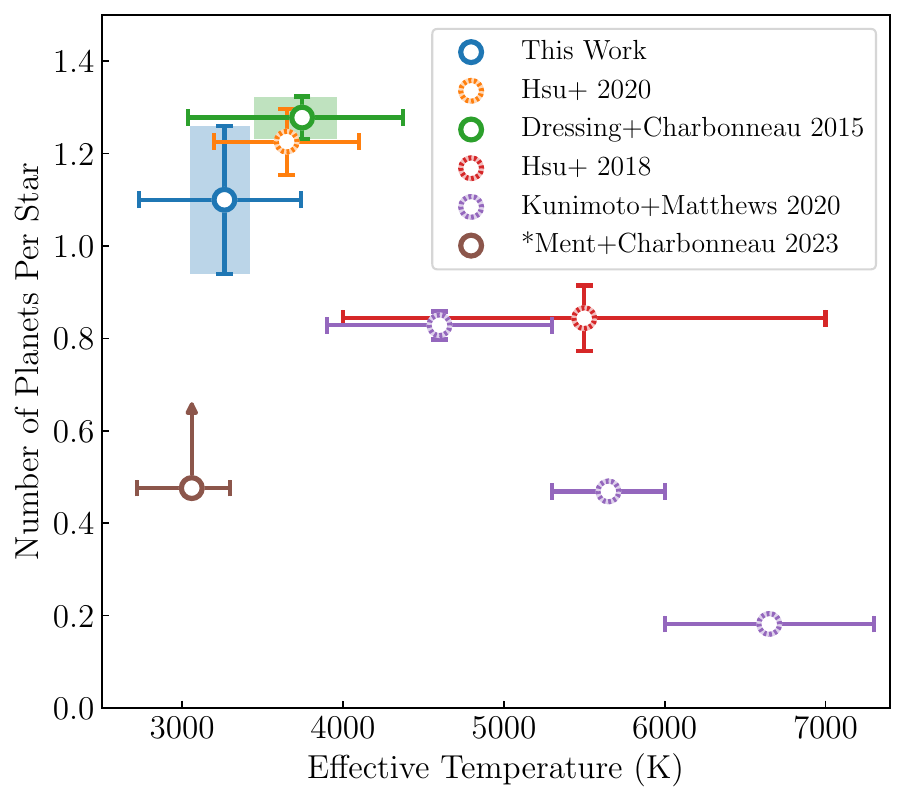}    
    \caption{Cumulative occurrence rates of planets with radii spanning $1-4\, R_\oplus$ and periods out to 30 days as a function of stellar effective temperature. The horizontal lines mark the recovered occurrence rate with the 1$\sigma$ uncertainty in each occurrence rate calculation marked by the errorbars. \rc{The full $T_\mathrm{eff}$ range of each work is marked by the horizontal extent of each marker, and the shaded region shows the $1\sigma$ $T_\mathrm{eff}$ range in the case of this work and \citetalias{Dressing_2015} when available.} The solid markers illustrate the mean effective temperatures, while the dashed markers mark the middle effective temperature in surveys where a mean effective temperature was not available. *\citetalias{ment_occurrence_2023} provides an lower limit as their survey extends out to 7 days.}
    \label{fig:cum_occ}
\end{figure}

In Figure~\ref{fig:cum_occ} we illustrate these occurrence rates with their associated $1\sigma$ errors, and mark the effective temperature range that each sample spans. The mean $T_\text{eff}$ is marked with a solid circle and we mark the middle $T_\text{eff}$ value with a dashed circle in cases where the mean $T_\text{eff}$ is not available. In general, we see an increasing trend of higher occurrence towards later stellar types, with M dwarfs, including early, mid, and late types, hosting the majority of planets larger than Earth and with periods $<30$ days. \rcnew{Contrasting our cumulative occurrence rate of \cumocc planets per star to the value around early M dwarfs \citepalias[$1.28\pm 0.05$;][]{Dressing_2015} confirms that early and mid-to-late M dwarfs host a consistent number of close-in planets larger than Earth and within 30 days. 
Our finding that the planet occurrence rate likely does not increase, and may decrease for the latest M dwarfs corroborates the predictions of \cite{brady2022}. We therefore conclude that the full suite of M dwarfs including early, mid, and late type M dwarfs are the most prolific hosts of small close-in planets. }

\subsection{Comparisons to Theoretical Predictions}
The TESS GI programs which secured the data used to complete our survey looked to establish the dominant driver of the Radius Valley by measuring the properties of the Radius valley around mid-to-late M dwarfs. While core-powered mass loss and photoevaporation make consistent predictions of the Radius Valley's location and slope for Sun-like stars, the Radius Valley locations they predict diverge towards mid-to-late M dwarfs \citep{wu_mass_2019, gupta_sculpting_2019}. Our finding that the Radius Valley does not exist around mid-to-late M dwarfs prevents our survey from directly disentangling their relative effects in sculpting the Radius Valley around higher-mass stars. This being said, the non-detection of a Radius Valley in this stellar mass regime still serves to support other Radius Valley emergence models.

\citet{burn_bern} uses the Bern New Generation Planetary Population Synthesis (NGPPS) models produce populations of planets around 0.3 $M_\odot$ and 0.1 $M_\odot$ stars. These stellar masses overlap with stars surveyed here. Their cumulative occurrence rate of planets with periods less than 100 days is dominated by sub-Earths and Earth-mass planets \rc{($0.5-2 M_\oplus$) with very few Neptunes ($10-30 M_\oplus$). 70\% of 0.1 $M_\odot$ stars and 88\% 0.3 $M_\odot$ stars host Earth-sized planets. Only 1\% of their $0.1 M_\odot$ star simulations host Neptunes, and 8\% of their $0.3 M_\odot$ stars host Neptunes.} Directly comparing these populations is difficult as we are not sensitive to planet mass in our transit survey. However, their finding that sub- and super-Earths dominate the planet population is consistent with our survey, with their simulations possibly producing more of these planets. Their higher occurrence of Neptunes is a discrepancy with our results, but our median stellar mass of 0.22M$_\odot$ and limited sensitivity to long-period planets may explain this difference. It bears noting that the Bern model's treatment of water's contribution to the radius of sub-Neptunes has been updated in \citet{burn_steam}, which they claim is needed to replicate the Radius Valley around Sun-like stars. However, the effect of this update on their planet population around mid-to-late M dwarfs is unknown because they have not yet considered these stars in their steam simulations.

A declining population of sub-Neptunes is predicted by \citet{chachan2023} in their work combining the theory pebble migration models with measured disk masses. They predict that as a function of disk dynamics, fewer sub-Neptunes should be expected around stars starting at $0.3–0.5M_\odot$. They also find that while late M disks are smaller, they have support greater pebble accretion efficiency, leading to a relatively flat occurrence rate of terrestrial planets towards smaller mass stars, which is also matched by our observations in comparison to \citetalias{cloutier_evolution_2020}.

The results of our survey also strongly agree with the prediction from \citet{venturini_radval} where a Radius Valley is carved out between rocky super-Earths and water-rich, sometimes steamy sub-Neptunes. This division fades towards lower mass stars where the two planet populations overlap in planet radius. Their simulated population of planets qualitatively matches the smooth occurrence rate that we find in our survey (Figure \ref{fig:radius_dist_err}), with a clear fall-off towards larger radii, with many fewer sub-Neptunes than higher mass stars. The \citet{venturini_radval} models suggest the Radius Valley, while clear around Sun-like stars, gets increasingly ``filled in" with decreasing stellar mass as the populations of super-Earths and water-rich worlds increasingly overlap in radius space. The Radius Valley detected around early M and late K type stars (\citetalias{cloutier_evolution_2020}) is shallower than the valley found around FGK stars \citep{fulton_california-kepler_2017} and we have shown that the Radius Valley effectively disappears around mid-to-late M dwarfs. These independent transit surveys targeting different host star spectral types support the theoretical picture presented in \citet{venturini_radval}, but deeper investigation is still necessary. If the planet population in our survey is explained by overlapping rocky and water worlds, a complete confirmation of their model's predictions is only possible with precise mass measurements and atmospheric characterization to confirm the water-rich nature of sub-Neptunes. Both of these measurements are beyond the scope of our paper, but are well suited to observations with JWST (e.g. \citealt{steam_jwst}).

\subsection{The Observed Radius Valley}
While our survey does not find a Radius Valley in this regime, we still find a peak in the super-Earth population. In comparing the population of planets across stellar types, it is useful to compare our observed population peak to the Radius Valley peaks seen around earlier type stars. \rc{Figure \ref{fig:radius_dist_compare} illustrates the falloff of the sub-Neptune peak in the Radius Valley across spectral types. This progression is qualitatively consistent with the planet populations produced by water rich pebble accretion scenario in \citet{venturini_radval}.}

\begin{figure}
    \centering
    \includegraphics[width=\linewidth]{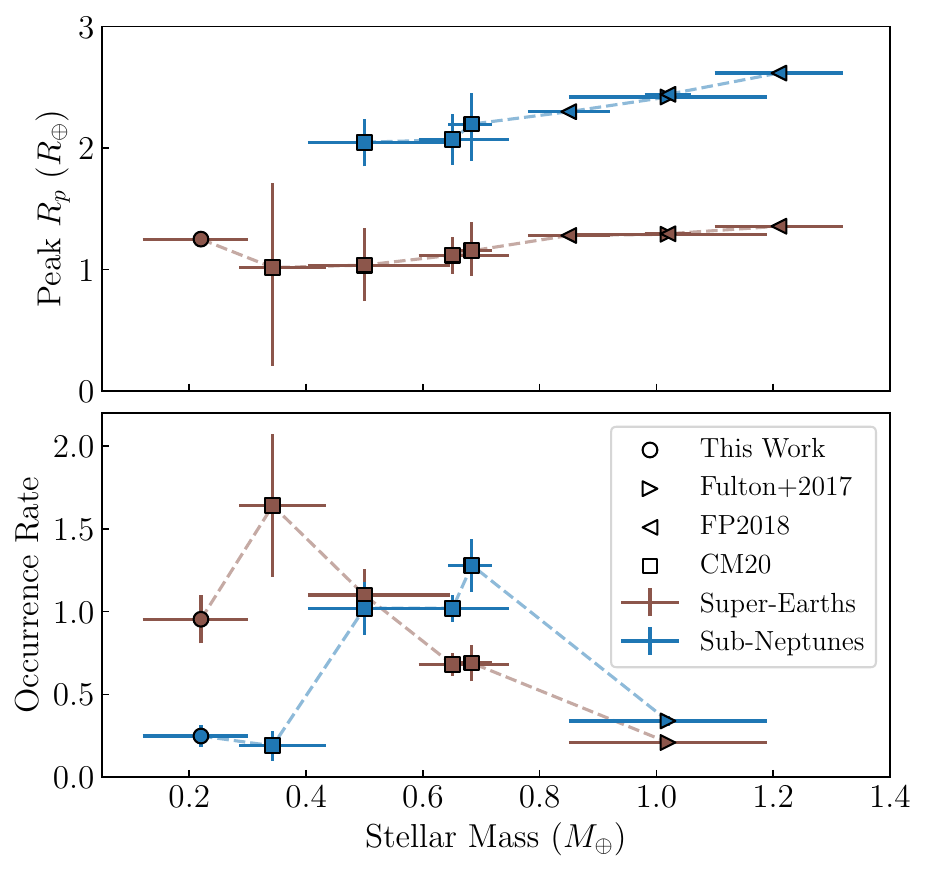}
    \caption{Evolution of the sub-Neptune and super-Earth populations across stellar mass. Top: The peak occurrence of super-Earths and sub-Neptunes plotted in brown and blue respectively. Bottom: The cumulative occurrence of super-Earths and sub-Neptunes plotted in brown and blue respectively. The data presented herein are taken from this work, \citetalias{fulton_california-kepler_2017}, \citetalias{FP2018}, and \citetalias{cloutier_evolution_2020}, and are outlined in Table~\ref{tab:radval_compare}.}
    \label{fig:occ_compare}
\end{figure}

In Figure \ref{fig:occ_compare}, we outline the super-Earth and sub-Neptune peaks found by \citet{FP2018} and \citetalias{cloutier_evolution_2020} to further compare the planet populations' properties across stellar types. We also supply the numerical values of the peak and occurrence rates in Table \ref{tab:radval_compare}.

\begin{table*}[ht]
    \centering
    \begin{tabular}{ccccccc}
    \toprule
     Source & Stellar Mass $(M_\odot)$ & SE Peak ($R_\oplus$) & SN Peak ($R_\oplus$) & $f_{SE}$ $[1R_\oplus,1.6R_\oplus]$ & $f_{SN}$ $[1.6R_\oplus,2.6R_\oplus]$  & $f_{SE}/f_{SN}$ \\ \midrule
     \citetalias{fulton_california-kepler_2017} & $1.02\pm0.17$ & 1.29* & 2.42* & $0.21\pm0.02^\dagger$ & $0.34\pm0.03^\dagger$ & $0.61 \pm 0.09^\dagger$ \\
     \citetalias{FP2018} & $1.21\pm0.11$ & $1.356\pm0.007$ & $2.617\pm0.017$ & - & - & - \\
     \citetalias{FP2018}  & $1.02\pm0.04$ & $1.294\pm0.009$ & $2.442\pm0.02$  & - & - & - \\
     \citetalias{FP2018}  & $0.85\pm0.07$ & $1.280\pm0.008$ & $2.299\pm0.014$ & - & - & - \\
     \citetalias{cloutier_evolution_2020} & ${0.684}^{+0.040}_{-0.035}$ & $1.154^{+0.205}_{-0.239}$ & $2.197^{+0.301}_{-0.256}$ & $0.69\pm0.11$ & $1.28\pm0.16$ & $0.54\pm0.11$ \\
     \citetalias{cloutier_evolution_2020} & ${0.651}^{+0.058}_{-0.096}$ & $1.118^{+0.151}_{-0.148}$ & $2.068_{-0.205}^{+0.211}$ & $0.68\pm 0.07$ & $1.02\pm0.08$ & $0.66\pm0.09$ \\  
     \citetalias{cloutier_evolution_2020} & ${0.500}^{+0.097}_{-0.146}$ & $1.036^{+0.297}_{-0.308}$ & $2.048_{-0.199}^{+0.191}$ & $1.10\pm0.16$ & $1.02\pm0.16$ & $1.08\pm0.23$ \\
     \citetalias{cloutier_evolution_2020} & ${0.343}^{+0.057}_{-0.092}$ & $1.016_{-0.807}^{+0.700}$ & - & $1.64\pm0.43$ & $0.19\pm0.09$ & $8.46\pm4.62$ \\
     This Work & $0.22^{+0.10}_{-0.08}$ & $1.25\pm0.05$ & -  & $0.954_{-0.146}^{+0.147}$ & $0.249_{-0.064}^{+0.064}$ & $3.62\pm0.52$\\
     \bottomrule
    \end{tabular}
    \caption{Comparison of the Radius Valley peaks observed by \citet{fulton_california-kepler_2017}, \citet{FP2018} (FP18), \citet{cloutier_evolution_2020} (CM20) and this work over stellar masses, with $1\sigma$ ranges. $^*$\cite{fulton_california-kepler_2017} values taken from their figure, $^\dagger f_{SE}$ and $f_{SN}$ computed on $[1.16R_\oplus,1.59R_\oplus]$ and $[1.59R_\oplus,2.7R_\oplus]$ respectively.} 
    \label{tab:radval_compare}
\end{table*}

Across the range of stellar masses surveyed, we consistently see the two peaks of the Radius Valley moving towards smaller radii, and moving closer together. We also see sub-Neptunes making up a smaller fraction of the planet population towards lower mass stars. This shift in the Radius Valley is overall consistent with the findings of \citet{venturini_radval}, suggesting that water-rich worlds may play a very strong role in shaping the Radius Valley. We also find that the trend in cumulative occurrence of super-Earths and sub-Neptunes seen past $0.5 M_\odot$ is consistent with the predictions of \citet{chachan2023}, where the observed super-Earth population is consistent with a flat line, and the sub-Neptune population drops off. 

Counter to this observed trend of the super-Earth peak moving towards smaller radii, we find that the peak in our super-Earth population is higher than the peak observed by \citetalias{cloutier_evolution_2020}. This shift of the super-Earth peak towards larger radii could be a sign that our population of planets is mixed super-Earths and compact water worlds which would otherwise form the sub-Neptune peak of the Radius Valley around earlier spectral types. This being said, we do not have sufficient evidence to present this as conclusive.

\section{Conclusions} \label{sec:concl}
In this work we have surveyed \ntargets mid-to-late M dwarfs ($0.1-0.4\, M_\odot$) for transiting planets using TESS with a custom built transit search pipeline. A detailed analysis of our survey's completeness through injection--recovery tests has allowed us to characterize the planet population with radii $>1\, R_\oplus$ out to 30 days. Our survey comprises the deepest ever systematic search for transiting planets around mid-to-late M dwarfs. Our main findings are as follows:

\begin{enumerate}
    \item We have confidently shown that the Radius Valley disappears around mid-to-late M dwarfs. The population of planets larger than Earth is unimodal in radius space, with a peak at $1.25\pm0.05\, R_\oplus$. 
    \item We find that there are \SEocc super-Earths ($1-1.6\, R_\oplus$) and \SNocc sub-Neptunes ($1.8-3\, R_\oplus$) per star respectively, such that super-Earths outnumber sub-Neptunes by a factor of $5.5\pm1.1$ to 1.
    \item We measure a cumulative occurrence of \cumocc planets per star with radii $>1\, R_\oplus$. \rcnew{When compared to the planet populations around earlier type stars, our work shows that M dwarfs, including early, mid, and late types, are the most prolific hosts of planets larger than Earth with $P<30$ days.}
    \item Our survey recovers one Jovian-sized object from which we measure a cumulative occurrence rate of $0.0028_{-0.0015}^{+0.0027}$ Jovian-sized objects per star within 10 days. Follow-up observations have shown that this object is an eclipsing brown dwarf \citep{Irwin_2018}, from which we set a 95\textsuperscript{th} percentile upper limit on the occurrence rate of hot Jupiters ($P<10$ days, $R_p>6.5\, R_\oplus$) around mid-to-late M dwarfs of 0.012 per star.
    \item We find an optimistic habitable zone ($0.2-2 S_\oplus$) cumulative occurrence rate of  $0.178_{-0.115}^{+0.116}$ planets per star and a terrestrial planet ($0.8-1.5\, R_\oplus$) occurrence rate of $0.142_{-0.092}^{+0.093}$ planets per star. With a more conservative habitable zone ($0.2-0.9 S_\oplus$), we can only place a 95\textsuperscript{th} percentile upper limit on the occurrence of terrestrial HZ planets at $<0.351$ per star.
\end{enumerate}

\begin{acknowledgments}
We acknowledge the support of the Natural Sciences and Engineering Research Council of Canada (NSERC).

This paper includes data collected with the TESS mission \citep{10.17909/fwdt-2x66,10.17909/t9-nmc8-f686,10.17909/0cp4-2j79}, obtained from the MAST data archive at the Space Telescope Science Institute (STScI). Funding for the TESS mission is provided by the NASA Explorer Program. STScI is operated by the Association of Universities for Research in Astronomy, Inc., under NASA contract NAS 5–26555. This research has made use of the NASA Exoplanet Archive \citep{10.26134/exofop3,10.26134/exofop5}, which is operated by the California Institute of Technology, under contract with the National Aeronautics and Space Administration under the Exoplanet Exploration Program. This research has made use of the NASA Exoplanet Archive, which is operated by the California Institute of Technology, under contract with the National Aeronautics and Space Administration under the Exoplanet Exploration Program. 
This research made use of Lightkurve, a Python package for Kepler and TESS data analysis \citep{lightkurve}. 
This research was enabled in part by support provided by McMaster University and the Digital Research Alliance of Canada (\href{https://www.alliancecan.ca/en}{alliancecan.ca}). 

We recognize that McMaster University is located on the traditional territories of the Mississauga and Haudenosaunee nations, and within the lands protected by the ‘Dish with One Spoon' wampum agreement.

\end{acknowledgments}

\facilities{Transiting Exoplanet Survey Satellite \citep[TESS;][]{tess_ricker}, Exoplanet Archive, Exofop.}

\software{
\texttt{astropy} \citep{astropy:2013, astropy:2018, astropy:2022},
\texttt{celerite2} \citep{celerite2},
\texttt{batman} \citep{batman},
\texttt{emcee} \citep{emcee},
\texttt{exovetter} \citep{exovetter}, 
\texttt{GLS} \citep{gls_py},
\texttt{lightkurve} \citep{lightkurve},
\texttt{numpy}, \citep{numpy},
\texttt{PyMC}, \citep{pymc2023},
\texttt{scipy} \citep{2020SciPy-NMeth},
\texttt{TLS} \citep{tls_hippke, tls_hippke_code},
\texttt{TRICERATOPS} \citep{triceratops_code, triceratops_paper}
}

\newpage
\begin{appendix} 
\section{Full Planet Parameters} \label{app:pc_tab}
Table~\ref{tab:pc_params_short} presents a shortened version of the the full set of planetary and stellar parameters for the PCs detected by our survey. A full version of this table is available in machine-readable form.
\begin{table*}[ht]
    \centering
    \begin{tabular}{ccccccccccc}
    \toprule
    TIC & TOI & Period (days) & Instel. ($S_\oplus$) & Radius ($R_\oplus$) & b & S/N & $T_\mathrm{mag}$ & $M_\star\,(M_\oplus)$ & $R_\star\,(R_\oplus)$ & $T_\mathrm{eff}$ (K) \\ \midrule
18318288 & 6086.01 & $1.3888677_{-0.0000046}^{+0.0000051}$ & $29.2_{-2.6}^{+2.8}$ & $1.11_{-0.08}^{+0.07}$ & $0.46_{-0.19}^{+0.13}$ & 7.89 & 12.41 & $0.23_{-0.01}^{+0.01}$ & $0.26_{-0.01}^{+0.01}$ & $3232_{-50}^{+49}$ \\
22233480 & 4438.01 & $7.446291_{-0.0000223}^{+0.0000204}$ & $5.9_{-0.5}^{+0.5}$ & $2.49_{-0.08}^{+0.08}$ & $0.16_{-0.11}^{+0.12}$ & 18.55 & 11.27 & $0.36_{-0.02}^{+0.02}$ & $0.38_{-0.01}^{+0.01}$ & $3387_{-50}^{+50}$ \\
34986694 & 6717.01 & $0.4707468_{-0.0000011}^{+0.0000010}$ & $97.6_{-8.8}^{+9.6}$ & $1.46_{-0.13}^{+0.14}$ & $0.58_{-0.22}^{+0.16}$ & 9.28 & 13.91 & $0.19_{-0.01}^{+0.01}$ & $0.22_{-0.01}^{+0.01}$ & $3189_{-50}^{+49}$ \\
36724087 & 732.01 & $0.7683797_{-0.0000003}^{+0.0000003}$ & $120.8_{-10.4}^{+11.2}$ & $1.27_{-0.05}^{+0.05}$ & $0.49_{-0.08}^{+0.07}$ & 16.68 & 10.58 & $0.36_{-0.02}^{+0.02}$ & $0.37_{-0.01}^{+0.01}$ & $3387_{-50}^{+51}$ \\
36724087 & 732.02 & $12.2522405_{-0.0000147}^{+0.0000148}$ & $3.0_{-0.3}^{+0.3}$ & $2.0_{-0.09}^{+0.09}$ & $0.79_{-0.02}^{+0.02}$ & 10.85 & 10.58 & $0.36_{-0.02}^{+0.02}$ & $0.37_{-0.01}^{+0.01}$ & $3387_{-50}^{+51}$ \\
\bottomrule
    \end{tabular}
    \caption{Parameters of the 77 planet candidates found by this survey. The complete table is available in machine-readable format in the online journal and as a csv file in the arXiv source code.}
    \label{tab:pc_params_short}
\end{table*} 
\section{TIC 149927512 Planet Candidate} \label{app:tic149927512}
Our planet search finds one transit-like signal not associated with a confirmed planet, TOI, or CTOI around TIC 149927512 ($M_\star = 0.25\,M_\odot, R_\star = 0.27\,R_\odot, T_{\rm eff} = 3269$ K) with an orbital period of $P=1.4646834^{+0.0000028}_{-0.0000024}$ days and a radius of $R_p=0.917^{+0.070}_{-0.067}\,R_\oplus$. In this appendix, we present the in-transit centroid offsets, TLS results, PDCSAP light curves, phase-folded light curve with fitted transit model, and transiting planet parameters of this PC. These figures are intended for diagnostic purposes. 

\begin{figure}[h]
    \centering
    \includegraphics[width=0.45\linewidth]{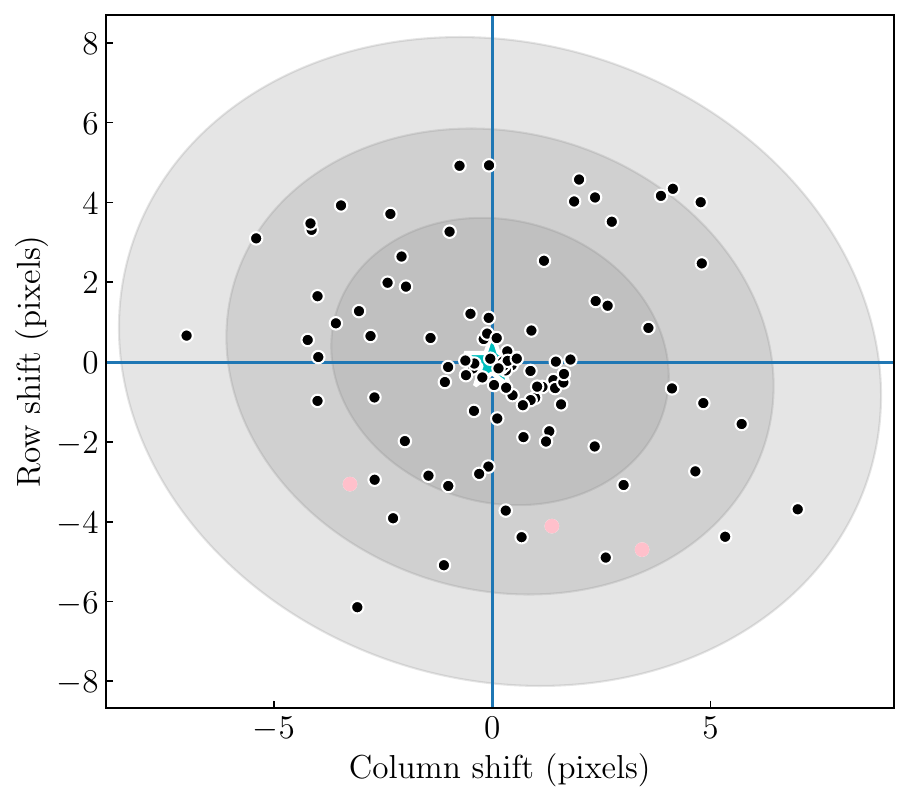}
    \includegraphics[width=0.45\linewidth]{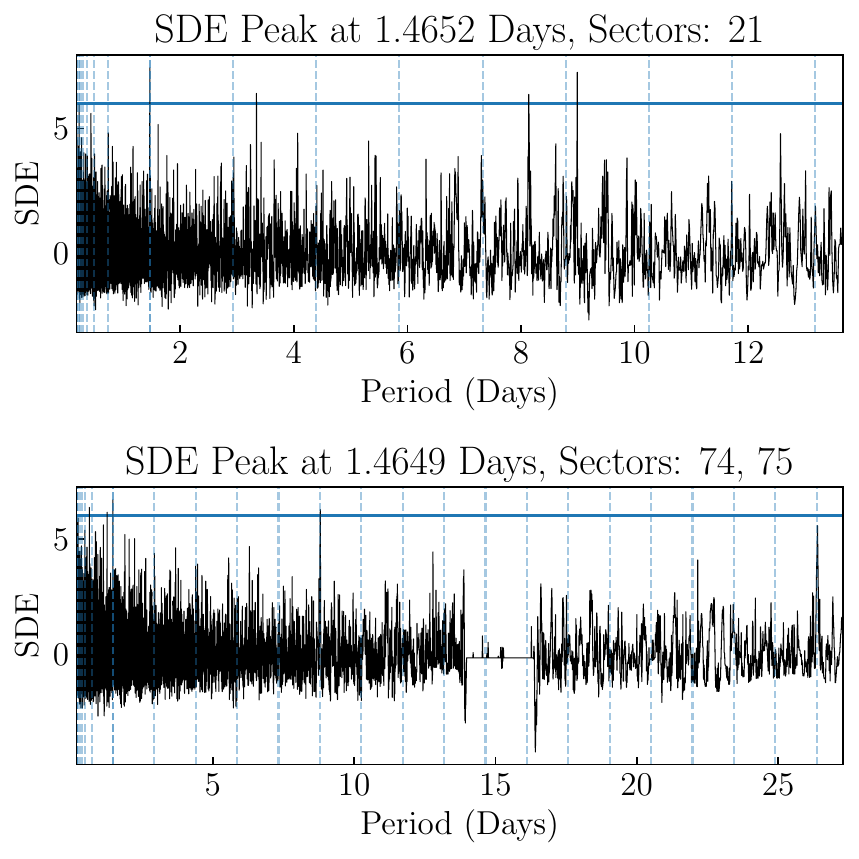}
    \caption{Left: In-transit centroid offsets of each of the 84 transit events observed. The star marker marks the out-of-transit centroid and the black points mark the relative in-transit centroid positions. We see no evidence of a systematic offset, but note that the shallow individual transit depths mean that some transits may have weakly constrained in-transit centroids, which drives the large scatter in centroid offsets of $\sim 4$ pixels. Right: TLS SDE spectra from the light curves where this planet candidate was detected (i.e. sectors 21 and the combined sectors $74,75$). The solid blue lines mark our SDE cutoff of 6 and the dashed lines mark harmonics of the most-likely orbital period of 1.46 days.}
    \label{fig:new_pc_TLS}
\end{figure}

\begin{figure}[h]
    \centering
    \includegraphics[width=\linewidth]{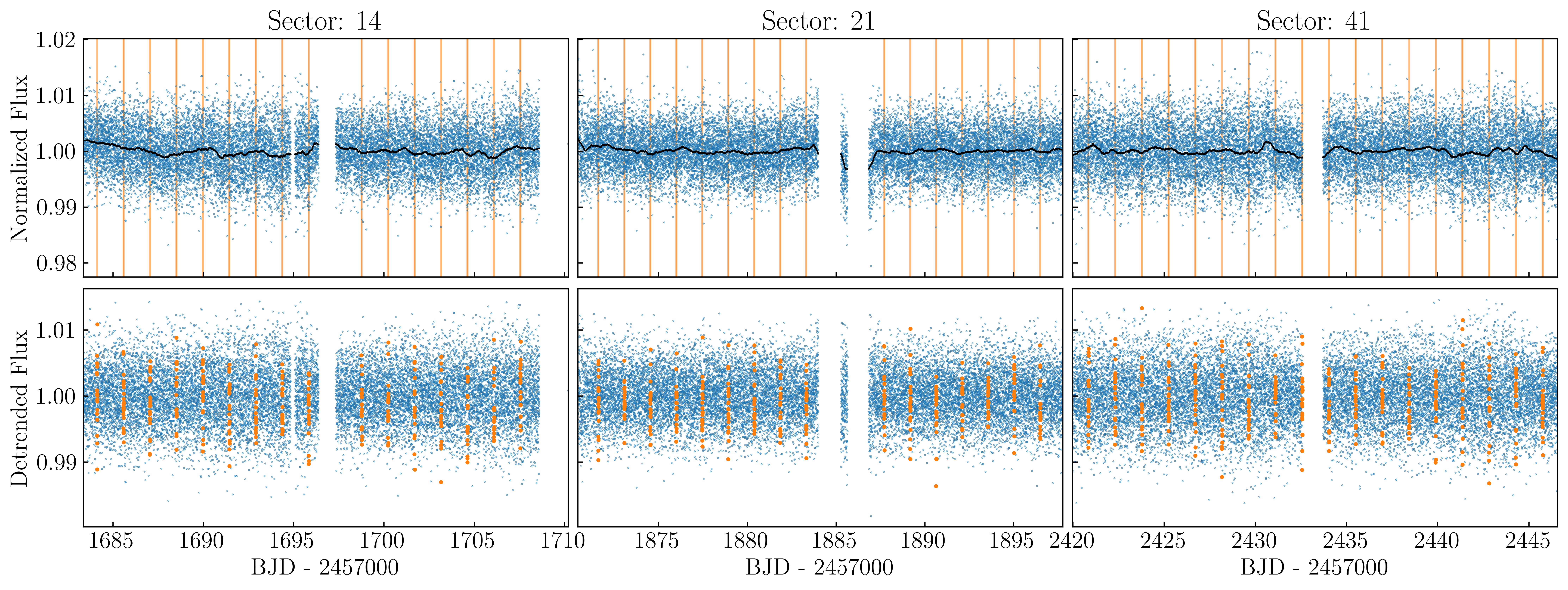}
    \includegraphics[width=0.38\linewidth]{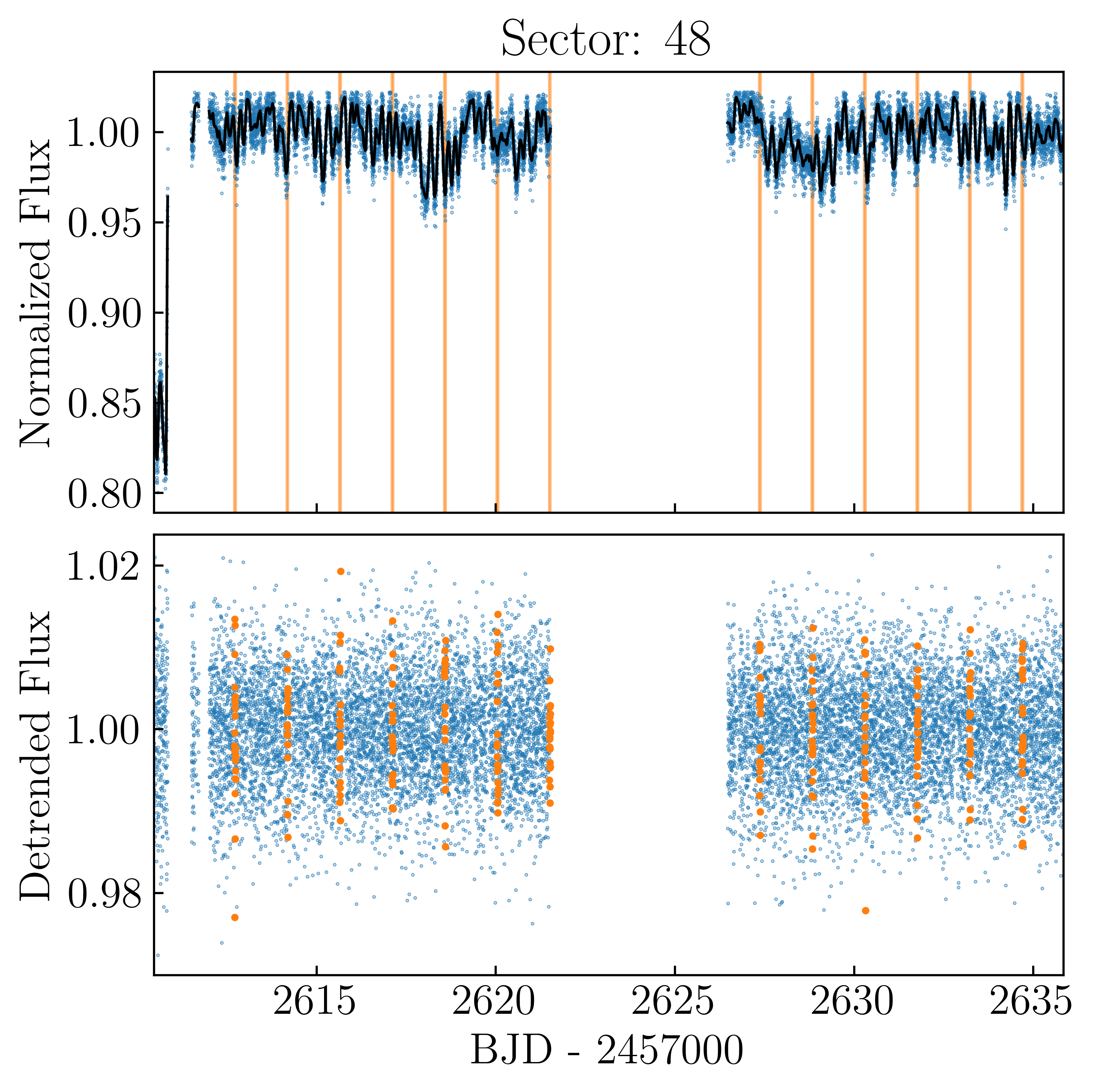}
    \includegraphics[width=0.38\linewidth]{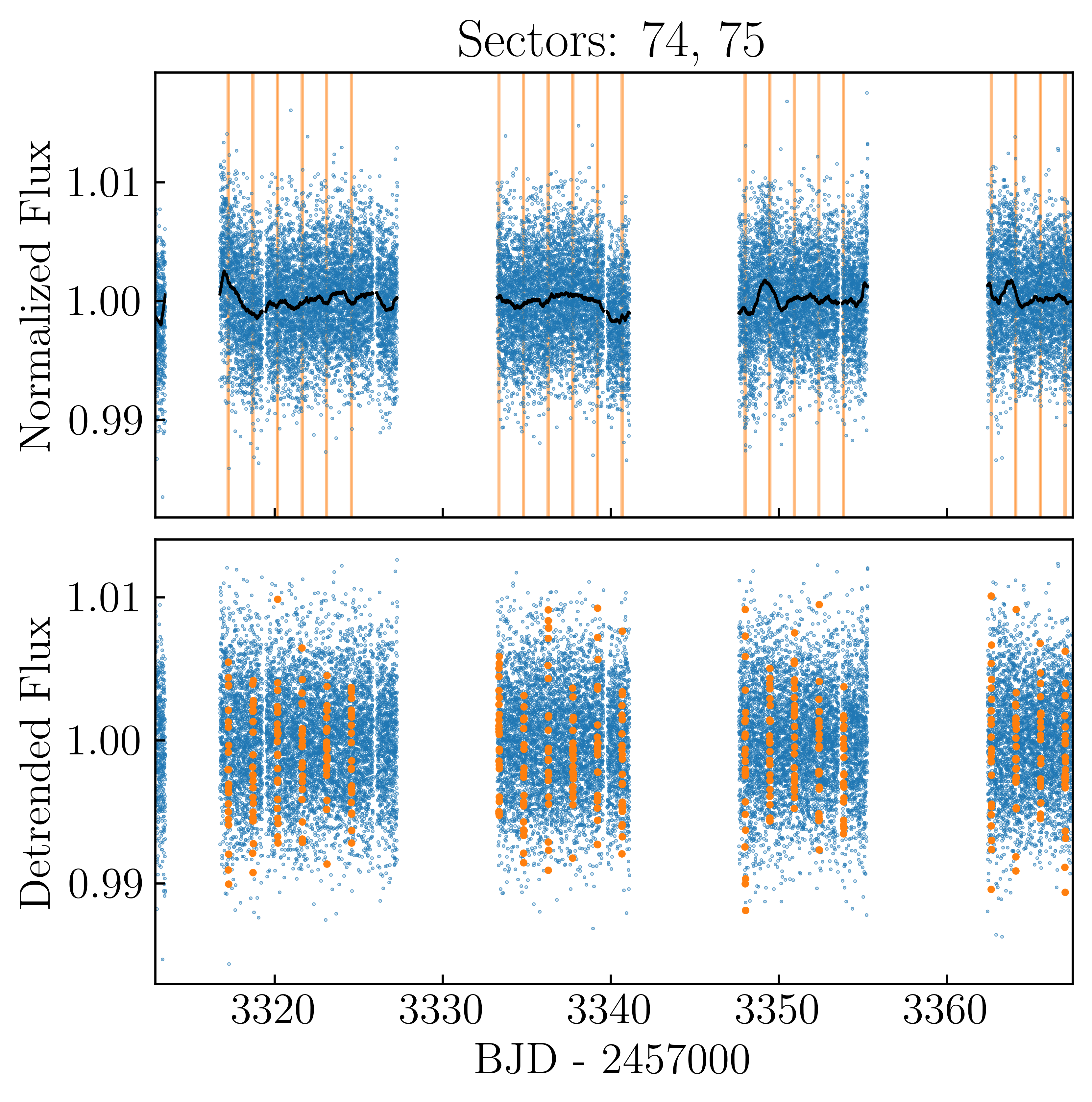}
    \caption{Pre-detrending (top) and post-detrending (bottom) PDCSAP light curves of TIC 149927512 in each of TESS sector of observation. In-transit points are highlighted in orange and the black curves represent the median detrending models in each light curve.}
    \label{fig:new_pc_lc}
\end{figure}

\begin{figure}[h]
    \centering
    \includegraphics[width=0.65\linewidth]{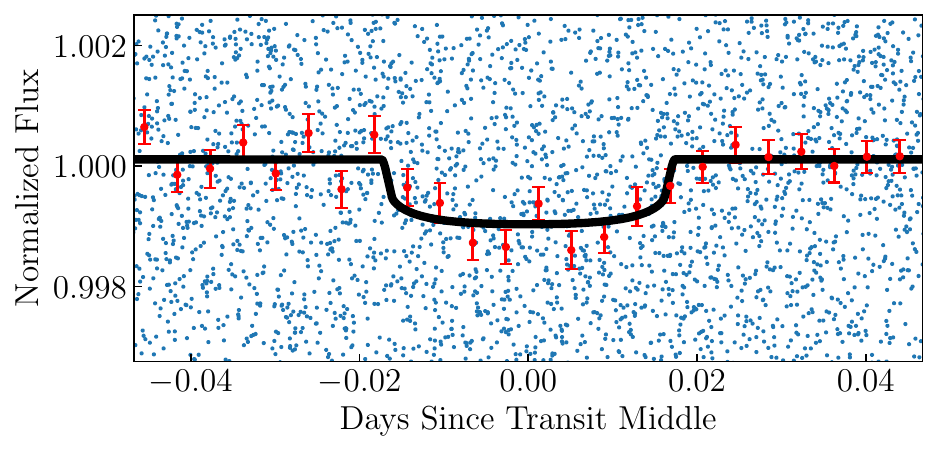}    
    \caption{The detrended light curve of TIC 149927512 phase-folded to the ephemeris of the planet candidate. The binned light curve is shown in red with our best-fit transit model in black. This fit has a transit S/N of 8.03.}
    \label{fig:new_pc_model}
\end{figure}

\begin{figure}[ht]
    \centering
    \includegraphics[width=0.8\linewidth]{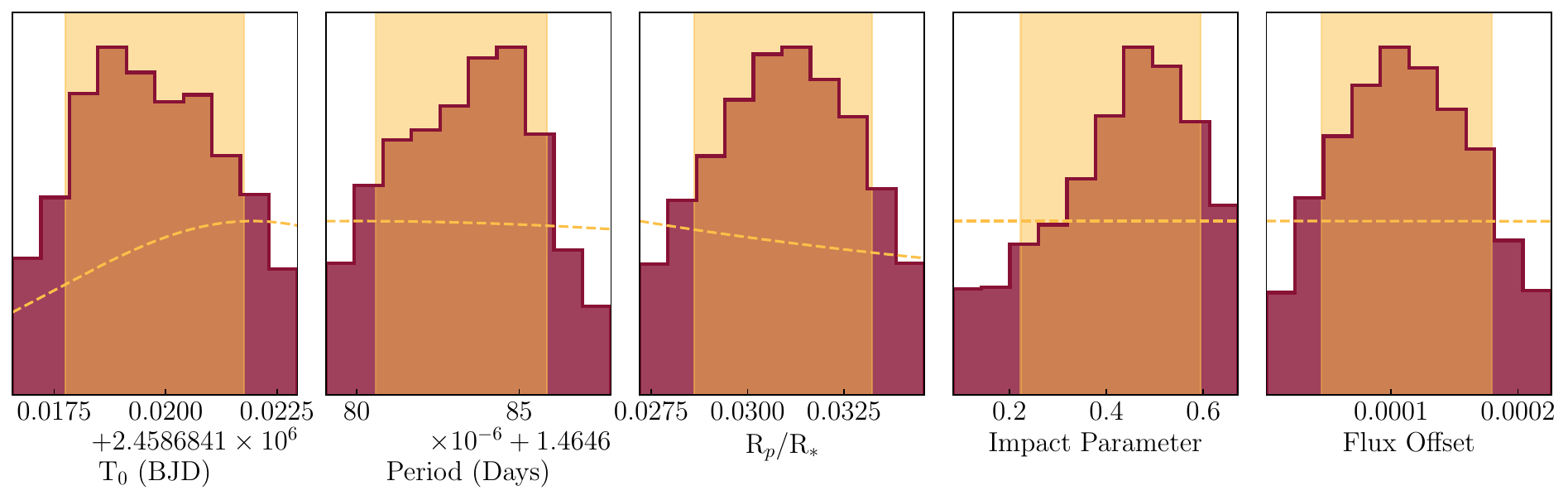}
    \caption{Posterior distributions for the parameters fit by our transit model. 1$\sigma$ confidence intervals are highlighted in orange, with each parameter's prior distribution marked by the dashed line.}
    \label{fig:new_pc_parameters}
\end{figure}

\end{appendix}

\clearpage

\bibliographystyle{aasjournalv7}
\bibliography{new.ms.bib}

@article{cloutier_evolution_2020,
	title = {Evolution of the {Radius} {Valley} around {Low}-mass {Stars} from {Kepler} and {K2}},
	volume = {159},
	issn = {0004-6256},
	url = {https://ui.adsabs.harvard.edu/abs/2020AJ....159..211C},
	doi = {10.3847/1538-3881/ab8237},
	urldate = {2024-02-02},
	journal = {The Astronomical Journal},
	author = {Cloutier, Ryan and Menou, Kristen},
	month = may,
	year = {2020},
	note = {ADS Bibcode: 2020AJ....159..211C},
	pages = {211},
}

@article{owen_evaporation_2017,
	title = {The {Evaporation} {Valley} in the {Kepler} {Planets}},
	volume = {847},
	issn = {0004-637X},
	url = {https://ui.adsabs.harvard.edu/abs/2017ApJ...847...29O},
	doi = {10.3847/1538-4357/aa890a},
	urldate = {2024-02-02},
	journal = {The Astrophysical Journal},
	author = {Owen, James E. and Wu, Yanqin},
	month = sep,
	year = {2017},
	note = {ADS Bibcode: 2017ApJ...847...29O},
	pages = {29},
}

@article{fulton_california-kepler_2017,
	title = {The {California}-{Kepler} {Survey}. {III}. {A} {Gap} in the {Radius} {Distribution} of {Small} {Planets}*},
	volume = {154},
	issn = {0004-6256, 1538-3881},
	url = {https://iopscience.iop.org/article/10.3847/1538-3881/aa80eb},
	doi = {10.3847/1538-3881/aa80eb},
	number = {3},
	urldate = {2024-02-02},
	journal = {The Astronomical Journal},
	author = {Fulton, Benjamin J. and Petigura, Erik A. and Howard, Andrew W. and Isaacson, Howard and Marcy, Geoffrey W. and Cargile, Phillip A. and Hebb, Leslie and Weiss, Lauren M. and Johnson, John Asher and Morton, Timothy D. and Sinukoff, Evan and Crossfield, Ian J. M. and Hirsch, Lea A.},
	month = sep,
	year = {2017},
	pages = {109},
}

@article{gupta_sculpting_2019,
	title = {Sculpting the valley in the radius distribution of small exoplanets as a by-product of planet formation: the core-powered mass-loss mechanism},
	volume = {487},
	issn = {0035-8711, 1365-2966},
	shorttitle = {Sculpting the valley in the radius distribution of small exoplanets as a by-product of planet formation},
	url = {https://academic.oup.com/mnras/article/487/1/24/5484904},
	doi = {10.1093/mnras/stz1230},
	language = {en},
	number = {1},
	urldate = {2024-02-02},
	journal = {Monthly Notices of the Royal Astronomical Society},
	author = {Gupta, Akash and Schlichting, Hilke E},
	month = jul,
	year = {2019},
	pages = {24--33},
}

@article{ment_occurrence_2023,
	title = {The {Occurrence} {Rate} of {Terrestrial} {Planets} {Orbiting} {Nearby} {Mid}-to-late {M} {Dwarfs} from {TESS} {Sectors} 1-42},
	volume = {165},
	issn = {0004-6256},
	url = {https://ui.adsabs.harvard.edu/abs/2023AJ....165..265M},
	doi = {10.3847/1538-3881/acd175},
	urldate = {2024-02-21},
	journal = {The Astronomical Journal},
	author = {Ment, Kristo and Charbonneau, David},
	month = jun,
	year = {2023},
	note = {ADS Bibcode: 2023AJ....165..265M},
	pages = {265},
}

@article{wu_mass_2019,
	title = {Mass and {Mass} {Scalings} of {Super}-{Earths}},
	volume = {874},
	issn = {0004-637X},
	url = {https://ui.adsabs.harvard.edu/abs/2019ApJ...874...91W},
	doi = {10.3847/1538-4357/ab06f8},
	urldate = {2024-02-21},
	journal = {The Astrophysical Journal},
	author = {Wu, Yanqin},
	month = mar,
	year = {2019},
	note = {ADS Bibcode: 2019ApJ...874...91W},
	pages = {91},
}

@misc{tls_hippke_code,
       author = {{Hippke}, Michael and {Heller}, Ren{\'e}},
        title = "{TLS: Transit Least Squares}",
 howpublished = {Astrophysics Source Code Library, record ascl:1910.007},
         year = 2019,
        month = oct,
          eid = {ascl:1910.007},
       adsurl = {https://ui.adsabs.harvard.edu/abs/2019ascl.soft10007H}
}

@ARTICLE{tls_hippke,
       author = {{Hippke}, Michael and {Heller}, Ren{\'e}},
        title = "{Optimized transit detection algorithm to search for periodic transits of small planets}",
      journal = {\aap},
         year = 2019,
        month = mar,
       volume = {623},
          eid = {A39},
        pages = {A39},
          doi = {10.1051/0004-6361/201834672},
archivePrefix = {arXiv},
       eprint = {1901.02015},
 primaryClass = {astro-ph.EP},
       adsurl = {https://ui.adsabs.harvard.edu/abs/2019A&A...623A..39H}
}

@ARTICLE{bls_kovaks,
       author = {{Kov{\'a}cs}, G. and {Zucker}, S. and {Mazeh}, T.},
        title = "{A box-fitting algorithm in the search for periodic transits}",
      journal = {AAP},
         year = 2002,
        month = aug,
       volume = {391},
        pages = {369-377},
          doi = {10.1051/0004-6361:20020802},
archivePrefix = {arXiv},
       eprint = {astro-ph/0206099},
 primaryClass = {astro-ph},
       adsurl = {https://ui.adsabs.harvard.edu/abs/2002A&A...391..369K}
}

@ARTICLE{emcee,
       author = {{Foreman-Mackey}, Daniel and {Hogg}, David W. and {Lang}, Dustin and {Goodman}, Jonathan},
        title = "{emcee: The MCMC Hammer}",
      journal = {PASP},
         year = 2013,
        month = mar,
       volume = {125},
       number = {925},
        pages = {306},
          doi = {10.1086/670067},
archivePrefix = {arXiv},
       eprint = {1202.3665},
 primaryClass = {astro-ph.IM},
       adsurl = {https://ui.adsabs.harvard.edu/abs/2013PASP..125..306F}
}

@ARTICLE{batman,
       author = {{Kreidberg}, Laura},
        title = "{batman: BAsic Transit Model cAlculatioN in Python}",
      journal = {PASP},
         year = 2015,
        month = nov,
       volume = {127},
       number = {957},
        pages = {1161},
          doi = {10.1086/683602},
archivePrefix = {arXiv},
       eprint = {1507.08285},
 primaryClass = {astro-ph.EP},
       adsurl = {https://ui.adsabs.harvard.edu/abs/2015PASP..127.1161K}
}

@ARTICLE{Dressing_2015,
       author = {{Dressing}, Courtney D. and {Charbonneau}, David},
        title = "{The Occurrence of Potentially Habitable Planets Orbiting M Dwarfs Estimated from the Full Kepler Dataset and an Empirical Measurement of the Detection Sensitivity}",
      journal = {APJ},
         year = 2015,
        month = jul,
       volume = {807},
       number = {1},
          eid = {45},
        pages = {45},
          doi = {10.1088/0004-637X/807/1/45},
archivePrefix = {arXiv},
       eprint = {1501.01623},
 primaryClass = {astro-ph.EP},
       adsurl = {https://ui.adsabs.harvard.edu/abs/2015ApJ...807...45D}
}

@ARTICLE{chang_flares,
       author = {{Chang}, S. -W. and {Byun}, Y. -I. and {Hartman}, J.~D.},
        title = "{Photometric Study on Stellar Magnetic Activity. I. Flare Variability of Red Dwarf Stars in the Open Cluster M37}",
      journal = {APJ},
         year = 2015,
        month = nov,
       volume = {814},
       number = {1},
          eid = {35},
        pages = {35},
          doi = {10.1088/0004-637X/814/1/35},
archivePrefix = {arXiv},
       eprint = {1510.01005},
 primaryClass = {astro-ph.SR},
       adsurl = {https://ui.adsabs.harvard.edu/abs/2015ApJ...814...35C}
}

@ARTICLE{2020SciPy-NMeth,
  author  = {Virtanen, Pauli and Gommers, Ralf and Oliphant, Travis E. and
            Haberland, Matt and Reddy, Tyler and Cournapeau, David and
            Burovski, Evgeni and Peterson, Pearu and Weckesser, Warren and
            Bright, Jonathan and {van der Walt}, St{\'e}fan J. and
            Brett, Matthew and Wilson, Joshua and Millman, K. Jarrod and
            Mayorov, Nikolay and Nelson, Andrew R. J. and Jones, Eric and
            Kern, Robert and Larson, Eric and Carey, C J and
            Polat, {\.I}lhan and Feng, Yu and Moore, Eric W. and
            {VanderPlas}, Jake and Laxalde, Denis and Perktold, Josef and
            Cimrman, Robert and Henriksen, Ian and Quintero, E. A. and
            Harris, Charles R. and Archibald, Anne M. and
            Ribeiro, Ant{\^o}nio H. and Pedregosa, Fabian and
            {van Mulbregt}, Paul and {SciPy 1.0 Contributors}},
  title   = {{{SciPy} 1.0: Fundamental Algorithms for Scientific
            Computing in Python}},
  journal = {Nature Methods},
  year    = {2020},
  volume  = {17},
  pages   = {261--272},
  adsurl  = {https://rdcu.be/b08Wh},
  doi     = {10.1038/s41592-019-0686-2},
}

@MISC{triceratops_code,
       author = {{Giacalone}, Steven and {Dressing}, Courtney D.},
        title = "{triceratops: Candidate exoplanet rating tool}",
         year = 2020,
        month = feb,
          eid = {ascl:2002.004},
        pages = {ascl:2002.004},
archivePrefix = {ascl},
       eprint = {2002.004},
       adsurl = {https://ui.adsabs.harvard.edu/abs/2020ascl.soft02004G}
}

@ARTICLE{triceratops_paper,
       author = {{Giacalone}, Steven and {Dressing}, Courtney D. and {Jensen}, Eric L.~N. and {Collins}, Karen A. and {Ricker}, George R. and {Vanderspek}, Roland and {Seager}, S. and {Winn}, Joshua N. and {Jenkins}, Jon M. and {Barclay}, Thomas and {Barkaoui}, Khalid and {Cadieux}, Charles and {Charbonneau}, David and {Collins}, Kevin I. and {Conti}, Dennis M. and {Doyon}, Ren{\'e} and {Evans}, Phil and {Ghachoui}, Mourad and {Gillon}, Micha{\"e}l and {Guerrero}, Natalia M. and {Hart}, Rhodes and {Jehin}, Emmanu{\"e}l and {Kielkopf}, John F. and {McLean}, Brian and {Murgas}, Felipe and {Palle}, Enric and {Parviainen}, Hannu and {Pozuelos}, Francisco J. and {Relles}, Howard M. and {Shporer}, Avi and {Socia}, Quentin and {Stockdale}, Chris and {Tan}, Thiam-Guan and {Torres}, Guillermo and {Twicken}, Joseph D. and {Waalkes}, William C. and {Waite}, Ian A.},
        title = "{Vetting of 384 TESS Objects of Interest with TRICERATOPS and Statistical Validation of 12 Planet Candidates}",
      journal = {Astronomy Journals},
         year = 2021,
        month = jan,
       volume = {161},
       number = {1},
          eid = {24},
        pages = {24},
          doi = {10.3847/1538-3881/abc6af},
archivePrefix = {arXiv},
       eprint = {2002.00691},
 primaryClass = {astro-ph.EP},
       adsurl = {https://ui.adsabs.harvard.edu/abs/2021AJ....161...24G}
}

@ARTICLE{venturini_radval,
       author = {{Venturini}, J. and {Ronco}, M.~P. and {Guilera}, O.~M. and {Haldemann}, J. and {Mordasini}, C. and {Miller Bertolami}, M.},
        title = "{A fading radius valley towards M dwarfs, a persistent density valley across stellar types}",
      journal = {AAP},
         year = 2024,
        month = jun,
       volume = {686},
          eid = {L9},
        pages = {L9},
          doi = {10.1051/0004-6361/202349088},
archivePrefix = {arXiv},
       eprint = {2404.01967},
 primaryClass = {astro-ph.EP},
       adsurl = {https://ui.adsabs.harvard.edu/abs/2024A&A...686L...9V}
}

@ARTICLE{burn_steam,
       author = {{Burn}, Remo and {Mordasini}, Christoph and {Mishra}, Lokesh and {Haldemann}, Jonas and {Venturini}, Julia and {Emsenhuber}, Alexandre and {Henning}, Thomas},
        title = "{A radius valley between migrated steam worlds and evaporated rocky cores}",
      journal = {Nature Astronomy},
         year = 2024,
        month = apr,
       volume = {8},
        pages = {463-471},
          doi = {10.1038/s41550-023-02183-7},
archivePrefix = {arXiv},
       eprint = {2401.04380},
 primaryClass = {astro-ph.EP},
       adsurl = {https://ui.adsabs.harvard.edu/abs/2024NatAs...8..463B}
}

@article{celerite2,
   author = {{Foreman-Mackey}, D.},
    title = "{Scalable Backpropagation for Gaussian Processes using Celerite}",
  journal = {Research Notes of the American Astronomical Society},
     year = 2018,
    month = feb,
   volume = 2,
   number = 1,
    pages = {31},
      doi = {10.3847/2515-5172/aaaf6c},
   adsurl = {http://adsabs.harvard.edu/abs/2018RNAAS...2a..31F}
}

@ARTICLE{nuance,
       author = {{Garcia}, Lionel J. and {Foreman-Mackey}, Daniel and {Murray}, Catriona A. and {Aigrain}, Suzanne and {Feliz}, Dax L. and {Pozuelos}, Francisco J.},
        title = "{nuance: Efficient Detection of Planets Transiting Active Stars}",
      journal = {Astronomy Journals},
         year = 2024,
        month = jun,
       volume = {167},
       number = {6},
          eid = {284},
        pages = {284},
          doi = {10.3847/1538-3881/ad3cd6},
archivePrefix = {arXiv},
       eprint = {2402.06835},
 primaryClass = {astro-ph.EP},
       adsurl = {https://ui.adsabs.harvard.edu/abs/2024AJ....167..284G}
}

@misc{gls_py,
  author = {Sebastian Schröter and Stefan Czesla and Mathias Zechmeister},
  title = {gls.py},
  url = {https://github.com/sczesla/PyAstronomy/blob/master/src/pyTiming/pyPeriod/gls.py},
  date = {2019-12-18},
  year = {2019}
}

@ARTICLE{burn_bern,
       author = {{Burn}, R. and {Schlecker}, M. and {Mordasini}, C. and {Emsenhuber}, A. and {Alibert}, Y. and {Henning}, T. and {Klahr}, H. and {Benz}, W.},
        title = "{The New Generation Planetary Population Synthesis (NGPPS). IV. Planetary systems around low-mass stars}",
      journal = {AAP},
         year = 2021,
        month = dec,
       volume = {656},
          eid = {A72},
        pages = {A72},
          doi = {10.1051/0004-6361/202140390},
archivePrefix = {arXiv},
       eprint = {2105.04596},
 primaryClass = {astro-ph.EP},
       adsurl = {https://ui.adsabs.harvard.edu/abs/2021A&A...656A..72B}
}

@ARTICLE{winters_most_common,
       author = {{Winters}, Jennifer G. and {Henry}, Todd J. and {Lurie}, John C. and {Hambly}, Nigel C. and {Jao}, Wei-Chun and {Bartlett}, Jennifer L. and {Boyd}, Mark R. and {Dieterich}, Sergio B. and {Finch}, Charlie T. and {Hosey}, Altonio D. and {Ianna}, Philip A. and {Riedel}, Adric R. and {Slatten}, Kenneth J. and {Subasavage}, John P.},
        title = "{The Solar Neighborhood. XXXV. Distances to 1404 m Dwarf Systems Within 25 pc in the Southern Sky}",
      journal = {The Astronomical Journal},
         year = 2015,
        month = jan,
       volume = {149},
       number = {1},
          eid = {5},
        pages = {5},
          doi = {10.1088/0004-6256/149/1/5},
archivePrefix = {arXiv},
       eprint = {1407.7837},
 primaryClass = {astro-ph.SR},
       adsurl = {https://ui.adsabs.harvard.edu/abs/2015AJ....149....5W}
}

@ARTICLE{kepler_mission,
       author = {{Koch}, David G. and {Borucki}, William J. and {Basri}, Gibor and {Batalha}, Natalie M. and {Brown}, Timothy M. and {Caldwell}, Douglas and {Christensen-Dalsgaard}, J{\o}rgen and {Cochran}, William D. and {DeVore}, Edna and {Dunham}, Edward W. and {Gautier}, Thomas N., III and {Geary}, John C. and {Gilliland}, Ronald L. and {Gould}, Alan and {Jenkins}, Jon and {Kondo}, Yoji and {Latham}, David W. and {Lissauer}, Jack J. and {Marcy}, Geoffrey and {Monet}, David and {Sasselov}, Dimitar and {Boss}, Alan and {Brownlee}, Donald and {Caldwell}, John and {Dupree}, Andrea K. and {Howell}, Steve B. and {Kjeldsen}, Hans and {Meibom}, S{\o}ren and {Morrison}, David and {Owen}, Tobias and {Reitsema}, Harold and {Tarter}, Jill and {Bryson}, Stephen T. and {Dotson}, Jessie L. and {Gazis}, Paul and {Haas}, Michael R. and {Kolodziejczak}, Jeffrey and {Rowe}, Jason F. and {Van Cleve}, Jeffrey E. and {Allen}, Christopher and {Chandrasekaran}, Hema and {Clarke}, Bruce D. and {Li}, Jie and {Quintana}, Elisa V. and {Tenenbaum}, Peter and {Twicken}, Joseph D. and {Wu}, Hayley},
        title = "{Kepler Mission Design, Realized Photometric Performance, and Early Science}",
      journal = {APJL},
         year = 2010,
        month = apr,
       volume = {713},
       number = {2},
        pages = {L79-L86},
          doi = {10.1088/2041-8205/713/2/L79},
archivePrefix = {arXiv},
       eprint = {1001.0268},
 primaryClass = {astro-ph.EP},
       adsurl = {https://ui.adsabs.harvard.edu/abs/2010ApJ...713L..79K}
}

@ARTICLE{tess_ricker,
       author = {{Ricker}, George R. and {Winn}, Joshua N. and {Vanderspek}, Roland and {Latham}, David W. and {Bakos}, G{\'a}sp{\'a}r {\'A}. and {Bean}, Jacob L. and {Berta-Thompson}, Zachory K. and {Brown}, Timothy M. and {Buchhave}, Lars and {Butler}, Nathaniel R. and {Butler}, R. Paul and {Chaplin}, William J. and {Charbonneau}, David and {Christensen-Dalsgaard}, J{\o}rgen and {Clampin}, Mark and {Deming}, Drake and {Doty}, John and {De Lee}, Nathan and {Dressing}, Courtney and {Dunham}, Edward W. and {Endl}, Michael and {Fressin}, Francois and {Ge}, Jian and {Henning}, Thomas and {Holman}, Matthew J. and {Howard}, Andrew W. and {Ida}, Shigeru and {Jenkins}, Jon M. and {Jernigan}, Garrett and {Johnson}, John Asher and {Kaltenegger}, Lisa and {Kawai}, Nobuyuki and {Kjeldsen}, Hans and {Laughlin}, Gregory and {Levine}, Alan M. and {Lin}, Douglas and {Lissauer}, Jack J. and {MacQueen}, Phillip and {Marcy}, Geoffrey and {McCullough}, Peter R. and {Morton}, Timothy D. and {Narita}, Norio and {Paegert}, Martin and {Palle}, Enric and {Pepe}, Francesco and {Pepper}, Joshua and {Quirrenbach}, Andreas and {Rinehart}, Stephen A. and {Sasselov}, Dimitar and {Sato}, Bun'ei and {Seager}, Sara and {Sozzetti}, Alessandro and {Stassun}, Keivan G. and {Sullivan}, Peter and {Szentgyorgyi}, Andrew and {Torres}, Guillermo and {Udry}, Stephane and {Villasenor}, Joel},
        title = "{Transiting Exoplanet Survey Satellite (TESS)}",
      journal = {Journal of Astronomical Telescopes, Instruments, and Systems},
         year = 2015,
        month = jan,
       volume = {1},
          eid = {014003},
        pages = {014003},
          doi = {10.1117/1.JATIS.1.1.014003},
       adsurl = {https://ui.adsabs.harvard.edu/abs/2015JATIS...1a4003R}
}

@ARTICLE{hab_zone_koppararu,
       author = {{Kopparapu}, Ravi Kumar and {Ramirez}, Ramses and {Kasting}, James F. and {Eymet}, Vincent and {Robinson}, Tyler D. and {Mahadevan}, Suvrath and {Terrien}, Ryan C. and {Domagal-Goldman}, Shawn and {Meadows}, Victoria and {Deshpande}, Rohit},
        title = "{Habitable Zones around Main-sequence Stars: New Estimates}",
      journal = {ApJ},
         year = 2013,
        month = mar,
       volume = {765},
       number = {2},
          eid = {131},
        pages = {131},
          doi = {10.1088/0004-637X/765/2/131},
archivePrefix = {arXiv},
       eprint = {1301.6674},
 primaryClass = {astro-ph.EP},
       adsurl = {https://ui.adsabs.harvard.edu/abs/2013ApJ...765..131K}
}

@ARTICLE{mann2019,
       author = {{Mann}, Andrew W. and {Dupuy}, Trent and {Kraus}, Adam L. and {Gaidos}, Eric and {Ansdell}, Megan and {Ireland}, Michael and {Rizzuto}, Aaron C. and {Hung}, Chao-Ling and {Dittmann}, Jason and {Factor}, Samuel and {Feiden}, Gregory and {Martinez}, Raquel A. and {Ru{\'\i}z-Rodr{\'\i}guez}, Dary and {Thao}, Pa Chia},
        title = "{How to Constrain Your M Dwarf. II. The Mass-Luminosity-Metallicity Relation from 0.075 to 0.70 Solar Masses}",
      journal = {\apj},
         year = 2019,
        month = jan,
       volume = {871},
       number = {1},
          eid = {63},
        pages = {63},
          doi = {10.3847/1538-4357/aaf3bc},
archivePrefix = {arXiv},
       eprint = {1811.06938},
 primaryClass = {astro-ph.SR},
       adsurl = {https://ui.adsabs.harvard.edu/abs/2019ApJ...871...63M}
}

@ARTICLE{mann2015,
       author = {{Mann}, Andrew W. and {Feiden}, Gregory A. and {Gaidos}, Eric and {Boyajian}, Tabetha and {von Braun}, Kaspar},
        title = "{How to Constrain Your M Dwarf: Measuring Effective Temperature, Bolometric Luminosity, Mass, and Radius}",
      journal = {\apj},
         year = 2015,
        month = may,
       volume = {804},
       number = {1},
          eid = {64},
        pages = {64},
          doi = {10.1088/0004-637X/804/1/64},
archivePrefix = {arXiv},
       eprint = {1501.01635},
 primaryClass = {astro-ph.SR},
       adsurl = {https://ui.adsabs.harvard.edu/abs/2015ApJ...804...64M}
}

@ARTICLE{pecault2013,
       author = {{Pecaut}, Mark J. and {Mamajek}, Eric E.},
        title = "{Intrinsic Colors, Temperatures, and Bolometric Corrections of Pre-main-sequence Stars}",
      journal = {\apjs},
         year = 2013,
        month = sep,
       volume = {208},
       number = {1},
          eid = {9},
        pages = {9},
          doi = {10.1088/0067-0049/208/1/9},
archivePrefix = {arXiv},
       eprint = {1307.2657},
 primaryClass = {astro-ph.SR},
       adsurl = {https://ui.adsabs.harvard.edu/abs/2013ApJS..208....9P}
}

@ARTICLE{sullivan_tess_noise,
       author = {{Sullivan}, Peter W. and {Winn}, Joshua N. and {Berta-Thompson}, Zachory K. and {Charbonneau}, David and {Deming}, Drake and {Dressing}, Courtney D. and {Latham}, David W. and {Levine}, Alan M. and {McCullough}, Peter R. and {Morton}, Timothy and {Ricker}, George R. and {Vanderspek}, Roland and {Woods}, Deborah},
        title = "{Erratum: {\textquotedblleft}The Transiting Exoplanet Survey Satellite: Simulations of Planet Detections and Astrophysical False Positives{\textquotedblright} <A href=``/abs/2015ApJ...809...77S''>(2015, ApJ, 809, 77)</A>}",
      journal = {\apj},
         year = 2017,
        month = mar,
       volume = {837},
       number = {1},
          eid = {99},
        pages = {99},
          doi = {10.3847/1538-4357/837/1/99},
       adsurl = {https://ui.adsabs.harvard.edu/abs/2017ApJ...837...99S}
}

@ARTICLE{evans_photometry,
       author = {{Evans}, D.~W. and {Riello}, M. and {De Angeli}, F. and {Carrasco}, J.~M. and {Montegriffo}, P. and {Fabricius}, C. and {Jordi}, C. and {Palaversa}, L. and {Diener}, C. and {Busso}, G. and {Cacciari}, C. and {van Leeuwen}, F. and {Burgess}, P.~W. and {Davidson}, M. and {Harrison}, D.~L. and {Hodgkin}, S.~T. and {Pancino}, E. and {Richards}, P.~J. and {Altavilla}, G. and {Balaguer-N{\'u}{\~n}ez}, L. and {Barstow}, M.~A. and {Bellazzini}, M. and {Brown}, A.~G.~A. and {Castellani}, M. and {Cocozza}, G. and {De Luise}, F. and {Delgado}, A. and {Ducourant}, C. and {Galleti}, S. and {Gilmore}, G. and {Giuffrida}, G. and {Holl}, B. and {Kewley}, A. and {Koposov}, S.~E. and {Marinoni}, S. and {Marrese}, P.~M. and {Osborne}, P.~J. and {Piersimoni}, A. and {Portell}, J. and {Pulone}, L. and {Ragaini}, S. and {Sanna}, N. and {Terrett}, D. and {Walton}, N.~A. and {Wevers}, T. and {Wyrzykowski}, {\L}.},
        title = "{Gaia Data Release 2. Photometric content and validation}",
      journal = {\aap},
         year = 2018,
        month = aug,
       volume = {616},
          eid = {A4},
        pages = {A4},
          doi = {10.1051/0004-6361/201832756},
archivePrefix = {arXiv},
       eprint = {1804.09368},
 primaryClass = {astro-ph.IM},
       adsurl = {https://ui.adsabs.harvard.edu/abs/2018A&A...616A...4E}
}

@ARTICLE{pdcsap_smith_2012,
       author = {{Smith}, Jeffrey C. and {Stumpe}, Martin C. and {Van Cleve}, Jeffrey E. and {Jenkins}, Jon M. and {Barclay}, Thomas S. and {Fanelli}, Michael N. and {Girouard}, Forrest R. and {Kolodziejczak}, Jeffery J. and {McCauliff}, Sean D. and {Morris}, Robert L. and {Twicken}, Joseph D.},
        title = "{Kepler Presearch Data Conditioning II - A Bayesian Approach to Systematic Error Correction}",
      journal = {\pasp},
         year = 2012,
        month = sep,
       volume = {124},
       number = {919},
        pages = {1000},
          doi = {10.1086/667697},
archivePrefix = {arXiv},
       eprint = {1203.1383},
 primaryClass = {astro-ph.IM},
       adsurl = {https://ui.adsabs.harvard.edu/abs/2012PASP..124.1000S}
}

@ARTICLE{pdcsap_stumpe_2012,
       author = {{Stumpe}, Martin C. and {Smith}, Jeffrey C. and {Van Cleve}, Jeffrey E. and {Twicken}, Joseph D. and {Barclay}, Thomas S. and {Fanelli}, Michael N. and {Girouard}, Forrest R. and {Jenkins}, Jon M. and {Kolodziejczak}, Jeffery J. and {McCauliff}, Sean D. and {Morris}, Robert L.},
        title = "{Kepler Presearch Data Conditioning I{\textemdash}Architecture and Algorithms for Error Correction in Kepler Light Curves}",
      journal = {\pasp},
         year = 2012,
        month = sep,
       volume = {124},
       number = {919},
        pages = {985},
          doi = {10.1086/667698},
archivePrefix = {arXiv},
       eprint = {1203.1382},
 primaryClass = {astro-ph.IM},
       adsurl = {https://ui.adsabs.harvard.edu/abs/2012PASP..124..985S}
}

@ARTICLE{pdscap_stumpe_2014,
       author = {{Stumpe}, Martin C. and {Smith}, Jeffrey C. and {Catanzarite}, Joseph H. and {Van Cleve}, Jeffrey E. and {Jenkins}, Jon M. and {Twicken}, Joseph D. and {Girouard}, Forrest R.},
        title = "{Multiscale Systematic Error Correction via Wavelet-Based Bandsplitting in Kepler Data}",
      journal = {\pasp},
         year = 2014,
        month = jan,
       volume = {126},
       number = {935},
        pages = {100},
          doi = {10.1086/674989},
       adsurl = {https://ui.adsabs.harvard.edu/abs/2014PASP..126..100S}
}

@ARTICLE{LP-271-72,
       author = {{Reid}, I. Neill and {Walkowicz}, Lucianne M.},
        title = "{LP 261-75/2MASSW J09510549+3558021: A Young, Wide M4.5/L6 Binary}",
      journal = {\pasp},
         year = 2006,
        month = may,
       volume = {118},
       number = {843},
        pages = {671-677},
          doi = {10.1086/503446},
       adsurl = {https://ui.adsabs.harvard.edu/abs/2006PASP..118..671R}
}

@ARTICLE{FP2018,
       author = {{Fulton}, Benjamin J. and {Petigura}, Erik A.},
        title = "{The California-Kepler Survey. VII. Precise Planet Radii Leveraging Gaia DR2 Reveal the Stellar Mass Dependence of the Planet Radius Gap}",
      journal = {\aj},
         year = 2018,
        month = dec,
       volume = {156},
       number = {6},
          eid = {264},
        pages = {264},
          doi = {10.3847/1538-3881/aae828},
archivePrefix = {arXiv},
       eprint = {1805.01453},
 primaryClass = {astro-ph.EP},
       adsurl = {https://ui.adsabs.harvard.edu/abs/2018AJ....156..264F}
}

@ARTICLE{km_fgk,
       author = {{Kunimoto}, Michelle and {Matthews}, Jaymie M.},
        title = "{Searching the Entirety of Kepler Data. II. Occurrence Rate Estimates for FGK Stars}",
      journal = {\aj},
         year = 2020,
        month = jun,
       volume = {159},
       number = {6},
          eid = {248},
        pages = {248},
          doi = {10.3847/1538-3881/ab88b0},
archivePrefix = {arXiv},
       eprint = {2004.05296},
 primaryClass = {astro-ph.EP},
       adsurl = {https://ui.adsabs.harvard.edu/abs/2020AJ....159..248K}
}

@ARTICLE{hsu2018,
       author = {{Hsu}, Danley C. and {Ford}, Eric B. and {Ragozzine}, Darin and {Morehead}, Robert C.},
        title = "{Improving the Accuracy of Planet Occurrence Rates from Kepler Using Approximate Bayesian Computation}",
      journal = {\aj},
         year = 2018,
        month = may,
       volume = {155},
       number = {5},
          eid = {205},
        pages = {205},
          doi = {10.3847/1538-3881/aab9a8},
archivePrefix = {arXiv},
       eprint = {1803.10787},
 primaryClass = {astro-ph.EP},
       adsurl = {https://ui.adsabs.harvard.edu/abs/2018AJ....155..205H}
}

@ARTICLE{hsu2020,
       author = {{Hsu}, Danley C. and {Ford}, Eric B. and {Terrien}, Ryan},
        title = "{Occurrence rates of planets orbiting M Stars: applying ABC to Kepler DR25, Gaia DR2, and 2MASS data}",
      journal = {\mnras},
         year = 2020,
        month = oct,
       volume = {498},
       number = {2},
        pages = {2249-2262},
          doi = {10.1093/mnras/staa2391},
archivePrefix = {arXiv},
       eprint = {2002.02573},
 primaryClass = {astro-ph.EP},
       adsurl = {https://ui.adsabs.harvard.edu/abs/2020MNRAS.498.2249H}
}

@MISC{lightkurve,
   author = {{Lightkurve Collaboration} and {Cardoso}, J.~V.~d.~M. and
             {Hedges}, C. and {Gully-Santiago}, M. and {Saunders}, N. and
             {Cody}, A.~M. and {Barclay}, T. and {Hall}, O. and
             {Sagear}, S. and {Turtelboom}, E. and {Zhang}, J. and
             {Tzanidakis}, A. and {Mighell}, K. and {Coughlin}, J. and
             {Bell}, K. and {Berta-Thompson}, Z. and {Williams}, P. and
             {Dotson}, J. and {Barentsen}, G.},
    title = "{Lightkurve: Kepler and TESS time series analysis in Python}",
howpublished = {Astrophysics Source Code Library},
     year = 2018,
    month = dec,
archivePrefix = "ascl",
   eprint = {1812.013},
   adsurl = {http://adsabs.harvard.edu/abs/2018ascl.soft12013L},
}

@ARTICLE{twicken2025,
       author = {{Twicken}, Joseph D. and {Jenkins}, Jon M. and {Caldwell}, Douglas A. and {Tofflemire}, Benjamin M. and {Jafariyazani}, Marziye and {Tenenbaum}, Peter and {Smith}, Jeffrey C. and {Striegel}, Stephanie L. and {Ting}, Eric and {Wohler}, Bill and {Rose}, Mark E. and {Rapetti}, David and {Fausnaugh}, Michael M. and {Vanderspek}, Roland},
        title = "{TESS Science Processing Operations Center Photometric Precision Archival Product}",
      journal = {Research Notes of the American Astronomical Society},
         year = 2025,
        month = jun,
       volume = {9},
       number = {6},
          eid = {132},
        pages = {132},
          doi = {10.3847/2515-5172/addec4},
archivePrefix = {arXiv},
       eprint = {2506.04136},
 primaryClass = {astro-ph.IM},
       adsurl = {https://ui.adsabs.harvard.edu/abs/2025RNAAS...9..132T}
}

@ARTICLE{steam_jwst,
       author = {{Piaulet-Ghorayeb}, Caroline and {Benneke}, Bj{\"o}rn and {Radica}, Michael and {Raul}, Eshan and {Coulombe}, Louis-Philippe and {Ahrer}, Eva-Maria and {Kubyshkina}, Daria and {Howard}, Ward S. and {Krissansen-Totton}, Joshua and {MacDonald}, Ryan J. and {Roy}, Pierre-Alexis and {Louca}, Amy and {Christie}, Duncan and {Fournier-Tondreau}, Marylou and {Allart}, Romain and {Miguel}, Yamila and {Schlichting}, Hilke E. and {Welbanks}, Luis and {Cadieux}, Charles and {Dorn}, Caroline and {Evans-Soma}, Thomas M. and {Fortney}, Jonathan J. and {Pierrehumbert}, Raymond and {Lafreni{\`e}re}, David and {Acu{\~n}a}, Lorena and {Komacek}, Thaddeus and {Innes}, Hamish and {Beatty}, Thomas G. and {Cloutier}, Ryan and {Doyon}, Ren{\'e} and {Gagnebin}, Anna and {Gapp}, Cyril and {Knutson}, Heather A.},
        title = "{JWST/NIRISS Reveals the Water-rich ``Steam World'' Atmosphere of GJ 9827 d}",
      journal = {\apjl},
         year = 2024,
        month = oct,
       volume = {974},
       number = {1},
          eid = {L10},
        pages = {L10},
          doi = {10.3847/2041-8213/ad6f00},
archivePrefix = {arXiv},
       eprint = {2410.03527},
 primaryClass = {astro-ph.EP},
       adsurl = {https://ui.adsabs.harvard.edu/abs/2024ApJ...974L..10P}
}

@ARTICLE{hardegree2019,
       author = {{Hardegree-Ullman}, Kevin K. and {Cushing}, Michael C. and {Muirhead}, Philip S. and {Christiansen}, Jessie L.},
        title = "{Kepler Planet Occurrence Rates for Mid-type M Dwarfs as a Function of Spectral Type}",
      journal = {\aj},
         year = 2019,
        month = aug,
       volume = {158},
       number = {2},
          eid = {75},
        pages = {75},
          doi = {10.3847/1538-3881/ab21d2},
archivePrefix = {arXiv},
       eprint = {1905.05900},
 primaryClass = {astro-ph.EP},
       adsurl = {https://ui.adsabs.harvard.edu/abs/2019AJ....158...75H}
}

@ARTICLE{howard2012,
       author = {{Howard}, Andrew W. and {Marcy}, Geoffrey W. and {Bryson}, Stephen T. and {Jenkins}, Jon M. and {Rowe}, Jason F. and {Batalha}, Natalie M. and {Borucki}, William J. and {Koch}, David G. and {Dunham}, Edward W. and {Gautier}, III, Thomas N. and {Van Cleve}, Jeffrey and {Cochran}, William D. and {Latham}, David W. and {Lissauer}, Jack J. and {Torres}, Guillermo and {Brown}, Timothy M. and {Gilliland}, Ronald L. and {Buchhave}, Lars A. and {Caldwell}, Douglas A. and {Christensen-Dalsgaard}, J{\o}rgen and {Ciardi}, David and {Fressin}, Francois and {Haas}, Michael R. and {Howell}, Steve B. and {Kjeldsen}, Hans and {Seager}, Sara and {Rogers}, Leslie and {Sasselov}, Dimitar D. and {Steffen}, Jason H. and {Basri}, Gibor S. and {Charbonneau}, David and {Christiansen}, Jessie and {Clarke}, Bruce and {Dupree}, Andrea and {Fabrycky}, Daniel C. and {Fischer}, Debra A. and {Ford}, Eric B. and {Fortney}, Jonathan J. and {Tarter}, Jill and {Girouard}, Forrest R. and {Holman}, Matthew J. and {Johnson}, John Asher and {Klaus}, Todd C. and {Machalek}, Pavel and {Moorhead}, Althea V. and {Morehead}, Robert C. and {Ragozzine}, Darin and {Tenenbaum}, Peter and {Twicken}, Joseph D. and {Quinn}, Samuel N. and {Isaacson}, Howard and {Shporer}, Avi and {Lucas}, Philip W. and {Walkowicz}, Lucianne M. and {Welsh}, William F. and {Boss}, Alan and {Devore}, Edna and {Gould}, Alan and {Smith}, Jeffrey C. and {Morris}, Robert L. and {Prsa}, Andrej and {Morton}, Timothy D. and {Still}, Martin and {Thompson}, Susan E. and {Mullally}, Fergal and {Endl}, Michael and {MacQueen}, Phillip J.},
        title = "{Planet Occurrence within 0.25 AU of Solar-type Stars from Kepler}",
      journal = {\apjs},
         year = 2012,
        month = aug,
       volume = {201},
       number = {2},
          eid = {15},
        pages = {15},
          doi = {10.1088/0067-0049/201/2/15},
archivePrefix = {arXiv},
       eprint = {1103.2541},
 primaryClass = {astro-ph.EP},
       adsurl = {https://ui.adsabs.harvard.edu/abs/2012ApJS..201...15H}
}

@MISC{exovetter,
      author = {Susan Mullally and others},
      year = 2025,
      url = {https://exovetter.readthedocs.io/en/latest/}}

@INPROCEEDINGS{Jenkins2010,
       author = {{Jenkins}, Jon M. and {Chandrasekaran}, Hema and {McCauliff}, Sean D. and {Caldwell}, Douglas A. and {Tenenbaum}, Peter and {Li}, Jie and {Klaus}, Todd C. and {Cote}, Miles T. and {Middour}, Christopher},
        title = "{Transiting planet search in the Kepler pipeline}",
    booktitle = {Software and Cyberinfrastructure for Astronomy},
         year = 2010,
       editor = {{Radziwill}, Nicole M. and {Bridger}, Alan},
       series = {Society of Photo-Optical Instrumentation Engineers (SPIE) Conference Series},
       volume = {7740},
        month = jul,
          eid = {77400D},
        pages = {77400D},
          doi = {10.1117/12.856764},
       adsurl = {https://ui.adsabs.harvard.edu/abs/2010SPIE.7740E..0DJ}
}

@ARTICLE{Jenkins2002,
       author = {{Jenkins}, Jon M.},
        title = "{The Impact of Solar-like Variability on the Detectability of Transiting Terrestrial Planets}",
      journal = {\apj},
         year = 2002,
        month = aug,
       volume = {575},
       number = {1},
        pages = {493-505},
          doi = {10.1086/341136},
       adsurl = {https://ui.adsabs.harvard.edu/abs/2002ApJ...575..493J}
}

@ARTICLE{GJ12-2024-1,
       author = {{Dholakia}, Shishir and {Palethorpe}, Larissa and {Venner}, Alexander and {Mortier}, Annelies and {Wilson}, Thomas G. and {Huang}, Chelsea X. and {Rice}, Ken and {Van Eylen}, Vincent and {Nabbie}, Emma and {Cloutier}, Ryan and {Boschin}, Walter and {Ciardi}, David and {Delrez}, Laetitia and {Dransfield}, Georgina and {Ducrot}, Elsa and {Essack}, Zahra and {Everett}, Mark E. and {Gillon}, Micha{\"e}l and {Hooton}, Matthew J. and {Kunimoto}, Michelle and {Latham}, David W. and {L{\'o}pez-Morales}, Mercedes and {Li}, Bin and {Li}, Fan and {McDermott}, Scott and {Murphy}, Simon J. and {Murray}, Catriona A. and {Seager}, Sara and {Timmermans}, Mathilde and {Triaud}, Amaury and {Turner}, Daisy A. and {Twicken}, Joseph D. and {Vanderburg}, Andrew and {Wang}, Su and {Wittenmyer}, Robert A. and {Wright}, Duncan},
        title = "{Gliese 12 b, a temperate Earth-sized planet at 12 parsecs discovered with TESS and CHEOPS}",
      journal = {\mnras},
         year = 2024,
        month = jun,
       volume = {531},
       number = {1},
        pages = {1276-1293},
          doi = {10.1093/mnras/stae1152},
archivePrefix = {arXiv},
       eprint = {2405.13118},
 primaryClass = {astro-ph.EP},
       adsurl = {https://ui.adsabs.harvard.edu/abs/2024MNRAS.531.1276D}
}

@ARTICLE{GJ12-2024-2,
       author = {{Kuzuhara}, Masayuki and {Fukui}, Akihiko and {Livingston}, John H. and {Caballero}, Jos{\'e} A. and {de Leon}, Jerome P. and {Hirano}, Teruyuki and {Kasagi}, Yui and {Murgas}, Felipe and {Narita}, Norio and {Omiya}, Masashi and {Orell-Miquel}, Jaume and {Palle}, Enric and {Changeat}, Quentin and {Esparza-Borges}, Emma and {Harakawa}, Hiroki and {Hellier}, Coel and {Hori}, Yasunori and {Ikuta}, Kai and {Ishikawa}, Hiroyuki Tako and {Kodama}, Takanori and {Kotani}, Takayuki and {Kudo}, Tomoyuki and {Morales}, Juan C. and {Mori}, Mayuko and {Nagel}, Evangelos and {Parviainen}, Hannu and {Perdelwitz}, Volker and {Reiners}, Ansgar and {Ribas}, Ignasi and {Sanz-Forcada}, Jorge and {Sato}, Bun'ei and {Schweitzer}, Andreas and {Tabernero}, Hugo M. and {Takarada}, Takuya and {Uyama}, Taichi and {Watanabe}, Noriharu and {Zechmeister}, Mathias and {Garc{\'\i}a}, N{\'e}stor Abreu and {Aoki}, Wako and {Beichman}, Charles and {B{\'e}jar}, V{\'\i}ctor J.~S. and {Brandt}, Timothy D. and {Calatayud-Borras}, Y{\'e}ssica and {Carleo}, Ilaria and {Charbonneau}, David and {Collins}, Karen A. and {Currie}, Thayne and {Doty}, John P. and {Dreizler}, Stefan and {Fern{\'a}ndez-Rodr{\'\i}guez}, Gareb and {Fukuda}, Izuru and {Gal{\'a}n}, Daniel and {Gerald{\'\i}a-Gonz{\'a}lez}, Samuel and {Gonz{\'a}lez-Rodr{\'\i}guez}, Josafat and {Hayashi}, Yuya and {Hedges}, Christina and {Henning}, Thomas and {Hodapp}, Klaus and {Ikoma}, Masahiro and {Isogai}, Keisuke and {Jacobson}, Shane and {Janson}, Markus and {Jenkins}, Jon M. and {Kagetani}, Taiki and {Kambe}, Eiji and {Kawai}, Yugo and {Kawauchi}, Kiyoe and {Kokubo}, Eiichiro and {Konishi}, Mihoko and {Korth}, Judith and {Krishnamurthy}, Vigneshwaran and {Kurokawa}, Takashi and {Kusakabe}, Nobuhiko and {Kwon}, Jungmi and {Laza-Ramos}, Andr{\'e}s and {Libotte}, Florence and {Luque}, Rafael and {Madrigal-Aguado}, Alberto and {Matsumoto}, Yuji and {Mawet}, Dimitri and {McElwain}, Michael W. and {Meni Gallardo}, Pedro Pablo and {Morello}, Giuseppe and {Mu{\~n}oz Torres}, Sara and {Nishikawa}, Jun and {Nugroho}, Stevanus K. and {Ogihara}, Masahiro and {Pel{\'a}ez-Torres}, Alberto and {Rapetti}, David and {S{\'a}nchez-Benavente}, Manuel and {Schlecker}, Martin and {Seager}, Sara and {Serabyn}, Eugene and {Serizawa}, Takuma and {Stangret}, Monika and {Takahashi}, Aoi and {Teng}, Huan-Yu and {Tamura}, Motohide and {Terada}, Yuka and {Ueda}, Akitoshi and {Usuda}, Tomonori and {Vanderspek}, Roland and {Vievard}, S{\'e}bastien and {Watanabe}, David and {Winn}, Joshua N. and {Zapatero Osorio}, Maria Rosa},
        title = "{Gliese 12 b: A Temperate Earth-sized Planet at 12 pc Ideal for Atmospheric Transmission Spectroscopy}",
      journal = {\apjl},
         year = 2024,
        month = jun,
       volume = {967},
       number = {2},
          eid = {L21},
        pages = {L21},
          doi = {10.3847/2041-8213/ad3642},
archivePrefix = {arXiv},
       eprint = {2405.14708},
 primaryClass = {astro-ph.EP},
       adsurl = {https://ui.adsabs.harvard.edu/abs/2024ApJ...967L..21K}
}

@ARTICLE{GJ12-2025-1,
       author = {{Brady}, Madison and {Bean}, Jacob and {Basant}, Ritvik and {Brown}, Nina and {Das}, Tanya and {Nixon}, Matthew and {Luque}, Rafael and {Piaulet-Ghorayeb}, Caroline and {Radica}, Michael and {Seifahrt}, Andreas and {St{\"u}rmer}, Julian and {Zhao}, Lily},
        title = "{An Earth-like Density for the Temperate Earth-sized Planet GJ 12b}",
      journal = {arXiv e-prints},
         year = 2025,
        month = jun,
          eid = {arXiv:2506.20561},
        pages = {arXiv:2506.20561},
          doi = {10.48550/arXiv.2506.20561},
archivePrefix = {arXiv},
       eprint = {2506.20561},
 primaryClass = {astro-ph.EP},
       adsurl = {https://ui.adsabs.harvard.edu/abs/2025arXiv250620561B}
}

@ARTICLE{GJ12-2025-2,
       author = {{Turner}, Daisy A. and {Eschen}, Yoshi Nike Emilia and {Murgas}, Felipe and {Mortier}, Annelies and {Wilson}, Thomas G and {Fern{\'a}ndez Fern{\'a}ndez}, Jorge and {Morello}, Giuseppe and {Vissapragada}, Shreyas and {Caballero}, Jos{\'e} A. and {Dreizler}, Stefan and {Egger}, Jo Ann and {Freckelton}, Alix Violet and {Gromek}, Nicole and {Hatzes}, Artie P. and {Lakeland}, Ben Scott and {Nagel}, Evangelos and {Naponiello}, Luca and {Tabernero}, Hugo M. and {Vanaverbeke}, Siegfried and {Venner}, Alexander and {Rosa Zapatero Osorio}, Mar{\'\i}a and {Amado}, Pedro J. and {B{\'e}jar}, V{\'\i}ctor J.~S. and {Bonomo}, Aldo Stefano and {Buchhave}, Lars A. and {Collier Cameron}, Andrew and {Carleo}, Ilaria and {Chaturvedi}, Priyanka and {Cloutier}, Ryan and {Damasso}, Mario and {Daspute}, Mangesh and {Dholakia}, Shishir and {Dufoer}, Sjoerd and {Dumusque}, Xavier and {Fabricio Martinez Fiorenzano}, Aldo and {Ghedina}, Adriano and {Harutyunyan}, Avet and {Herrero}, Enrique and {John}, Ancy Anna and {Lillo-Box}, Jorge and {Lodieu}, Nicolas and {L{\'o}pez-Morales}, Mercedes and {Malavolta}, Luca and {Mancini}, Luigi and {Mantovan}, Giacomo and {Montes}, David and {Morales}, Juan Carlos and {Nicholson}, Belinda and {Orell-Miquel}, Jaume and {Palethorpe}, Larissa and {Palle}, Enric and {Quirrenbach}, Andreas and {Reffert}, Sabine and {Reiners}, Ansgar and {Ribas}, Ignasi and {Rice}, Ken and {Silva}, Andr{\'e} M. and {Sozzetti}, Alessandro and {Stalport}, Manu and {Tal-Or}, Lev and {Trifonov}, Trifon and {Udry}, St{\'e}phane and {Zechmeister}, Mathias},
        title = "{The mass of the exo-Venus Gliese 12 b, as revealed by HARPS-N, ESPRESSO, and CARMENES}",
      journal = {arXiv e-prints},
         year = 2025,
        month = jun,
          eid = {arXiv:2506.20564},
        pages = {arXiv:2506.20564},
          doi = {10.48550/arXiv.2506.20564},
archivePrefix = {arXiv},
       eprint = {2506.20564},
 primaryClass = {astro-ph.EP},
       adsurl = {https://ui.adsabs.harvard.edu/abs/2025arXiv250620564T}
}

@ARTICLE{tess_giant_occurrence,
       author = {{Bryant}, Edward M. and {Bayliss}, Daniel and {Van Eylen}, Vincent},
        title = "{The occurrence rate of giant planets orbiting low-mass stars with TESS}",
      journal = {\mnras},
         year = 2023,
        month = may,
       volume = {521},
       number = {3},
        pages = {3663-3681},
          doi = {10.1093/mnras/stad626},
archivePrefix = {arXiv},
       eprint = {2303.00659},
 primaryClass = {astro-ph.EP},
       adsurl = {https://ui.adsabs.harvard.edu/abs/2023MNRAS.521.3663B}
}

@ARTICLE{HU2025,
       author = {{Hardegree-Ullman}, Kevin K. and {Bergsten}, Galen J. and {Christiansen}, Jessie L. and {Zink}, Jon K. and {Bhure}, Sakhee and {Boley}, Kiersten M. and {Fernandes}, Rachel B. and {Giacalone}, Steven and {Karpoor}, Preethi R.},
        title = "{Scaling K2. VIII. Short-period Sub-Neptune Occurrence Rates Peak Around Early-type M Dwarfs}",
      journal = {\aj},
         year = 2025,
        month = sep,
       volume = {170},
       number = {3},
          eid = {183},
        pages = {183},
          doi = {10.3847/1538-3881/adf633},
archivePrefix = {arXiv},
       eprint = {2508.05734},
 primaryClass = {astro-ph.EP},
       adsurl = {https://ui.adsabs.harvard.edu/abs/2025AJ....170..183H}
}

@ARTICLE{chachan2023,
       author = {{Chachan}, Yayaati and {Lee}, Eve J.},
        title = "{Small Planets around Cool Dwarfs: Enhanced Formation Efficiency of Super-Earths around M Dwarfs}",
      journal = {\apjl},
         year = 2023,
        month = jul,
       volume = {952},
       number = {1},
          eid = {L20},
        pages = {L20},
          doi = {10.3847/2041-8213/ace257},
archivePrefix = {arXiv},
       eprint = {2305.00803},
 primaryClass = {astro-ph.EP},
       adsurl = {https://ui.adsabs.harvard.edu/abs/2023ApJ...952L..20C}
}

@ARTICLE{cherubim_2023,
       author = {{Cherubim}, Collin and {Cloutier}, Ryan and {Charbonneau}, David and {Stockdale}, Chris and {Stassun}, Keivan G. and {Schwarz}, Richard P. and {Safonov}, Boris and {Mortier}, Annelies and {Lewin}, Pablo and {Latham}, David W. and {Horne}, Keith and {Haywood}, Rapha{\"e}lle D. and {Gonzales}, Erica and {Goliguzova}, Maria V. and {Collins}, Karen A. and {Ciardi}, David R. and {Bieryla}, Allyson and {Belinski}, Alexandre A. and {Wohler}, Bill and {Watson}, Christopher A. and {Vanderspek}, Roland and {Udry}, St{\'e}phane and {Sozzetti}, Alessandro and {S{\'e}gransan}, Damien and {Sasselov}, Dimitar and {Ricker}, George R. and {Rice}, Ken and {Poretti}, Ennio and {Piotto}, Giampaolo and {Pepe}, Francesco and {Molinari}, Emilio and {Micela}, Giuseppina and {Mayor}, Michel and {Lovis}, Christophe and {L{\'o}pez-Morales}, Mercedes and {Jenkins}, Jon M. and {Essack}, Zahra and {Dumusque}, Xavier and {Doty}, John P. and {Col{\'o}n}, Knicole D. and {Cameron}, Andrew Collier and {Buchhave}, Lars A.},
        title = "{TOI-1695 b: A Water World Orbiting an Early-M Dwarf in the Planet Radius Valley}",
      journal = {\aj},
         year = 2023,
        month = apr,
       volume = {165},
       number = {4},
          eid = {167},
        pages = {167},
          doi = {10.3847/1538-3881/acbdfd},
archivePrefix = {arXiv},
       eprint = {2211.06445},
 primaryClass = {astro-ph.EP},
       adsurl = {https://ui.adsabs.harvard.edu/abs/2023AJ....165..167C}
}

@ARTICLE{diamond-lowe_2022,
       author = {{Diamond-Lowe}, Hannah and {Kreidberg}, Laura and {Harman}, C.~E. and {Kempton}, Eliza M. -R. and {Rogers}, Leslie A. and {Joyce}, Simon R.~G. and {Eastman}, Jason D. and {King}, George W. and {Kopparapu}, Ravi and {Youngblood}, Allison and {Kosiarek}, Molly R. and {Livingston}, John H. and {Hardegree-Ullman}, Kevin K. and {Crossfield}, Ian J.~M.},
        title = "{The K2-3 System Revisited: Testing Photoevaporation and Core-powered Mass Loss with Three Small Planets Spanning the Radius Valley}",
      journal = {\aj},
         year = 2022,
        month = nov,
       volume = {164},
       number = {5},
          eid = {172},
        pages = {172},
          doi = {10.3847/1538-3881/ac7807},
archivePrefix = {arXiv},
       eprint = {2207.12755},
 primaryClass = {astro-ph.EP},
       adsurl = {https://ui.adsabs.harvard.edu/abs/2022AJ....164..172D}
}

@ARTICLE{piaulet_2023,
       author = {{Piaulet}, Caroline and {Benneke}, Bj{\"o}rn and {Almenara}, Jose M. and {Dragomir}, Diana and {Knutson}, Heather A. and {Thorngren}, Daniel and {Peterson}, Merrin S. and {Crossfield}, Ian J.~M. and {Kempton}, Eliza M. -R. and {Kubyshkina}, Daria and {Howard}, Andrew W. and {Angus}, Ruth and {Isaacson}, Howard and {Weiss}, Lauren M. and {Beichman}, Charles A. and {Fortney}, Jonathan J. and {Fossati}, Luca and {Lammer}, Helmut and {McCullough}, P.~R. and {Morley}, Caroline V. and {Wong}, Ian},
        title = "{Evidence for the volatile-rich composition of a 1.5-Earth-radius planet}",
      journal = {Nature Astronomy},
         year = 2023,
        month = feb,
       volume = {7},
        pages = {206-222},
          doi = {10.1038/s41550-022-01835-4},
archivePrefix = {arXiv},
       eprint = {2212.08477},
 primaryClass = {astro-ph.EP},
       adsurl = {https://ui.adsabs.harvard.edu/abs/2023NatAs...7..206P}
}

@ARTICLE{icy_ho_2024,
       author = {{Ho}, Cynthia S.~K. and {Rogers}, James G. and {Van Eylen}, Vincent and {Owen}, James E. and {Schlichting}, Hilke E.},
        title = "{Shallower radius valley around low-mass hosts: evidence for icy planets, collisions, or high-energy radiation scatter}",
      journal = {\mnras},
         year = 2024,
        month = jul,
       volume = {531},
       number = {3},
        pages = {3698-3714},
          doi = {10.1093/mnras/stae1376},
archivePrefix = {arXiv},
       eprint = {2401.12378},
 primaryClass = {astro-ph.EP},
       adsurl = {https://ui.adsabs.harvard.edu/abs/2024MNRAS.531.3698H}
}

@ARTICLE{luque_2022,
       author = {{Luque}, Rafael and {Pall{\'e}}, Enric},
        title = "{Density, not radius, separates rocky and water-rich small planets orbiting M dwarf stars}",
      journal = {Science},
         year = 2022,
        month = sep,
       volume = {377},
       number = {6611},
        pages = {1211-1214},
          doi = {10.1126/science.abl7164},
archivePrefix = {arXiv},
       eprint = {2209.03871},
 primaryClass = {astro-ph.EP},
       adsurl = {https://ui.adsabs.harvard.edu/abs/2022Sci...377.1211L}
}

@ARTICLE{gaidos_2024,
       author = {{Gaidos}, Eric and {Ali}, Aleezah and {Kraus}, Adam L. and {Rowe}, Jason F.},
        title = "{The radius distribution of M dwarf-hosted planets and its evolution}",
      journal = {\mnras},
         year = 2024,
        month = nov,
       volume = {534},
       number = {4},
        pages = {3277-3290},
          doi = {10.1093/mnras/stae2207},
archivePrefix = {arXiv},
       eprint = {2404.11022},
 primaryClass = {astro-ph.EP},
       adsurl = {https://ui.adsabs.harvard.edu/abs/2024MNRAS.534.3277G}
}

@ARTICLE{neptune_1,
       author = {{Jones}, Sinclaire E. and {Stef{\'a}nsson}, Gu{\dj}mundur and {Masuda}, Kento and {Libby-Roberts}, Jessica E. and {Gardner}, Cristilyn N. and {Holcomb}, Rae and {Beard}, Corey and {Robertson}, Paul and {Ca{\~n}as}, Caleb I. and {Mahadevan}, Suvrath and {Kanodia}, Shubham and {Lin}, Andrea S.~J. and {Kobulnicky}, Henry A. and {Parker}, Brock A. and {Bender}, Chad F. and {Cochran}, William D. and {Diddams}, Scott A. and {Fernandes}, Rachel B. and {Gupta}, Arvind F. and {Halverson}, Samuel and {Hawley}, Suzanne L. and {Hearty}, Fred R. and {Hebb}, Leslie and {Kowalski}, Adam and {Lubin}, Jack and {Monson}, Andrew and {Ninan}, Joe P. and {Ramsey}, Lawrence and {Roy}, Arpita and {Schwab}, Christian and {Terrien}, Ryan C. and {Wisniewski}, John},
        title = "{TOI-2015 b: A Warm Neptune with Transit Timing Variations Orbiting an Active Mid-type M Dwarf}",
      journal = {\aj},
         year = 2024,
        month = aug,
       volume = {168},
       number = {2},
          eid = {93},
        pages = {93},
          doi = {10.3847/1538-3881/ad55ea},
archivePrefix = {arXiv},
       eprint = {2310.11775},
 primaryClass = {astro-ph.EP},
       adsurl = {https://ui.adsabs.harvard.edu/abs/2024AJ....168...93J}
}

@ARTICLE{neptune_2,
       author = {{Almenara}, J.~M. and {Bonfils}, X. and {Forveille}, T. and {Astudillo-Defru}, N. and {Ciardi}, D.~R. and {Schwarz}, R.~P. and {Collins}, K.~A. and {Cointepas}, M. and {Lund}, M.~B. and {Bouchy}, F. and {Charbonneau}, D. and {D{\'\i}az}, R.~F. and {Delfosse}, X. and {Kidwell}, R.~C. and {Kunimoto}, M. and {Latham}, D.~W. and {Lissauer}, J.~J. and {Murgas}, F. and {Ricker}, G. and {Seager}, S. and {Vezie}, M. and {Watanabe}, D.},
        title = "{TOI-3884 b: A rare 6-R$_{E}$ planet that transits a low-mass star with a giant and likely polar spot}",
      journal = {\aap},
         year = 2022,
        month = nov,
       volume = {667},
          eid = {L11},
        pages = {L11},
          doi = {10.1051/0004-6361/202244791},
archivePrefix = {arXiv},
       eprint = {2210.10909},
 primaryClass = {astro-ph.EP},
       adsurl = {https://ui.adsabs.harvard.edu/abs/2022A&A...667L..11A}
}

@ARTICLE{neptune_3,
       author = {{Libby-Roberts}, Jessica E. and {Schutte}, Maria and {Hebb}, Leslie and {Kanodia}, Shubham and {Ca{\~n}as}, Caleb I. and {Stef{\'a}nsson}, Gu{\dh}mundur and {Lin}, Andrea S.~J. and {Mahadevan}, Suvrath and {Parts}, Winter and {Powers}, Luke and {Wisniewski}, John and {Bender}, Chad F. and {Cochran}, William D. and {Diddams}, Scott A. and {Everett}, Mark E. and {Gupta}, Arvind F. and {Halverson}, Samuel and {Kobulnicky}, Henry A. and {Kowalski}, Adam F. and {Larsen}, Alexander and {Monson}, Andrew and {Ninan}, Joe P. and {Parker}, Brock A. and {Ramsey}, Lawrence W. and {Robertson}, Paul and {Schwab}, Christian and {Swaby}, Tera N. and {Terrien}, Ryan C.},
        title = "{An In-depth Look at TOI-3884b: A Super-Neptune Transiting an M4Dwarf with Persistent Starspot Crossings}",
      journal = {\aj},
         year = 2023,
        month = jun,
       volume = {165},
       number = {6},
          eid = {249},
        pages = {249},
          doi = {10.3847/1538-3881/accc2f},
       adsurl = {https://ui.adsabs.harvard.edu/abs/2023AJ....165..249L}
}

@ARTICLE{early_venus,
       author = {{Solomon}, S.~C. and {Head}, J.~W.},
        title = "{Fundamental Issues in the Geology and Geophysics of Venus}",
      journal = {Science},
         year = 1991,
        month = apr,
       volume = {252},
       number = {5003},
        pages = {252-260},
          doi = {10.1126/science.252.5003.252},
       adsurl = {https://ui.adsabs.harvard.edu/abs/1991Sci...252..252S}
}

@article{astropy:2013,
Adsurl = {http://adsabs.harvard.edu/abs/2013A%26A...558A..33A},
Archiveprefix = {arXiv},
Author = {{Astropy Collaboration} and {Robitaille}, T.~P. and {Tollerud}, E.~J. and {Greenfield}, P. and {Droettboom}, M. and {Bray}, E. and {Aldcroft}, T. and {Davis}, M. and {Ginsburg}, A. and {Price-Whelan}, A.~M. and {Kerzendorf}, W.~E. and {Conley}, A. and {Crighton}, N. and {Barbary}, K. and {Muna}, D. and {Ferguson}, H. and {Grollier}, F. and {Parikh}, M.~M. and {Nair}, P.~H. and {Unther}, H.~M. and {Deil}, C. and {Woillez}, J. and {Conseil}, S. and {Kramer}, R. and {Turner}, J.~E.~H. and {Singer}, L. and {Fox}, R. and {Weaver}, B.~A. and {Zabalza}, V. and {Edwards}, Z.~I. and {Azalee Bostroem}, K. and {Burke}, D.~J. and {Casey}, A.~R. and {Crawford}, S.~M. and {Dencheva}, N. and {Ely}, J. and {Jenness}, T. and {Labrie}, K. and {Lim}, P.~L. and {Pierfederici}, F. and {Pontzen}, A. and {Ptak}, A. and {Refsdal}, B. and {Servillat}, M. and {Streicher}, O.},
Doi = {10.1051/0004-6361/201322068},
Eid = {A33},
Eprint = {1307.6212},
Journal = {\aap},
Month = oct,
Pages = {A33},
Primaryclass = {astro-ph.IM},
Title = {{Astropy: A community Python package for astronomy}},
Volume = 558,
Year = 2013}

@ARTICLE{astropy:2018,
       author = {{Astropy Collaboration} and {Price-Whelan}, A.~M. and
         {Sip{\H{o}}cz}, B.~M. and {G{\"u}nther}, H.~M. and {Lim}, P.~L. and
         {Crawford}, S.~M. and {Conseil}, S. and {Shupe}, D.~L. and
         {Craig}, M.~W. and {Dencheva}, N. and {Ginsburg}, A. and {Vand
        erPlas}, J.~T. and {Bradley}, L.~D. and {P{\'e}rez-Su{\'a}rez}, D. and
         {de Val-Borro}, M. and {Aldcroft}, T.~L. and {Cruz}, K.~L. and
         {Robitaille}, T.~P. and {Tollerud}, E.~J. and {Ardelean}, C. and
         {Babej}, T. and {Bach}, Y.~P. and {Bachetti}, M. and {Bakanov}, A.~V. and
         {Bamford}, S.~P. and {Barentsen}, G. and {Barmby}, P. and
         {Baumbach}, A. and {Berry}, K.~L. and {Biscani}, F. and {Boquien}, M. and
         {Bostroem}, K.~A. and {Bouma}, L.~G. and {Brammer}, G.~B. and
         {Bray}, E.~M. and {Breytenbach}, H. and {Buddelmeijer}, H. and
         {Burke}, D.~J. and {Calderone}, G. and {Cano Rodr{\'\i}guez}, J.~L. and
         {Cara}, M. and {Cardoso}, J.~V.~M. and {Cheedella}, S. and {Copin}, Y. and
         {Corrales}, L. and {Crichton}, D. and {D'Avella}, D. and {Deil}, C. and
         {Depagne}, {\'E}. and {Dietrich}, J.~P. and {Donath}, A. and
         {Droettboom}, M. and {Earl}, N. and {Erben}, T. and {Fabbro}, S. and
         {Ferreira}, L.~A. and {Finethy}, T. and {Fox}, R.~T. and
         {Garrison}, L.~H. and {Gibbons}, S.~L.~J. and {Goldstein}, D.~A. and
         {Gommers}, R. and {Greco}, J.~P. and {Greenfield}, P. and
         {Groener}, A.~M. and {Grollier}, F. and {Hagen}, A. and {Hirst}, P. and
         {Homeier}, D. and {Horton}, A.~J. and {Hosseinzadeh}, G. and {Hu}, L. and
         {Hunkeler}, J.~S. and {Ivezi{\'c}}, {\v{Z}}. and {Jain}, A. and
         {Jenness}, T. and {Kanarek}, G. and {Kendrew}, S. and {Kern}, N.~S. and
         {Kerzendorf}, W.~E. and {Khvalko}, A. and {King}, J. and {Kirkby}, D. and
         {Kulkarni}, A.~M. and {Kumar}, A. and {Lee}, A. and {Lenz}, D. and
         {Littlefair}, S.~P. and {Ma}, Z. and {Macleod}, D.~M. and
         {Mastropietro}, M. and {McCully}, C. and {Montagnac}, S. and
         {Morris}, B.~M. and {Mueller}, M. and {Mumford}, S.~J. and {Muna}, D. and
         {Murphy}, N.~A. and {Nelson}, S. and {Nguyen}, G.~H. and
         {Ninan}, J.~P. and {N{\"o}the}, M. and {Ogaz}, S. and {Oh}, S. and
         {Parejko}, J.~K. and {Parley}, N. and {Pascual}, S. and {Patil}, R. and
         {Patil}, A.~A. and {Plunkett}, A.~L. and {Prochaska}, J.~X. and
         {Rastogi}, T. and {Reddy Janga}, V. and {Sabater}, J. and
         {Sakurikar}, P. and {Seifert}, M. and {Sherbert}, L.~E. and
         {Sherwood-Taylor}, H. and {Shih}, A.~Y. and {Sick}, J. and
         {Silbiger}, M.~T. and {Singanamalla}, S. and {Singer}, L.~P. and
         {Sladen}, P.~H. and {Sooley}, K.~A. and {Sornarajah}, S. and
         {Streicher}, O. and {Teuben}, P. and {Thomas}, S.~W. and
         {Tremblay}, G.~R. and {Turner}, J.~E.~H. and {Terr{\'o}n}, V. and
         {van Kerkwijk}, M.~H. and {de la Vega}, A. and {Watkins}, L.~L. and
         {Weaver}, B.~A. and {Whitmore}, J.~B. and {Woillez}, J. and
         {Zabalza}, V. and {Astropy Contributors}},
        title = "{The Astropy Project: Building an Open-science Project and Status of the v2.0 Core Package}",
      journal = {\aj},
         year = 2018,
        month = sep,
       volume = {156},
       number = {3},
          eid = {123},
        pages = {123},
          doi = {10.3847/1538-3881/aabc4f},
archivePrefix = {arXiv},
       eprint = {1801.02634},
 primaryClass = {astro-ph.IM},
       adsurl = {https://ui.adsabs.harvard.edu/abs/2018AJ....156..123A}
}

@ARTICLE{astropy:2022,
       author = {{Astropy Collaboration} and {Price-Whelan}, Adrian M. and {Lim}, Pey Lian and {Earl}, Nicholas and {Starkman}, Nathaniel and {Bradley}, Larry and {Shupe}, David L. and {Patil}, Aarya A. and {Corrales}, Lia and {Brasseur}, C.~E. and {N{"o}the}, Maximilian and {Donath}, Axel and {Tollerud}, Erik and {Morris}, Brett M. and {Ginsburg}, Adam and {Vaher}, Eero and {Weaver}, Benjamin A. and {Tocknell}, James and {Jamieson}, William and {van Kerkwijk}, Marten H. and {Robitaille}, Thomas P. and {Merry}, Bruce and {Bachetti}, Matteo and {G{"u}nther}, H. Moritz and {Aldcroft}, Thomas L. and {Alvarado-Montes}, Jaime A. and {Archibald}, Anne M. and {B{'o}di}, Attila and {Bapat}, Shreyas and {Barentsen}, Geert and {Baz{'a}n}, Juanjo and {Biswas}, Manish and {Boquien}, M{'e}d{'e}ric and {Burke}, D.~J. and {Cara}, Daria and {Cara}, Mihai and {Conroy}, Kyle E. and {Conseil}, Simon and {Craig}, Matthew W. and {Cross}, Robert M. and {Cruz}, Kelle L. and {D'Eugenio}, Francesco and {Dencheva}, Nadia and {Devillepoix}, Hadrien A.~R. and {Dietrich}, J{"o}rg P. and {Eigenbrot}, Arthur Davis and {Erben}, Thomas and {Ferreira}, Leonardo and {Foreman-Mackey}, Daniel and {Fox}, Ryan and {Freij}, Nabil and {Garg}, Suyog and {Geda}, Robel and {Glattly}, Lauren and {Gondhalekar}, Yash and {Gordon}, Karl D. and {Grant}, David and {Greenfield}, Perry and {Groener}, Austen M. and {Guest}, Steve and {Gurovich}, Sebastian and {Handberg}, Rasmus and {Hart}, Akeem and {Hatfield-Dodds}, Zac and {Homeier}, Derek and {Hosseinzadeh}, Griffin and {Jenness}, Tim and {Jones}, Craig K. and {Joseph}, Prajwel and {Kalmbach}, J. Bryce and {Karamehmetoglu}, Emir and {Ka{l}uszy{'n}ski}, Miko{l}aj and {Kelley}, Michael S.~P. and {Kern}, Nicholas and {Kerzendorf}, Wolfgang E. and {Koch}, Eric W. and {Kulumani}, Shankar and {Lee}, Antony and {Ly}, Chun and {Ma}, Zhiyuan and {MacBride}, Conor and {Maljaars}, Jakob M. and {Muna}, Demitri and {Murphy}, N.~A. and {Norman}, Henrik and {O'Steen}, Richard and {Oman}, Kyle A. and {Pacifici}, Camilla and {Pascual}, Sergio and {Pascual-Granado}, J. and {Patil}, Rohit R. and {Perren}, Gabriel I. and {Pickering}, Timothy E. and {Rastogi}, Tanuj and {Roulston}, Benjamin R. and {Ryan}, Daniel F. and {Rykoff}, Eli S. and {Sabater}, Jose and {Sakurikar}, Parikshit and {Salgado}, Jes{'u}s and {Sanghi}, Aniket and {Saunders}, Nicholas and {Savchenko}, Volodymyr and {Schwardt}, Ludwig and {Seifert-Eckert}, Michael and {Shih}, Albert Y. and {Jain}, Anany Shrey and {Shukla}, Gyanendra and {Sick}, Jonathan and {Simpson}, Chris and {Singanamalla}, Sudheesh and {Singer}, Leo P. and {Singhal}, Jaladh and {Sinha}, Manodeep and {Sip{H{o}}cz}, Brigitta M. and {Spitler}, Lee R. and {Stansby}, David and {Streicher}, Ole and {{{S}}umak}, Jani and {Swinbank}, John D. and {Taranu}, Dan S. and {Tewary}, Nikita and {Tremblay}, Grant R. and {Val-Borro}, Miguel de and {Van Kooten}, Samuel J. and {Vasovi{'c}}, Zlatan and {Verma}, Shresth and {de Miranda Cardoso}, Jos{'e} Vin{'i}cius and {Williams}, Peter K.~G. and {Wilson}, Tom J. and {Winkel}, Benjamin and {Wood-Vasey}, W.~M. and {Xue}, Rui and {Yoachim}, Peter and {Zhang}, Chen and {Zonca}, Andrea and {Astropy Project Contributors}},
        title = "{The Astropy Project: Sustaining and Growing a Community-oriented Open-source Project and the Latest Major Release (v5.0) of the Core Package}",
      journal = {\apj},
         year = 2022,
        month = aug,
       volume = {935},
       number = {2},
          eid = {167},
        pages = {167},
          doi = {10.3847/1538-4357/ac7c74},
archivePrefix = {arXiv},
       eprint = {2206.14220},
 primaryClass = {astro-ph.IM},
       adsurl = {https://ui.adsabs.harvard.edu/abs/2022ApJ...935..167A}
}

@article{pymc2023,
  title = {{PyMC}: A Modern and Comprehensive Probabilistic Programming Framework in {P}ython},
  author = {Oriol Abril-Pla and Virgile Andreani and Colin Carroll and Larry Dong and Christopher J. Fonnesbeck and Maxim Kochurov and Ravin Kumar and Junpeng Lao and Christian C. Luhmann and Osvaldo A. Martin and Michael Osthege and Ricardo Vieira and Thomas Wiecki and Robert Zinkov },
  journal = {{PeerJ} Computer Science},
  volume = {9},
  number = {e1516},
  doi = {10.7717/peerj-cs.1516},
  year = {2023}
}

@Article{         numpy,
 title         = {Array programming with {NumPy}},
 author        = {Charles R. Harris and K. Jarrod Millman and St{\'{e}}fan J.
                 van der Walt and Ralf Gommers and Pauli Virtanen and David
                 Cournapeau and Eric Wieser and Julian Taylor and Sebastian
                 Berg and Nathaniel J. Smith and Robert Kern and Matti Picus
                 and Stephan Hoyer and Marten H. van Kerkwijk and Matthew
                 Brett and Allan Haldane and Jaime Fern{\'{a}}ndez del
                 R{\'{i}}o and Mark Wiebe and Pearu Peterson and Pierre
                 G{\'{e}}rard-Marchant and Kevin Sheppard and Tyler Reddy and
                 Warren Weckesser and Hameer Abbasi and Christoph Gohlke and
                 Travis E. Oliphant},
 year          = {2020},
 month         = sep,
 journal       = {Nature},
 volume        = {585},
 number        = {7825},
 pages         = {357--362},
 doi           = {10.1038/s41586-020-2649-2},
 publisher     = {Springer Science and Business Media {LLC}},
 url           = {https://doi.org/10.1038/s41586-020-2649-2}
}

@ARTICLE{LHS1140,
       author = {{Dittmann}, Jason A. and {Irwin}, Jonathan M. and {Charbonneau}, David and {Bonfils}, Xavier and {Astudillo-Defru}, Nicola and {Haywood}, Rapha{\"e}lle D. and {Berta-Thompson}, Zachory K. and {Newton}, Elisabeth R. and {Rodriguez}, Joseph E. and {Winters}, Jennifer G. and {Tan}, Thiam-Guan and {Almenara}, Jose-Manuel and {Bouchy}, Fran{\c{c}}ois and {Delfosse}, Xavier and {Forveille}, Thierry and {Lovis}, Christophe and {Murgas}, Felipe and {Pepe}, Francesco and {Santos}, Nuno C. and {Udry}, Stephane and {W{\"u}nsche}, Ana{\"e}l and {Esquerdo}, Gilbert A. and {Latham}, David W. and {Dressing}, Courtney D.},
        title = "{A temperate rocky super-Earth transiting a nearby cool star}",
      journal = {\nat},
         year = 2017,
        month = apr,
       volume = {544},
       number = {7650},
        pages = {333-336},
          doi = {10.1038/nature22055},
archivePrefix = {arXiv},
       eprint = {1704.05556},
 primaryClass = {astro-ph.EP},
       adsurl = {https://ui.adsabs.harvard.edu/abs/2017Natur.544..333D}
}

@ARTICLE{LHS1140c,
       author = {{Ment}, Kristo and {Dittmann}, Jason A. and {Astudillo-Defru}, Nicola and {Charbonneau}, David and {Irwin}, Jonathan and {Bonfils}, Xavier and {Murgas}, Felipe and {Almenara}, Jose-Manuel and {Forveille}, Thierry and {Agol}, Eric and {Ballard}, Sarah and {Berta-Thompson}, Zachory K. and {Bouchy}, Fran{\c{c}}ois and {Cloutier}, Ryan and {Delfosse}, Xavier and {Doyon}, Ren{\'e} and {Dressing}, Courtney D. and {Esquerdo}, Gilbert A. and {Haywood}, Rapha{\"e}lle D. and {Kipping}, David M. and {Latham}, David W. and {Lovis}, Christophe and {Newton}, Elisabeth R. and {Pepe}, Francesco and {Rodriguez}, Joseph E. and {Santos}, Nuno C. and {Tan}, Thiam-Guan and {Udry}, Stephane and {Winters}, Jennifer G. and {W{\"u}nsche}, Ana{\"e}l},
        title = "{A Second Terrestrial Planet Orbiting the Nearby M Dwarf LHS 1140}",
      journal = {\aj},
         year = 2019,
        month = jan,
       volume = {157},
       number = {1},
          eid = {32},
        pages = {32},
          doi = {10.3847/1538-3881/aaf1b1},
archivePrefix = {arXiv},
       eprint = {1808.00485},
 primaryClass = {astro-ph.EP},
       adsurl = {https://ui.adsabs.harvard.edu/abs/2019AJ....157...32M}
}

@ARTICLE{LHS1140RV,
       author = {{Cadieux}, Charles and {Plotnykov}, Mykhaylo and {Doyon}, Ren{\'e} and {Valencia}, Diana and {Jahandar}, Farbod and {Dang}, Lisa and {Turbet}, Martin and {Fauchez}, Thomas J. and {Cloutier}, Ryan and {Cherubim}, Collin and {Artigau}, {\'E}tienne and {Cook}, Neil J. and {Edwards}, Billy and {Hallatt}, Tim and {Charnay}, Benjamin and {Bouchy}, Fran{\c{c}}ois and {Allart}, Romain and {Mignon}, Lucile and {Baron}, Fr{\'e}d{\'e}rique and {Barros}, Susana C.~C. and {Benneke}, Bj{\"o}rn and {Canto Martins}, B.~L. and {Cowan}, Nicolas B. and {De Medeiros}, J.~R. and {Delfosse}, Xavier and {Delgado-Mena}, Elisa and {Dumusque}, Xavier and {Ehrenreich}, David and {Frensch}, Yolanda G.~C. and {Gonz{\'a}lez Hern{\'a}ndez}, J.~I. and {Hara}, Nathan C. and {Lafreni{\`e}re}, David and {Lo Curto}, Gaspare and {Malo}, Lison and {Melo}, Claudio and {Mounzer}, Dany and {Passeger}, Vera Maria and {Pepe}, Francesco and {Poulin-Girard}, Anne-Sophie and {Santos}, Nuno C. and {Sosnowska}, Danuta and {Su{\'a}rez Mascare{\~n}o}, Alejandro and {Thibault}, Simon and {Vaulato}, Valentina and {Wade}, Gregg A. and {Wildi}, Fran{\c{c}}ois},
        title = "{New Mass and Radius Constraints on the LHS 1140 Planets: LHS 1140 b Is either a Temperate Mini-Neptune or a Water World}",
      journal = {\apjl},
         year = 2024,
        month = jan,
       volume = {960},
       number = {1},
          eid = {L3},
        pages = {L3},
          doi = {10.3847/2041-8213/ad1691},
archivePrefix = {arXiv},
       eprint = {2310.15490},
 primaryClass = {astro-ph.EP},
       adsurl = {https://ui.adsabs.harvard.edu/abs/2024ApJ...960L...3C}
}

@ARTICLE{trappist1h,
       author = {{Gillon}, Micha{\"e}l and {Triaud}, Amaury H.~M.~J. and {Demory}, Brice-Olivier and {Jehin}, Emmanu{\"e}l and {Agol}, Eric and {Deck}, Katherine M. and {Lederer}, Susan M. and {de Wit}, Julien and {Burdanov}, Artem and {Ingalls}, James G. and {Bolmont}, Emeline and {Leconte}, Jeremy and {Raymond}, Sean N. and {Selsis}, Franck and {Turbet}, Martin and {Barkaoui}, Khalid and {Burgasser}, Adam and {Burleigh}, Matthew R. and {Carey}, Sean J. and {Chaushev}, Aleksander and {Copperwheat}, Chris M. and {Delrez}, Laetitia and {Fernandes}, Catarina S. and {Holdsworth}, Daniel L. and {Kotze}, Enrico J. and {Van Grootel}, Val{\'e}rie and {Almleaky}, Yaseen and {Benkhaldoun}, Zouhair and {Magain}, Pierre and {Queloz}, Didier},
        title = "{Seven temperate terrestrial planets around the nearby ultracool dwarf star TRAPPIST-1}",
      journal = {\nat},
         year = 2017,
        month = feb,
       volume = {542},
       number = {7642},
        pages = {456-460},
          doi = {10.1038/nature21360},
archivePrefix = {arXiv},
       eprint = {1703.01424},
 primaryClass = {astro-ph.EP},
       adsurl = {https://ui.adsabs.harvard.edu/abs/2017Natur.542..456G}
}

@ARTICLE{k1512b,
       author = {{Torres}, Guillermo and {Kane}, Stephen R. and {Rowe}, Jason F. and {Batalha}, Natalie M. and {Henze}, Christopher E. and {Ciardi}, David R. and {Barclay}, Thomas and {Borucki}, William J. and {Buchhave}, Lars A. and {Crepp}, Justin R. and {Everett}, Mark E. and {Horch}, Elliott P. and {Howard}, Andrew W. and {Howell}, Steve B. and {Isaacson}, Howard T. and {Jenkins}, Jon M. and {Latham}, David W. and {Petigura}, Erik A. and {Quintana}, Elisa V.},
        title = "{Validation of Small Kepler Transiting Planet Candidates in or near the Habitable Zone}",
      journal = {\aj},
         year = 2017,
        month = dec,
       volume = {154},
       number = {6},
          eid = {264},
        pages = {264},
          doi = {10.3847/1538-3881/aa984b},
archivePrefix = {arXiv},
       eprint = {1711.01267},
 primaryClass = {astro-ph.EP},
       adsurl = {https://ui.adsabs.harvard.edu/abs/2017AJ....154..264T}
}

@ARTICLE{k54d,
       author = {{Dressing}, Courtney D. and {Charbonneau}, David},
        title = "{The Occurrence Rate of Small Planets around Small Stars}",
      journal = {\apj},
         year = 2013,
        month = apr,
       volume = {767},
       number = {1},
          eid = {95},
        pages = {95},
          doi = {10.1088/0004-637X/767/1/95},
archivePrefix = {arXiv},
       eprint = {1302.1647},
 primaryClass = {astro-ph.EP},
       adsurl = {https://ui.adsabs.harvard.edu/abs/2013ApJ...767...95D}
}

@ARTICLE{k272e,
       author = {{Crossfield}, Ian J.~M. and {Ciardi}, David R. and {Petigura}, Erik A. and {Sinukoff}, Evan and {Schlieder}, Joshua E. and {Howard}, Andrew W. and {Beichman}, Charles A. and {Isaacson}, Howard and {Dressing}, Courtney D. and {Christiansen}, Jessie L. and {Fulton}, Benjamin J. and {L{\'e}pine}, S{\'e}bastien and {Weiss}, Lauren and {Hirsch}, Lea and {Livingston}, John and {Baranec}, Christoph and {Law}, Nicholas M. and {Riddle}, Reed and {Ziegler}, Carl and {Howell}, Steve B. and {Horch}, Elliott and {Everett}, Mark and {Teske}, Johanna and {Martinez}, Arturo O. and {Obermeier}, Christian and {Benneke}, Bj{\"o}rn and {Scott}, Nic and {Deacon}, Niall and {Aller}, Kimberly M. and {Hansen}, Brad M.~S. and {Mancini}, Luigi and {Ciceri}, Simona and {Brahm}, Rafael and {Jord{\'a}n}, Andr{\'e}s and {Knutson}, Heather A. and {Henning}, Thomas and {Bonnefoy}, Micha{\"e}l and {Liu}, Michael C. and {Crepp}, Justin R. and {Lothringer}, Joshua and {Hinz}, Phil and {Bailey}, Vanessa and {Skemer}, Andrew and {Defrere}, Denis},
        title = "{197 Candidates and 104 Validated Planets in K2{\textquoteright}s First Five Fields}",
      journal = {\apjs},
         year = 2016,
        month = sep,
       volume = {226},
       number = {1},
          eid = {7},
        pages = {7},
          doi = {10.3847/0067-0049/226/1/7},
archivePrefix = {arXiv},
       eprint = {1607.05263},
 primaryClass = {astro-ph.EP},
       adsurl = {https://ui.adsabs.harvard.edu/abs/2016ApJS..226....7C}
}

@ARTICLE{toi1227b,
       author = {{Mann}, Andrew W. and {Wood}, Mackenna L. and {Schmidt}, Stephen P. and {Barber}, Madyson G. and {Owen}, James E. and {Tofflemire}, Benjamin M. and {Newton}, Elisabeth R. and {Mamajek}, Eric E. and {Bush}, Jonathan L. and {Mace}, Gregory N. and {Kraus}, Adam L. and {Thao}, Pa Chia and {Vanderburg}, Andrew and {Llama}, Joe and {Johns-Krull}, Christopher M. and {Prato}, L. and {Stahl}, Asa G. and {Tang}, Shih-Yun and {Fields}, Matthew J. and {Collins}, Karen A. and {Collins}, Kevin I. and {Gan}, Tianjun and {Jensen}, Eric L.~N. and {Kamler}, Jacob and {Schwarz}, Richard P. and {Furlan}, Elise and {Gnilka}, Crystal L. and {Howell}, Steve B. and {Lester}, Kathryn V. and {Owens}, Dylan A. and {Suarez}, Olga and {Mekarnia}, Djamel and {Guillot}, Tristan and {Abe}, Lyu and {Triaud}, Amaury H.~M.~J. and {Johnson}, Marshall C. and {Milburn}, Reilly P. and {Rizzuto}, Aaron C. and {Quinn}, Samuel N. and {Kerr}, Ronan and {Ricker}, George R. and {Vanderspek}, Roland and {Latham}, David W. and {Seager}, Sara and {Winn}, Joshua N. and {Jenkins}, Jon M. and {Guerrero}, Natalia M. and {Shporer}, Avi and {Schlieder}, Joshua E. and {McLean}, Brian and {Wohler}, Bill},
        title = "{TESS Hunt for Young and Maturing Exoplanets (THYME). VI. An 11 Myr Giant Planet Transiting a Very-low-mass Star in Lower Centaurus Crux}",
      journal = {\aj},
         year = 2022,
        month = apr,
       volume = {163},
       number = {4},
          eid = {156},
        pages = {156},
          doi = {10.3847/1538-3881/ac511d},
archivePrefix = {arXiv},
       eprint = {2110.09531},
 primaryClass = {astro-ph.EP},
       adsurl = {https://ui.adsabs.harvard.edu/abs/2022AJ....163..156M}
}

@ARTICLE{Irwin_2018,
       author = {{Irwin}, Jonathan M. and {Charbonneau}, David and {Esquerdo}, Gilbert A. and {Latham}, David W. and {Winters}, Jennifer G. and {Dittmann}, Jason A. and {Newton}, Elisabeth R. and {Berta-Thompson}, Zachory K. and {Berlind}, Perry and {Calkins}, Michael L.},
        title = "{Four New Eclipsing Mid M-dwarf Systems from the New Luyten Two Tenths Catalog}",
      journal = {\aj},
         year = 2018,
        month = oct,
       volume = {156},
       number = {4},
          eid = {140},
        pages = {140},
          doi = {10.3847/1538-3881/aad9a3},
archivePrefix = {arXiv},
       eprint = {1808.03243},
 primaryClass = {astro-ph.SR},
       adsurl = {https://ui.adsabs.harvard.edu/abs/2018AJ....156..140I}
}

@ARTICLE{neptune_4,
       author = {{Scott}, Madison G. and {Dransfield}, Georgina and {Timmermans}, Mathilde and {Triaud}, Amaury H.~M.~J. and {Rackham}, Benjamin V. and {Barkaoui}, Khalid and {Burgasser}, Adam J. and {Collins}, Karen A. and {Gillon}, Micha{\"e}l and {Howell}, Steve B. and {Levine}, Alan M. and {Pozuelos}, Francisco J. and {Stassun}, Keivan G. and {Ziegler}, Carl and {Chew}, Yilen Gomez Maqueo and {Clark}, Catherine A. and {Davis}, Yasmin and {Davoudi}, Fatemeh and {Daylan}, Tansu and {Demory}, Brice-Olivier and {Feliz}, Dax and {Fukui}, Akihiko and {G{\"u}nther}, Maximilian N. and {Jehin}, Emmanu{\"e}l and {Lienhard}, Florian and {Mann}, Andrew W. and {Mu{\~n}oz}, Cl{\`a}udia Jan{\'o} and {Narita}, Norio and {Pedersen}, Peter P. and {Schwarz}, Richard P. and {Shporer}, Avi and {Soubkiou}, Abderahmane and {Z{\'u}{\~n}iga-Fern{\'a}ndez}, Sebasti{\'a}n},
        title = "{Two temperate Earth- and Neptune-sized planets orbiting fully convective M dwarfs}",
      journal = {\mnras},
         year = 2026,
        month = jan,
          doi = {10.1093/mnras/stag070},
archivePrefix = {arXiv},
       eprint = {2601.05799},
 primaryClass = {astro-ph.EP},
       adsurl = {https://ui.adsabs.harvard.edu/abs/2026MNRAS.tmp...63S}
}

@ARTICLE{LTT_3780,
       author = {{Cloutier}, Ryan and {Eastman}, Jason D. and {Rodriguez}, Joseph E. and {Astudillo-Defru}, Nicola and {Bonfils}, Xavier and {Mortier}, Annelies and {Watson}, Christopher A. and {Stalport}, Manu and {Pinamonti}, Matteo and {Lienhard}, Florian and {Harutyunyan}, Avet and {Damasso}, Mario and {Latham}, David W. and {Collins}, Karen A. and {Massey}, Robert and {Irwin}, Jonathan and {Winters}, Jennifer G. and {Charbonneau}, David and {Ziegler}, Carl and {Matthews}, Elisabeth and {Crossfield}, Ian J.~M. and {Kreidberg}, Laura and {Quinn}, Samuel N. and {Ricker}, George and {Vanderspek}, Roland and {Seager}, Sara and {Winn}, Joshua and {Jenkins}, Jon M. and {Vezie}, Michael and {Udry}, St{\'e}phane and {Twicken}, Joseph D. and {Tenenbaum}, Peter and {Sozzetti}, Alessandro and {S{\'e}gransan}, Damien and {Schlieder}, Joshua E. and {Sasselov}, Dimitar and {Santos}, Nuno C. and {Rice}, Ken and {Rackham}, Benjamin V. and {Poretti}, Ennio and {Piotto}, Giampaolo and {Phillips}, David and {Pepe}, Francesco and {Molinari}, Emilio and {Mignon}, Lucile and {Micela}, Giuseppina and {Melo}, Claudio and {de Medeiros}, Jos{\'e} R. and {Mayor}, Michel and {Matson}, Rachel A. and {Martinez Fiorenzano}, Aldo F. and {Mann}, Andrew W. and {Magazz{\'u}}, Antonio and {Lovis}, Christophe and {L{\'o}pez-Morales}, Mercedes and {Lopez}, Eric and {Lissauer}, Jack J. and {L{\'e}pine}, S{\'e}bastien and {Law}, Nicholas and {Kielkopf}, John F. and {Johnson}, John A. and {Jensen}, Eric L.~N. and {Howell}, Steve B. and {Gonzales}, Erica and {Ghedina}, Adriano and {Forveille}, Thierry and {Figueira}, Pedro and {Dumusque}, Xavier and {Dressing}, Courtney D. and {Doyon}, Ren{\'e} and {D{\'\i}az}, Rodrigo F. and {Fabrizio}, Luca Di and {Delfosse}, Xavier and {Cosentino}, Rosario and {Conti}, Dennis M. and {Collins}, Kevin I. and {Cameron}, Andrew Collier and {Ciardi}, David and {Caldwell}, Douglas A. and {Burke}, Christopher and {Buchhave}, Lars and {Brice{\~n}o}, C{\'e}sar and {Boyd}, Patricia and {Bouchy}, Fran{\c{c}}ois and {Beichman}, Charles and {Artigau}, {\'E}tienne and {Almenara}, Jose M.},
        title = "{A Pair of TESS Planets Spanning the Radius Valley around the Nearby Mid-M Dwarf LTT 3780}",
      journal = {\aj},
         year = 2020,
        month = jul,
       volume = {160},
       number = {1},
          eid = {3},
        pages = {3},
          doi = {10.3847/1538-3881/ab91c2},
archivePrefix = {arXiv},
       eprint = {2003.01136},
 primaryClass = {astro-ph.EP},
       adsurl = {https://ui.adsabs.harvard.edu/abs/2020AJ....160....3C}
}

@ARTICLE{exoplanet_archive,
       author = {{Christiansen}, Jessie L. and {McElroy}, Douglas L. and {Harbut}, Marcy and {Ciardi}, David R. and {Crane}, Megan and {Good}, John and {Hardegree-Ullman}, Kevin K. and {Kesseli}, Aurora Y. and {Lund}, Michael B. and {Lynn}, Meca and et al.},
        title = "{The NASA Exoplanet Archive and Exoplanet Follow-up Observing Program: Data, Tools, and Usage}",
      journal = {\psj},
         year = 2025,
        month = aug,
       volume = {6},
       number = {8},
          eid = {186},
        pages = {186},
          doi = {10.3847/PSJ/ade3c2},
archivePrefix = {arXiv},
       eprint = {2506.03299},
 primaryClass = {astro-ph.EP},
       adsurl = {https://ui.adsabs.harvard.edu/abs/2025PSJ.....6..186C}
}

@misc{10.17909/fwdt-2x66,
  author = {{TESS Team}},
  title = {TESS Input Catalog and Candidate Target List},
  year = {2018},
  doi = {10.17909/fwdt-2x66},
  publisher = {STScI/MAST},
  url = {https://doi.org/10.17909/fwdt-2x66}
}

@misc{10.17909/t9-nmc8-f686,
  author = {{TESS Team}},
  title = {TESS Light Curves - All Sectors},
  year = {2021},
  doi = {10.17909/t9-nmc8-f686},
  publisher = {STScI/MAST},
  url = {https://doi.org/10.17909/t9-nmc8-f686}
}

@misc{10.17909/0cp4-2j79,
  author = {{TESS Team}},
  title = {TESS Calibrated Full Frame Images: All Sectors},
  year = {2022},
  doi = {10.17909/0cp4-2j79},
  publisher = {STScI/MAST},
  url = {https://doi.org/10.17909/0cp4-2j79}
}

@misc{10.26134/exofop5,
  author = {{NExScI}},
  title = {Exoplanet Follow-up Observing Program Web Service},
  year = {2022},
  doi = {10.26134/EXOFOP5},
  publisher = {IPAC},
  url = {https://doi.org/10.26134/EXOFOP5}
}

@misc{10.26134/exofop3,
  author = {{ExoFOP}},
  title = {Exoplanet Follow-up Observing Program - TESS},
  year = {2019},
  doi = {10.26134/EXOFOP3},
  publisher = {IPAC},
  url = {https://doi.org/10.26134/EXOFOP3}
}

@ARTICLE{brady2022,
       author = {{Brady}, Madison T. and {Bean}, Jacob L.},
        title = "{Assessing the Transiting Exoplanet Survey Satellite's Yield of Rocky Planets Around Nearby M Dwarfs}",
      journal = {\aj},
         year = 2022,
        month = jun,
       volume = {163},
       number = {6},
          eid = {255},
        pages = {255},
          doi = {10.3847/1538-3881/ac64a0},
archivePrefix = {arXiv},
       eprint = {2112.08337},
 primaryClass = {astro-ph.EP},
       adsurl = {https://ui.adsabs.harvard.edu/abs/2022AJ....163..255B}
}

\end{document}